\documentclass[11pt,a4paper]{article}
\usepackage{jcappub}
\usepackage{verbatim}
\usepackage{cleveref}

\usepackage[T1]{fontenc}
\usepackage[utf8]{inputenc}
\usepackage[american]{babel}
\usepackage{epsfig}
\usepackage{graphicx}
\usepackage{booktabs}
\usepackage{multirow}
\usepackage{dcolumn}
\usepackage{amsmath}
\usepackage{mathtools}
\usepackage{amsfonts}
\usepackage{amssymb}
\usepackage{epstopdf}
\usepackage{bm}
\usepackage{siunitx}
\usepackage{braket}
\usepackage{enumitem}
\usepackage{soul}
\usepackage[table]{xcolor}
\usepackage{color}
\usepackage{transparent}
\usepackage{comment}
\usepackage{pifont}
\usepackage{soul}

\definecolor{navyblue}{rgb}{0.0, 0.0, 0.5}
\definecolor{royalblue}{rgb}{0.25, 0.41, 0.88}
\definecolor{cadmiumgreen}{rgb}{0.0, 0.42, 0.24}
\definecolor{blue-violet}{rgb}{0.54, 0.17, 0.89}
\definecolor{darkviolet}{rgb}{0.58, 0.0, 0.83}
\definecolor{orange(colorwheel)}{rgb}{1.0, 0.5, 0.0}
\definecolor{burgundy}{rgb}{0.5, 0.0, 0.13}

\usepackage{hyperref}
\hypersetup{
    colorlinks=true, 
    linkcolor=royalblue, 
    citecolor=magenta}

\newcommand\ee{\end{equation}}
\newcommand\be{\begin{equation}}
\newcommand\eea{\end{eqnarray}}
\newcommand\bea{\begin{eqnarray}}

\usepackage{booktabs}
\usepackage{multirow}
\usepackage{dcolumn}
\usepackage{colortbl}

\definecolor{magenta(process)}{rgb}{1.0, 0.0, 0.56}

\definecolor{darkspringgreen}{rgb}{0.09, 0.45, 0.27}

\definecolor{royalblue(web)}{rgb}{0.25, 0.41, 0.88}

\begin{document}

\title{Robust Preference for Dynamical Dark Energy in DESI BAO and SN Measurements}

\author[a,1]{William Giar\`e, \note{Corresponding author.}}
\author[b,c]{Mahdi Najafi,}
\author[d,e]{Supriya Pan,}
\author[a]{Eleonora Di Valentino,}
\author[b,c,f]{Javad T. Firouzjaee}

\affiliation[a]{School of Mathematics and Statistics, University of Sheffield, Hounsfield Road, Sheffield S3 7RH, United Kingdom}
\affiliation[b]{Department of Physics, K.N. Toosi University of Technology, P.O. Box 15875-4416, Tehran, Iran}
\affiliation[c]{PDAT Laboratory, Department of Physics, K.N. Toosi University of Technology, P.O. Box 15875-4416, Tehran, Iran}
\affiliation[d]{Department of Mathematics, Presidency University, 86/1 College Street,  Kolkata 700073, India}
\affiliation[e]{Institute of Systems Science, Durban University of Technology, PO Box 1334, Durban 4000, Republic of South Africa}
\affiliation[f]{School of Physics, Institute for Research in Fundamental Sciences (IPM), P.O. Box 19395-5531, Tehran, Iran}

\emailAdd{w.giare@sheffield.ac.uk}
\emailAdd{mahdinajafi12676@yahoo.com}
\emailAdd{supriya.maths@presiuniv.ac.in}
\emailAdd{e.divalentino@sheffield.ac.uk}
\emailAdd{firouzjaee@kntu.ac.ir}

\date{\today}

\abstract{Recent Baryon Acoustic Oscillation (BAO) measurements released by DESI, when combined with Cosmic Microwave Background (CMB) data from Planck and two different samples of Type Ia supernovae (Pantheon-Plus and DESY5) reveal a preference for Dynamical Dark Energy (DDE) characterized by a present-day quintessence-like equation of state that crossed into the phantom regime in the past. A core \textit{ansatz} for this result is assuming a linear Chevallier-Polarski-Linder (CPL) parameterization $w(a) = w_0 + w_a (1-a)$ to describe the evolution of the DE equation of state (EoS). In this paper, we test if and to what extent this assumption impacts the results. To prevent broadening uncertainties in cosmological parameter inference and facilitate direct comparison with the baseline CPL case, we focus on 4 alternative well-known models that, just like CPL, consist of only two free parameters: the present-day DE EoS ($w_0$) and a parameter quantifying its dynamical evolution ($w_a$). We demonstrate that the preference for DDE remains robust regardless of the parameterization: $w_0$ consistently remains in the quintessence regime, while $w_a$ consistently indicates a preference for a dynamical evolution towards the phantom regime. This tendency is significantly strengthened by DESY5 SN measurements. By comparing the best-fit $\chi^2$ obtained within each DDE model, we notice that the linear CPL parameterization is not the best-fitting case. Among the models considered, the EoS proposed by Barboza and Alcaniz consistently leads to the most significant improvement.}

\maketitle

\section{Introduction}

One of the most undoubtedly fascinating and unforeseen discoveries of the past three decades is that the Universe is undergoing an accelerated phase of expansion. This was first argued in 1998 through observations of distant Type Ia Supernovae~\cite{SupernovaSearchTeam:1998fmf,SupernovaCosmologyProject:1998vns}, and has since been corroborated by a wide variety of other probes~\cite{SupernovaSearchTeam:2001qse,SDSS:2003eyi,SDSS:2003lnz,SupernovaSearchTeam:2003cyd,SupernovaCosmologyProject:2003dcn,SDSS:2004kqt,Feng:2004ad,SupernovaSearchTeam:2004lze,SNLS:2005qlf,SDSS:2005xqv,Eisenstein:2006nk,SDSS:2006lmn,Sahni:2006pa,ESSENCE:2007acn,Vikhlinin:2008ym,Stern:2009ep,Sherwin:2011gv,WMAP:2012fli,WMAP:2012nax,BOSS:2012dmf,deJong:2012zb,BOSS:2013rlg,Weinberg:2013agg,BOSS:2013uda,BOSS:2014hwf,SDSS:2014iwm,BOSS:2014hhw,Ross:2014qpa,Moresco:2016mzx,Moresco:2016nqq,Rubin:2016iqe,BOSS:2016wmc,DES:2016jjg,Haridasu:2017lma,DES:2017qwj,Pan-STARRS1:2017jku,Planck:2018nkj,Planck:2018vyg,Gomez-Valent:2018gvm, Yang:2019fjt,ACT:2020frw,ACT:2020gnv,eBOSS:2020yzd,Nadathur:2020kvq,Rose:2020shp,DiValentino:2020evt,eBOSS:2020yzd,KiDS:2020suj,KiDS:2020ghu,SPT-3G:2021eoc,DES:2021wwk,Moresco:2022phi,DES:2022ccp,Brout:2022vxf,ACT:2023kun,Kilo-DegreeSurvey:2023gfr,DESI:2024uvr,DESI:2024kob,DES:2024tys,DES:2024upw,DES:2024hip}.\footnote{For a few caveats, objections, and discussions surrounding  this conclusion raised over the years, see Refs.~\cite{Buchert:1999er,Buchert:2001sa,Buchert:2007ik,Hunt:2008wp,Nielsen:2015pga,Tutusaus:2017ibk,Dam:2017xqs,Colin:2019opb,Desgrange:2019npu,Koksbang:2019cen,Koksbang:2019glb,Heinesen:2022lqs}.}

Since all known forces and components in nature would decelerate the expansion rate of the Universe, acceleration itself requires a physical mechanism beyond the Standard Model of fundamental interactions, able to counteract deceleration, inducing instead a Dark Energy (DE) phase where the dynamics is characterized by negative pressure with an effective Equation of State (EoS) $w < -1/3$, a condition that directly follows from the second Friedmann equation (also known as the acceleration equation)~\cite{Sahni:1999gb,Carroll:2000fy,Peebles:2002gy,Padmanabhan:2002ji,Copeland:2006wr,Caldwell:2009ix,Li:2011sd,Martin:2012bt}.

In the theoretical framework described by the standard $\Lambda$CDM model of cosmology, DE is parametrized by a positive cosmological constant term ($\Lambda$) in the Einstein equation with its energy density comprising the vast majority of the Universe's energy budget. Despite its apparent simplicity, this interpretation is not free from conceptual problems and limitations~\cite{Weinberg:1988cp,Krauss:1995yb,Sahni:1999gb,Weinberg:2000yb,Padmanabhan:2002ji,Sahni:2002kh,Yokoyama:2003ii,Nobbenhuis:2004wn,Burgess:2013ara,Joyce:2014kja,Bull:2015stt,Wang:2016lxa}. Foremost, plugging a positive cosmological constant component into the gravitational field equations described by General Relativity (GR) implies living in an asymptotically de Sitter universe, which seems to contrast with several theories/models of quantum gravity proposing instead an asymptotically anti-de Sitter universe~\cite{Brustein:1992nk,Witten:2000zk,Kachru:2003aw,Polchinski:2006gy,Danielsson:2018ztv}. Secondly, it seems quite natural to question what made us so lucky to live precisely in the cosmic epoch when such a constant component came to be not only relevant but even dominant compared to other contributions such as matter, altering the expansion history of the Universe so prominently as to allow us to become easily aware of its existence and implications~\cite{Zlatev:1998tr,Pavon:2005yx,Martin:2012bt,Velten:2014nra}.
Finally, and most importantly, when it comes to the physical interpretations, anything that contributes to the energy density of the vacuum behaves akin to a cosmological constant, summing up within the energy-momentum tensor as $T_{\mu\nu} \propto g_{\mu\nu} \rho$ due to Lorentz invariance. Based on standard quantum field theory calculations, one would expect a zero-point energy density contribution of $\rho_{\rm vac}$, which, depending on the ultraviolet wavelength cutoff scale, is found to be somewhere between $10^{50}$ to $10^{120}$ orders of magnitude larger than what is inferred by cosmological data~\cite{Weinberg:1988cp}. This leads to one of the biggest disagreements between theoretical predictions and observations, arguably requiring a level of fine-tuning that appears to be well beyond what any current theories can realistically explain~\cite{Dolgov:1997za,Weinberg:1988cp,Straumann:1999ia,Padmanabhan:2002ji,Sola:2013gha}. As a result, from a theoretical perspective, the nature of DE remains one of the biggest puzzles in modern physics, sustaining significant research interest in the high-energy physics community.\footnote{This diffuse interest is reflected in the wide range of models -- both within and beyond the standard cosmological constant -- that have been proposed over the years. These include, for example, new (ultra)light fields and modifications to gravity. With no claims to completeness see, e.g., Refs.~\cite{Amendola:1999er,Kamenshchik:2001cp,Capozziello:2002rd,Bento:2002ps,Mangano:2002gg,Farrar:2003uw,Khoury:2003aq,Li:2004rb,Amendola:2006we,Hu:2007nk,Cognola:2007zu,Nojiri:2010pw,Zhang:2011uv,Rinaldi:2014yta,Luongo:2014nld,Rinaldi:2015iza,DeFelice:2016yws,Wang:2016lxa,Josset:2016vrq,Burrage:2016bwy,Sebastiani:2016ras,Nojiri:2017ncd,Burrage:2017qrf,Capozziello:2017buj,Benisty:2018qed,Casalino:2018tcd,Yang:2018euj,Saridakis:2018unr,Visinelli:2018utg,Langlois:2018dxi,Benisty:2018oyy,Boshkayev:2019qcx,Heckman:2019dsj,DAgostino:2019wko,Mukhopadhyay:2019wrw,Mukhopadhyay:2019cai,Mukhopadhyay:2019jla,Vagnozzi:2019kvw,Akarsu:2019hmw,Saridakis:2020zol,Ruchika:2020avj,Odintsov:2020zct,Odintsov:2020vjb,Oikonomou:2020qah,Oikonomou:2020oex,Vagnozzi:2021quy,Solanki:2021qni,Saridakis:2021qxb,Arora:2021tuh,Capozziello:2022jbw,Narawade:2022jeg,DAgostino:2022fcx,Oikonomou:2022wuk,Belfiglio:2022egm,Luciano:2022hhy,Kadam:2022yrj,Ong:2022wrs,Bernui:2023byc,Luciano:2023wtx,Giani:2023tai,Belfiglio:2023rxb,Frion:2023xwq,Adil:2023ara,Halder:2024gag,Fischer:2024eic}.}

From an observational standpoint, investigating the nature of DE has sparked research interest comparable to that driven by the theoretical problems surrounding its physical interpretation~\cite{Cooray:1999da,Efstathiou:1999tm,Chevallier:2000qy,Melchiorri:2002ux,Linder:2002et,Wetterich:2004pv,Feng:2004ff,Xia:2004rw,Hannestad:2004cb,Gong:2005de,Jassal:2005qc,Nesseris:2005ur,Liu:2008vy,Barboza:2008rh,Ma:2011nc,Sendra:2011pt,DeFelice:2012vd,Li:2012vn,Feng:2012gf,Magana:2014voa,Pantazis:2016nky,DiValentino:2016hlg,Yang:2017alx,Pan:2017zoh,Mortsell:2018mfj,Dutta:2018vmq,Yang:2018prh,Yang:2018qmz,Li:2019yem,Vagnozzi:2019ezj,Visinelli:2019qqu,DiValentino:2019ffd,Dutta:2019pio,DiValentino:2019jae,Pan:2019hac,Teixeira:2019hil,Teixeira:2019tfi,Martinelli:2019dau,Ruchika:2020avj,Zumalacarregui:2020cjh,Hogg:2020rdp,Alestas:2020mvb,DiValentino:2020naf,Alestas:2020zol,Rezaei:2020mrj,Perkovic:2020mph,Benaoum:2020qsi,Kumar:2021eev,Vagnozzi:2021tjv,Escamilla:2021uoj,Bag:2021cqm,Theodoropoulos:2021hkk,Alestas:2021luu,Sen:2021wld,Yang:2021eud,Hogg:2021yiz,Roy:2022fif,Heisenberg:2022lob,Chudaykin:2022rnl,Akarsu:2022typ,Santos:2022atq,Schiavone:2022wvq,vandeBruck:2022xbk,Ozulker:2022slu,Teixeira:2022sjr,Ben-Dayan:2023rgt,Ballardini:2023mzm,Yang:2022kho,deCruzPerez:2023wzd,Patil:2023rqy,Zhai:2023yny,Adil:2023exv,Montani:2023xpd,Akarsu:2023mfb,Vagnozzi:2023nrq,Avsajanishvili:2023jcl,Giani:2023aor,Lazkoz:2023oqc,Escamilla:2023oce,Escamilla:2023shf,Rezaei:2023xkj,Teixeira:2023zjt,Forconi:2023hsj,Sebastianutti:2023dbt,Wolf:2023uno,Giare:2024sdl,Giare:2024smz,Giare:2024ytc,Menci:2024rbq,Adil:2023ara,Akarsu:2024qiq,Teixeira:2024wsw,Benisty:2024lmj,Najafi:2024qzm,Moshafi:2024guo,Silva:2024ift,Reyhani:2024cnr,Escamilla:2024olw,Wang:2024sgo,Montani:2024xys,Li:2024qso,Yang:2024kdo,Dwivedi:2024okk}. Increasingly precise observations of the Cosmic Microwave Background (CMB) radiation, obtained from experiments such as WMAP~\cite{WMAP:2012fli,WMAP:2012nax} and, more recently, the Planck satellite~\cite{Planck:2018nkj,Planck:2018vyg}, as well as the Atacama Cosmology Telescope (ACT)~\cite{ACT:2020frw,ACT:2020gnv,ACT:2023kun,ACT:2023dou} and the South Pole Telescope (SPT)~\cite{SPT-3G:2014dbx,SPT-3G:2021eoc,SPT-3G:2022hvq}, have provided extremely accurate measurements of the angular power spectra of temperature and polarization anisotropies, revealing a precise snapshot of the Universe at the last scattering surface $z \sim 1100$.\footnote{Notably, the gravitational deflection, or lensing, experienced by CMB photons due to their interactions with the large-scale structure of the Universe imprints a distinctive non-Gaussian four-point correlation function (trispectrum) in both temperature and polarization anisotropies~\cite{Lewis:2006fu}. This signal provides complementary information about late-time processes affecting structure formation, from neutrinos and thermal relics~\cite{Giare:2020vzo,DiValentino:2021imh,DEramo:2022nvb,DiValentino:2022edq,Giare:2023aix} to dark energy and its dynamical properties~\cite{Planck:2018lbu,Ye:2023zel,ACT:2023skz,ACT:2023ipp,ACT:2023kun,ACT:2024okh,Sailer:2024coh,ACT:2024npz}.} Concurrently, progress in observational astronomy and astrophysics has culminated in a series of present surveys aimed at determining the properties of the Universe at low redshift through a multitude of probes, including -- but not limited to -- Baryon Acoustic Oscillations (BAO) and Type Ia Supernovae (SN) measurements. \footnote{Excitingly, upcoming Stage-IV astronomical surveys such as future data releases from DESI, Euclid \cite{EUCLID:2011zbd}, the Large Synoptic Survey Telescope (LSST) \cite{LSST:2008ijt}, the Wide-Field InfraRed Survey Telescope (WFIRST)~\cite{Spergel:2013tha}, and the Square Kilometre Array (SKA)~\cite{SKA:2018ckk}, are expected to improve upon current sensitivity and are forecasted to constrain DE parameters to near-percent precision, offering new insights into the dark sector of the Universe.}
These collective efforts have ushered in a new era of precision cosmology, eventually allowing percentage-level precision in cosmological parameter inference and enabling precise tests of physics within and beyond the $\Lambda$CDM framework~\cite{DiValentino:2021izs,Perivolaropoulos:2021jda,Abdalla:2022yfr}.

Despite these remarkable achievements, it is no exaggeration to say that at the time of writing, observations are inconclusive about the physical nature of DE.  While too far-away deviations from the canonical $\Lambda$CDM model appear severely constrained~\cite{Gariazzo:2024sil}, several alternative theoretical frameworks and phenomenological avenues featuring new physics in the dark sector of the model remain at the very least plausible.

Trying to summarize an otherwise very articulated debate, we can adhere to Occam's razor principle and start considering one of the simplest hypotheses beyond the cosmological constant. This involves assuming that DE can be modeled as a generic fluid with a constant EoS, $w_0$. By leaving $w_0$ as a free parameter in the theoretical model, cosmological observations can constrain deviations from the cosmological constant value ($w_0=-1$).\footnote{See, e.g., Refs~\cite{Bean:2001xy,Hannestad:2002ur,Said:2013jxa,Shafer:2013pxa,Zhang:2015uhk,Moresco:2016nqq,Xu:2016grp,Wang:2016tsz,Vagnozzi:2017ovm,Zhang:2017rbg,Feng:2017mfs,Pan-STARRS1:2017jku,Wang:2018ahw,Sprenger:2018tdb,Poulin:2018zxs,RoyChoudhury:2018vnm,DES:2018ufa,Planck:2018vyg,Wang:2019acf,RoyChoudhury:2019hls,eBOSS:2020yzd,Chudaykin:2020ghx,DAmico:2020kxu,Vagnozzi:2020dfn,Yang:2021flj,DiValentino:2020vnx,DES:2022ccp,KiDS:2020ghu,Brieden:2022lsd,Grillo:2020yvj,Cao:2021cix,Zhang:2021yof,Colgain:2021pmf,Teng:2021cvy,Krishnan:2021dyb,Nunes:2021ipq,Bernardo:2021cxi,Vagnozzi:2021tjv,Bargiacchi:2021hdp,Moresco:2022phi,Semenaite:2022unt,Carrilho:2022mon,Wang:2022xdw,Koussour:2022jyk,Bernardo:2022pyz,Narawade:2022cgb,Hou:2022rvk,Escamilla:2023oce,Kumar:2023bqj,Bhagat:2023ych,Mussatayeva:2023aoa,Hogg:2023khs,DESI:2024mwx} for recent and not-so-recent discussions and constraints on the DE EoS from a variety of astrophysical and cosmological probes.} In this regard, focusing exclusively on the Planck CMB data one might speculate about a preference for a phantom-like DE component ($w_0 < -1$)~\cite{Planck:2018vyg,Escamilla:2023oce}.\footnote{In recent years, the possibility that the DE EoS can be phantom in nature has gained substantial interest, as in principle a shift of $\sim20\%$ towards $w_0<-1$ could already be enough to address the well-known Hubble tension~\cite{Verde:2019ivm,DiValentino:2021izs,Perivolaropoulos:2021jda,Abdalla:2022yfr,Hu:2023jqc} -- see, e.g., Refs.~\cite{Vagnozzi:2019ezj,Giare:2024akf} for an overview, as well as for caveats surrounding this possibility.} However, as extensively documented in the literature, this preference is not confirmed by independent CMB experiments such as ACT and SPT~\cite{ACT:2020gnv,SPT-3G:2021eoc,SPT-3G:2022hvq,Giare:2023xoc}, and  -- most importantly -- it lacks consistent support from observations of the local universe. When combining CMB, BAO and SN data altogether, no convincing deviation from $w_0=-1$ is seen, possibly lending weight to the cosmological constant interpretation~\cite{Escamilla:2023oce}.

However, one may argue that simplicity may not always be a prerogative of nature. Pushing this approach forward, we can relax the assumption of a constant EoS and consider models where $w(a)$ varies with the universe's expansion -- here and after known as Dynamical Dark Energy (DDE).  This possibility, along with the various proposed physical realizations, has also undergone extensive testing; see, e.g., Refs.~\cite{Mainini:2003uf,Alam:2004jy,Sola:2005nh,Antoniadis:2006wq,Yang:2021flj,Zhao:2012aw,SolaPeracaula:2016qlq,Sola:2016hnq,Zhao:2017cud,Yang:2017yme,Yang:2018prh,Yang:2018qmz,SolaPeracaula:2018wwm,Pan:2019gop,Colgain:2021pmf,Escamilla-Rivera:2021boq,Zhao:2020ole}. From an observational standpoint, CMB data alone have limited capacity to constrain DDE models due to minimal effects left at the epoch of the last scattering surface and the increased number of cosmological parameters~\cite{DiValentino:2017zyq,DiValentino:2019dzu,DiValentino:2022oon}.\footnote{This difficulty is often referred to as geometrical degeneracy. At its core, the problem is that different combinations of late-time cosmic parameters can be adjusted in such a way that the acoustic angular scale $\theta_s$  -- determined by the ratio of the comoving sound horizon at recombination to the comoving distance to last scattering -- remains constant if both quantities change proportionally. Consequently, measurements based solely on this scale cannot provide strong constraints on (dynamical) DE parameters by themselves, unless perturbation-level effects and late-time data are also incorporated to break this degeneracy.} Even in simple parametrizations aimed at minimizing the number of free degrees of freedom, CMB experiments produce constraints that are typically too broad to provide informative results. Therefore, local universe observations acquire primary importance. 

A significant turning point in the study of DDE models is marked by the very recent BAO release from the Dark Energy Spectroscopic Instrument (DESI) \cite{DESI:2024uvr,DESI:2024kob}, and -- albeit to a lesser extent -- by SN distance moduli measurements from the Union3 compilation~\cite{Rubin:2023ovl} first and, more recently, from the five-year observations of the Dark Energy Survey (DESY5)~\cite{DES:2024tys,DES:2024upw,DES:2024hip}. Before these updated data, no significant preference had ever emerged in favor of DDE models, certainly not to the extent required to challenge the baseline cosmological constant interpretation~\cite{Gariazzo:2024sil}. Excitingly, when DESI BAO observations are combined with Planck CMB data and SN distance moduli measurements (whether from the Pantheon-plus catalog~\cite{Scolnic:2021amr, Brout:2022vxf}, the Union3 compilation~\cite{Rubin:2023ovl}, or DESY5~\cite{DES:2024tys,DES:2024upw,DES:2024hip}), they produce strong indications for DDE. Specifically, within the linear Chevallier-Polarski-Linder (CPL) parameterization of the DE EoS, $w(a) = w_0 + w_a(1 - a)$~\cite{Chevallier:2000qy,Linder:2002et} -- where $w_0$ represents the present-day DE EoS and $w_a$ quantifies the dynamical evolution -- we observe a preference for $w_0 > -1$ and $w_a < 0$ at a statistical level ranging between 2.5 and 3.9$\sigma$, depending on the specific combination of SN data used~\cite{DESI:2024mwx}.

Unexpectedly, these results have heated up the recent debate, fueling a multitude of re-analyses and re-interpretations~\cite{Cortes:2024lgw,Patel:2024odo,Orchard:2024bve,Liu:2024gfy,Chudaykin:2024gol,Notari:2024rti,Gialamas:2024lyw,Wang:2024hwd,DESI:2024kob,Wang:2024dka,Giare:2024smz,Carloni:2024zpl,Colgain:2024xqj,Tada:2024znt,Yin:2024hba,Luongo:2024fww,Park:2024jns,Wang:2024sgo,Shlivko:2024llw,Ye:2024ywg,Li:2024qso,Yang:2024kdo}. As ``extraordinary claims require extraordinary evidence'', we certainly advise and exercise caution. However, even taking in mind all the possible caveats and limitations surrounding the first DESI data release, it is undeniable that a high statistical preference for DDE holds intrinsic interest -- if confirmed, this would represent the first concrete evidence of new physics beyond the standard model of cosmology. 

Given the potential implications of this result, it is natural to question its robustness. Barring any possible systematic issues in these datasets\footnote{As argued in various recent works, the DESI BAO measurement at $ z = 0.71 $ (which is in $\sim 3\sigma$ tension with Planck) can play a crucial role in deriving many of the DESI signals for new physics, partially including the preference for DDE~\cite{Wang:2024pui,Colgain:2024xqj,Naredo-Tuero:2024sgf}. }, we examine if and to what extent assuming a linear parameterization for the DE EoS might impact the current findings. Although the CPL parameterization has been demonstrated to match the background evolution of distances arising from exact DE equations of motion to an accuracy of approximately 0.1\% for viable cosmologies over a wide range of physics, including scalar fields, modified gravity, and phase transitions (see, e.g., Refs.~\cite{Linder:2002et,dePutter:2008wt}), other parameterizations proposed over the years (which may deviate from CPL at both $z \ll 1$ and $z \gtrsim 1$) remain allowed by current observations. Testing these alternative models against new data can certainly represent a useful exercise to shed light on the role played by the parameterization itself. To be fair, the process has already begun with several independent groups actively engaged in this activity~\cite{Hernandez-Almada:2024ost,Pourojaghi:2024tmw,Ramadan:2024kmn,Berghaus:2024kra,Qu:2024lpx,Notari:2024rti,Adolf:2024twn}.

Given the vast number of parameterizations proposed over the years and recently analyzed in relation to the DESI data, a few warnings are in order. First and foremost, alternative parameterizations often introduce extra parameters compared to CPL. This is both a blessing and a curse: on the one hand, accounting for more degrees of freedom allows more flexibility in $w(a)$. On the other hand, this typically implies relaxing the overall constraining power due to the combined effects of degeneracies and correlations among parameters. Secondly, the physical interpretations of the parameters involved may differ from the two employed in the CPL model. This further compounds the comparison of the results, making it difficult to derive general guidelines on the preference towards DDE.

To overcome these difficulties, in this article, we test different DDE models while allowing for a fair comparison of the results. We restrict our attention to five well-known parameterizations that satisfy the following criteria: \textit{(i)} they consist of two parameters to describe the evolution of $w(a)$, $w_0$ and $w_a$; \textit{(ii)} these two parameters retain the same physical meaning as in the CPL parameterization; \textit{(iii)} for the same combinations of pairs $w_0$ - $w_a$, the resulting shape of $w(a)$ deviates from CPL either near the present epoch or in the past, depending on the specific case.

The paper is organized as follows. In Sec.~\ref{sec-2} we sketch the theoretical set-up of the gravitational equations and propose the DE parametrizations we wish to study. In Sec.~\ref{sec-data} we describe the observational data and the methodology to constrain the proposed DE parametrizations. In Sec.~\ref{sec-results} we present the constraints on the resulting DE scenarios. Finally, in Sec.~\ref{sec-summary} we draw our general conclusions.

\section{Dynamical Dark Energy Models}
\label{sec-2}

Considering a DDE component in cosmological models produces changes both in the background dynamics and in the dynamics of cosmological perturbations. 

Focusing on flat Friedmann-Lema\^{i}tre-Robertson-Walker cosmology and  models where the DE EoS $w(a)$ can be described by a continuous function of the scale factor, the first Friedman equation reads
\begin{align} 
\label{eq:background}
    H^2(a) = \frac{8\pi G}{3} \bigg[&\rho_{\rm r,0} \, a^{-4} + \rho_{\rm m,0} \, a^{-3} + \nonumber \\ &
    + \rho_{\rm DE,0} \, a^{-3} \exp \left(-3 \int_{a_0=1}^{a} \frac{w(a')}{a'} \, da' \right) \bigg].
\end{align}
Here $H(a) = \dot{a}/a$ is the Hubble parameter, the dot denotes the derivative with respect to physical time $t$, the subscript $0$ indicates quantities evaluate at present while $\rho_{\rm r}$, $\rho_{\rm m}$, and $\rho_{\rm DE}$ are the energy densities in radiation, matter, and DE, respectively.

When it comes to cosmological perturbations, diffeomorphism invariance of GR requires fixing a gauge. In the synchronous gauge, the line element reads~\cite{Ma:1995ey}
\begin{equation}
\label{eq12}
ds^2 = a^2(\tau) \left[-d\tau^2 + (\delta_{ij} + h_{ij}) dx^i dx^j \right], 
\end{equation}
where $d\tau=dt/a(t)$ is the conformal time, $\delta_{ij}$ and $h_{ij}$ are the unperturbed and perturbed spatial part of the metric tensors.
For a fluid component $i$, the equations governing the dimensionless density perturbations $\delta_i = \delta \rho_i / \rho_i$ and the divergence of the $i$-th fluid velocity $\theta_i = i \kappa^j v_j$ in the Fourier space are~\cite{Ma:1995ey}:
\begin{eqnarray}
\delta'_{i} & = & - (1+ w_{i})\, \left(\theta_{i}+ \frac{h'}{2}\right) - 
3\mathcal{H}\left(\frac{\delta P_i}{\delta \rho_i} - w_{i} \right)\delta_i \nonumber \\ 
& & -  9 \mathcal{H}^2\left(\frac{\delta P_i}{\delta \rho_i} - c^2_{a,i} \right) (1+w_i) 
\frac{\theta_i}
{{\kappa}^2}, \label{per1} \\
\theta'_{i} & = & - \mathcal{H} \left(1- 3 \frac{\delta P_i}{\delta
\rho_i}\right)\theta_{i} 
+ \frac{\delta P_i/\delta \rho_i}{1+w_{i}}\, {\kappa}^2\, \delta_{i} 
-{\kappa}^2\sigma_i,\label{per2}
\end{eqnarray}
where the primes denote the derivative with respect to conformal time $\tau$, $h$ is the usual synchronous gauge metric perturbation, $\mathcal{H}(a) = a'/a$ is the conformal Hubble parameter, $\kappa$ is the wavenumber in Fourier space, $\sigma_i$ stands for the anisotropic stress of the $i$-th fluid, and $c^2_{a,i}$ represents the adiabatic sound speed of the $i$-th fluid defined as $c_{a,i}^2 = \dot{P}_i/\dot{\rho}_i$. In this work, we fix the squared sound speed of the DE component in the rest frame $c^2_{\rm s,DE} = \frac{\delta P_{\rm DE}}{\delta \rho_{\rm DE}} = 1$ as broadly expected for the simplest DE models based on a single light minimally coupled scalar field with a canonical kinetic term.

Having outlined the background and perturbation dynamics, we now list and describe the five different two-parameter models employed for $w(a)$.

\subsection{Chevallier-Polarski-Linder parametrization}
The model proposed by Chevallier, Polarski and Linder~\cite{Chevallier:2000qy,Linder:2002et} (CPL hereafter) can be regarded as the baseline parameterization used in most analyses focusing on DDE, including this work. In this scenario, the DE EoS reads
\begin{equation}
\label{cpl}
w(a) = w_0 + w_a \times \left(1 - a \right). 
\end{equation}
As already mentioned in the introduction, its advantages include a manageable 2-dimensional phase space, reduction to linear redshift behavior at low redshift, bounded behavior at high redshift, high accuracy in reconstructing various scalar field equations of state and the resulting distance-redshift relations up to 0.1\% accuracy, good sensitivity to observational data, and a straightforward physical interpretation. The latter arises from its representation as the Taylor expansion of $w(a)$ around the present epoch $a \simeq a_0 \equiv 1$ up to the first order: $w_0 = w(a_0)$ and $w_a = -\frac{dw(a)}{da}\bigg|_{a=a_0}$ which is the coefficient for the dynamical term.

\subsection{Exponential parametrization}
As a next step, we consider the exponential form for the DE EoS~\cite{Dimakis:2016mip,Pan:2019brc}
\begin{equation}\label{exp}
w(a) = (w_0 - w_a) + w_a \times \exp\left(1 - a \right).
\end{equation}
Up to the first order of the Taylor expansion, this description reduces to the CPL parameterization around $a \simeq a_0 \equiv 1$. However, as $a$ moves far away from 1, the exponential form can introduce (small) deviations from the linear CPL regime without increasing the dimensionality of the parameter space~\cite{Najafi:2024qzm}.

\subsection{Jassal-Bagla-Padmanabhan parametrization}
The third parameterization studied in this work is the model proposed by Jassal-Bagla-Padmanabhan in Ref.~\cite{Jassal:2005qc} (JBP hereafter). In this case the DE EoS is
\begin{equation}\label{jbp}
w(a) = w_0 + w_a \times a \left(1 - a \right).
\end{equation}
It is characterized by the sum of a linear and a quadratic term in the scale factor. When $a^2$ is close to $1$, the term $-w_a a^2$ becomes comparable to $w_a a$, thereby leading to expected differences at low redshift compared to CPL.

\subsection{Logarithmic parametrization}
We consider the following logarithmic form for the EoS:
\begin{equation}\label{log}
w(a) = w_0 - w_a \times \ln a.
\end{equation}
To the best of our knowledge, this parameterization was originally introduced by G. Efstathiou in Ref.~\cite{Efstathiou:1999tm} to capture the behaviour of a wide class of potential scalar field models of DE at low redshift $z \lesssim 4$. Here, with a fair amount of courage and following thorough stability tests, we extend this parameterization all the way up to $z \to \infty$. In principle, for some combinations of parameters, the logarithmic term can actually grow in absolute value and cause instabilities.\footnote{To address this problem, in Ref.~\cite{Yang:2021flj}, some of us adopted flat priors $w_a \in [-3,0]$, thus removing positive values of $w_a$ \textit{a priori}. In this work, we have performed additional stability tests, which revealed that numerical instabilities arise only when CMB data are considered on their own (we do not report these results for any of the models under study, as they are not informative). Without late-time data, DE parameters remain essentially unconstrained, and $w_a$ can acquire large positive values. Despite the logarithmic nature of the equation of state, for these values, the DE energy density does not remain negligible in the early Universe, triggering warnings and errors in the Boltzmann solver code \texttt{CAMB}~\cite{Lewis:1999bs, Howlett:2012mh}. In contrast, when late-time data are included, the DE parameters are significantly constrained in the region $w_a < 0$, and no large positive values are allowed. As a result, in the allowed region of parameter space, the DE energy density remains negligible at early times and does not affect early Universe cosmology.} However, given the current data constraints and the slow logarithmic growth, this is not the case. We find that the parameterization can be safely extended to high redshift because the DE contribution remains largely negligible compared to other components in the Universe's energy budget.

\subsection{Barboza-Alcaniz parametrization}
The last (but as we shall see, not least) model involved in our analysis is the one proposed by Barboza and Alcaniz in Ref.~\cite{Barboza:2008rh} (referred to as BA hereafter). In this case, the DE EoS is characterized by the following functional form:
\begin{equation}\label{ba}
w(a) = w_0 + w_a \times \frac{1 - a}{a^2 + (1 - a)^2}.
\end{equation}
This parameterization shows a linear behavior at low redshifts and remains well-behaved as $z \to \infty$, while allowing for deviations from the baseline CPL scenario.

\begin{table}
\centering
\renewcommand{\arraystretch}{1.5}
\begin{tabular}{l @{\hspace{2cm}} c}
\toprule
\textbf{Parameter} & \textbf{Prior} \\
\hline \hline
$\Omega_\mathrm{b} h^2$ & $[0.005, 0.1]$ \\
$\Omega_\mathrm{c} h^2$ & $[0.01, 0.99]$ \\
$\log(10^{10} A_\mathrm{s})$ & $[1.61, 3.91]$ \\
$n_\mathrm{s}$ & $[0.8, 1.2]$ \\
$\tau$ & $[0.01, 0.8]$ \\
$100\theta_\mathrm{MC}$ & $[0.5, 10]$ \\
$w_0$ & $[-3, 1]$ \\
$w_a$ & $[-3, 2]$ \\
\bottomrule
\end{tabular}
\caption{Ranges for the flat prior distributions imposed on the free cosmological parameters in the analysis.}
\label{tab-priors}
\end{table}

\section{Methods}
\label{sec-data}
In this section, we describe the statistical methodologies and observational datasets employed in our analysis.

\subsection{Statistical Analyses}
The cosmology resulting from all the five DDE models listed in Sec.~\ref{sec-2} can be characterized by 8 free parameters: the physical baryon energy density $\Omega_\mathrm{b} h^2$, the physical cold dark matter energy density $\Omega_\mathrm{c} h^2$, the amplitude of the primordial scalar spectrum $A_\mathrm{s}$, its spectral index $n_s$, the optical depth to reionization $\tau$, the angular size of the sound horizon $\theta_{\rm{MC}}$, and the two free parameters describing the DE sector — i.e., the present-day value of the DE EoS $w_0$ and the parameter describing its dynamical evolution $w_a$. To compare the theoretical predictions against observations, we implement these models in five different modified versions of the publicly available cosmological code \texttt{CAMB}~\cite{Lewis:1999bs,Howlett:2012mh} and explore the posterior distributions of the 8-dimensional parameter space by performing Markov Chain Monte Carlo (MCMC) analyses via the publicly available sampler \texttt{Cobaya}~\cite{Lewis:2002ah,Lewis:2013hha} that employs the fast dragging speed hierarchy implementation~\cite{Neal:2005}. The convergence of the generated MCMC chains is assessed via the Gelman-Rubin parameter $R-1$~\cite{Gelman:1992zz}. For \textit{all} models and datasets, we require $R - 1 < 0.01$ for the chains to be considered converged. In Tab.~\ref{tab-priors}, we present the flat prior ranges on which the parameters are left to freely vary.

\subsection{Datasets}

\noindent The datasets involved in our analyses are: 

\begin{itemize}[leftmargin=*]   
\item \textbf{Planck:} Measurements of the Planck CMB temperature anisotropy and polarization power spectra, their cross-spectra, and the combination of the ACT and Planck lensing power spectrum. All CMB likelihoods employed in this work are listed below:
\begin{itemize}
\item[(i)] Measurements of the power spectra of temperature and polarization anisotropies, $C_{\ell}^{TT}$, $C_{\ell}^{TE}$, and $C_{\ell}^{EE}$, at small scales ($\ell>30$), obtained by the Planck \texttt{plik} likelihood~\cite{Planck:2018vyg,Planck:2019nip};
\item[(ii)] Measurements of the spectrum of temperature anisotropies, $C_{\ell}^{TT}$, at large scales ($2 \leq \ell \leq 30$), obtained by the Planck \texttt{Commander} likelihood~\cite{Planck:2018vyg,Planck:2019nip};
\item[(iii)] Measurements of the spectrum of E-mode polarization, $C_{\ell}^{EE}$, at large scales ($2 \leq \ell \leq 30$), obtained by the Planck \texttt{SimAll} likelihood~\cite{Planck:2018vyg,Planck:2019nip};
\item[(iv)] Reconstruction of the spectrum of the lensing potential, obtained by the Planck PR4 \texttt{NPIPE} data release~\cite{Carron:2022eyg} used in combination with \texttt{ACT-DR6} lensing likelihood~\cite{ACT:2023kun,ACT:2023dou}.\footnote{The \texttt{NPIPE} lensing map~\cite{Carron:2022eyg} covers CMB angular scales in the range $100 \le \ell \le 2048$ using the quadratic estimator and re-processing Planck time-ordered data with several improvements, including around 8\% more data compared to the \texttt{plik-lensing} likelihood. Notice also that \texttt{NPIPE} and \texttt{ACT-DR6} measurements explore distinct angular scales, as ACT uses only CMB multipoles $600 < \ell < 3000$ and has only partial overlap with the 67\% sky fraction used in the Planck analysis~\cite{ACT:2023dou}. Additionally they have different noise levels and instrument-related systematics. Therefore they can be regarded as nearly independent lensing measurements.}
\end{itemize} 

\item \textbf{DESI:} Baryon acoustic oscillations (BAO) measurements extracted by observations of galaxies \& quasars~\cite{DESI:2024uvr}, and Lyman-$\alpha$~\cite{DESI:2024lzq} tracers from the first year of observations using the Dark Energy Spectroscopic Instrument (DESI). These include measurements of the transverse comoving distance, the Hubble horizon, and the angle-averaged distance as summarized in Tab.~I of Ref.~\cite{DESI:2024mwx}.
    
\item \textbf{PantheonPlus:} Distance moduli measurements of 1701 light curves of 1550 spectroscopically confirmed type Ia SN sourced from eighteen different surveys, gathered from the Pantheon-plus sample~\cite{Scolnic:2021amr, Brout:2022vxf}.
    
\item \textbf{DESY5:} Distance moduli measurements of 1635 Type Ia SN covering the redshift range of $0.10 < z < 1.13$ that have been collected during the full five years of the Dark Energy Survey (DES) Supernova Program~\cite{DES:2024tys,DES:2024upw,DES:2024hip}, along with 194 low-redshift SN in the redshift range of $0.025 < z < 0.1$ which are in common with the Pantheon-plus sample~\cite{Scolnic:2021amr, Brout:2022vxf}.
\end{itemize} 

We conclude this subsection with a final remark. Our analysis focuses on two samples of Type-Ia SN: PantheonPlus and DESY5, excluding the Union3 sample. As highlighted in the DESI paper~\cite{DESI:2024mwx}, among these three SN samples, PantheonPlus (which uses spectroscopically confirmed SN) produces the smallest, yet significant, preference for DDE, deviating by about 2.5$\sigma$ from the cosmological constant scenario. In contrast, DESY5 (which uses photometry) shows the largest shift towards DDE, at $\sim 3.9 \sigma$. The Union3 sample (which also uses spectroscopically confirmed SN) shows a preference for DDE around 3.5$\sigma$, falling between PantheonPlus and DESY5. Although Union3 provides valuable confirmation of these results, here we focus on the two samples that represent the smallest and largest deviations from the cosmological constant.

\section{Results}
\label{sec-results}
In this section, we present the observational constraints on the five DDE models considered in this article. We discuss the results model by model, testing each case against three different data combinations: Planck+DESI, Planck+DESI+PantheonPlus, and Planck+DESI+DESY5. We make no secret that due to the large number of analyzed models and the similarity of the results obtained, the following discussion may appear somewhat repetitive (though necessary). Therefore, readers interested in the results of specific models can find the numerical constraints, two-dimensional correlations, and one-dimensional posterior distribution functions of key parameters as follows:

\begin{itemize}[leftmargin=*]
\item Table~\ref{tab:cpl} and Figure~\ref{fig:cpl} summarize the numerical constraints and parameter correlations for the baseline CPL case~\eqref{cpl}. The results for this case are detailed in Sec.~\ref{sec.results.CPL}.
\item Table~\ref{tab:exp} and Figure~\ref{fig:exp} present the results for the exponential parameterization~\eqref{exp}, discussed in Sec.~\ref{sec.results.epx}.
\item Table~\ref{tab:jbp} and Figure~\ref{fig:jbp} provide the results for the JBP EoS~\eqref{jbp}, discussed in Sec.~\ref{sec.results.JBP}.
\item Table~\ref{tab:log} and Figure~\ref{fig:log} summarize the results for the logarithmic parameterization~\eqref{log}, detailed in Sec.~\ref{sec.results.log}.
\item Table~\ref{tab:ba} and Figure~\ref{fig:ba} present the results for the BA parameterization~\eqref{ba}, discussed in Sec.~\ref{sec.results.BA}.
\end{itemize}

Conversely, readers interested in a comprehensive overview of the results, their interpretation, and implications can refer directly to Sec.~\ref{sec-summary}. Instead, a comprehensive discussion of the behavior of the EoS inferred from the different datasets employed in the analysis across various models, as well as constraints on crucial quantities that aid in interpreting the results discussed in this section -- such as the pivot redshift (i.e., the redshifts where the equation of state is better constrained by current data) and phantom crossing -- is presented in Appendix~\ref{app:A} and summarized in Fig.~\ref{fig:7} and Tab.~\ref{tab.Results.z}. Interested readers can refer to this appendix for further details.

\newpage

\subsection{Results for the CPL parameterization}
\label{sec.results.CPL}
\begin{figure}[ht!]
    \centering
    \includegraphics[width=0.75\linewidth]{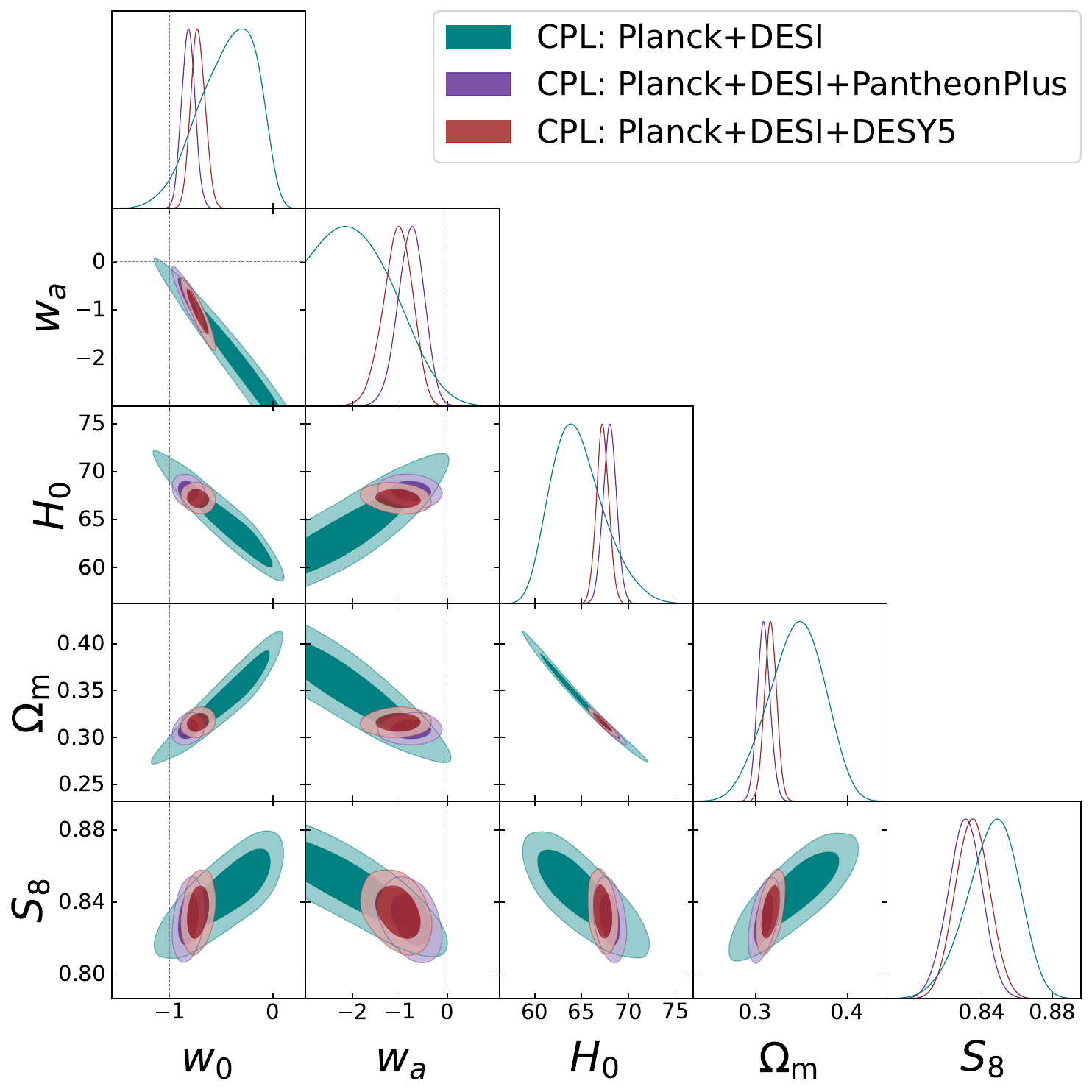}
    \caption{\textbf{CPL parametrization \eqref{cpl} --} one-dimensional posterior distributions and two-dimensional marginalized contours for the main key parameters as obtained from the Planck+DESI, Planck+DESI+PantheonPlus, and Planck+DESI+DESY5 dataset combinations.
    }
    \label{fig:cpl}
\end{figure}

The numerical constraints obtained by adopting a baseline CPL parametrization are given in Tab.~\ref{tab:cpl}. Fig.~\ref{fig:cpl} displays key parameters that characterize this model (i.e., the present-day value of the EoS $w_0$ and the parameter quantifying its redshift evolution $w_a$) as well as their correlations with other cosmological parameters of intrinsic interest for the late-time expansion history of the Universe such as the Hubble constant $H_0$, the present-day matter fractional energy density $\Omega_m$, and the matter clustering parameter $S_8$.

\begin{table}[h]
\centering
\renewcommand{\arraystretch}{1.2}
\resizebox{0.9\textwidth}{!}{\begin{tabular}{lcccccc}
\toprule
 \textbf{Parameter} & \textbf{Planck+DESI} & \textbf{Planck+DESI+PantheonPlus} & \textbf{Planck+DESI+DESY5} \\
\hline\hline
$\Omega_\mathrm{c} h^2$ & $0.11993\pm 0.00098$ & $0.11962\pm 0.00099$ & $0.11978\pm 0.00098$\\
$\Omega_\mathrm{b} h^2$ & $0.02238\pm 0.00014$ & $0.02241\pm 0.00014$ & $0.02239\pm 0.00014$\\
$100\theta_\mathrm{MC}$ & $1.04092\pm 0.00029$ & $1.04098\pm 0.00030$ & $1.04094\pm 0.00030$\\
$\tau$ & $0.0530\pm 0.0073$ & $0.0547\pm 0.0073$ & $0.0538\pm 0.0074$ \\
$n_\mathrm{s}$ & $0.9655\pm 0.0038$ & $0.9663\pm 0.0038$ & $0.9659\pm 0.0038$\\
$\log(10^{10} A_\mathrm{s})$ & $3.040\pm 0.013$ & $3.044\pm 0.013$ & $3.042\pm 0.013$\\
$w_0$ & $-0.44^{+0.34}_{-0.21}$ & $-0.820\pm 0.064$ & $-0.726\pm 0.069$\\
$w_a$ & $-1.81^{+0.37}_{-1.1}$ & $-0.77\pm 0.28$ & $-1.05^{+0.34}_{-0.28}$\\
$\Omega_\mathrm{m}$ & $0.344^{+0.033}_{-0.027}$ & $0.3088\pm 0.0070$ & $0.3161\pm 0.0066$\\
$\sigma_8$ & $0.791^{+0.021}_{-0.028}$ & $0.8186\pm 0.0094$ & $0.8130\pm 0.0091$\\
$S_8$ & $0.846^{+0.016}_{-0.013}$ & $0.8304\pm 0.0097$ & $0.8345\pm 0.0097$\\
$H_0 \mathrm{\,[km/s/Mpc]}$ & $64.6^{+2.2}_{-3.3}$ & $67.98\pm 0.72$ & $67.22\pm 0.66$\\
$r_\mathrm{drag} \mathrm{\, [Mpc]}$ & $147.11\pm 0.24$ & $147.16\pm 0.23$ & $147.14\pm 0.23$\\
$\Delta\chi^2$ & $-6.8$ & $-8.4$ & $-15.2$\\
\bottomrule
\end{tabular} }
\caption{\textbf{CPL Parametrization \eqref{cpl} --} 68\% CL constraints on the free and derived cosmological parameters for 3 different data combinations detailed in Sec.~\ref{sec-data}. Negative values of $\Delta\chi^2 = \chi^2_{\text{CPL}} - \chi^2_{\Lambda\text{CDM}}$ indicate an improvement in the fit to the data compared to $\Lambda$CDM.}
\label{tab:cpl}
\end{table}

We recover all the results discussed in the recent DESI release paper~\cite{DESI:2024mwx}. For Planck+DESI, the constraints favor a present-day quintessence EoS with $w_0 = -0.44^{+0.34}_{-0.21}$ at 68\% CL,\footnote{Hereafter, constraints will always be quoted at 68\% CL unless otherwise specified.} showing a notable shift away from $w_0 = -1$. On the other hand, $w_a = -1.81^{+0.37}_{-1.1}$ provides hints of dynamical evolution towards the phantom regime.

The addition of PantheonPlus significantly refines the constraints on the parameter space, reducing the error bars on the DE parameters by up to a factor of 5. Although $w_0$ shifts towards $-1$, it remains strictly in the quintessence regime: $w_0 = -0.820\pm 0.064$. Consistent with DESI 2024 findings~\cite{DESI:2024kob}, the mean value of $w_a$ increases compared to Planck+DESI, now reading $w_a = -0.77\pm 0.28$. This boosts the evidence for a past phantom-like DDE component to approximately 2.5$\sigma$, 
see also Fig.~\ref{fig:cpl}.

When replacing PantheonPlus with DESY5 type Ia SN, we observe a shift of $w_0$ away from $-1$, resulting in $w_0 = -0.726 \pm 0.069$. This places $w_0$ deep in the quintessence regime, see also Fig.~\ref{fig:cpl}. Similarly, $w_a = -1.05^{+0.34}_{-0.28}$ is found to be non-zero at high statistical significance. Thus, the evidence for DDE remains stronger in the Planck+DESI+DESY5 case compared to Planck+DESI+PantheonPlus, and the cosmological constant case falls well outside the joint probability contours in the $w_0$-$w_a$ plane, as seen in Fig.~\ref{fig:cpl}.

\newpage
\subsection{Results for the Exponential parametrization} 
\label{sec.results.epx}
\begin{figure}[h]
    \centering
    \includegraphics[width=0.75\linewidth]{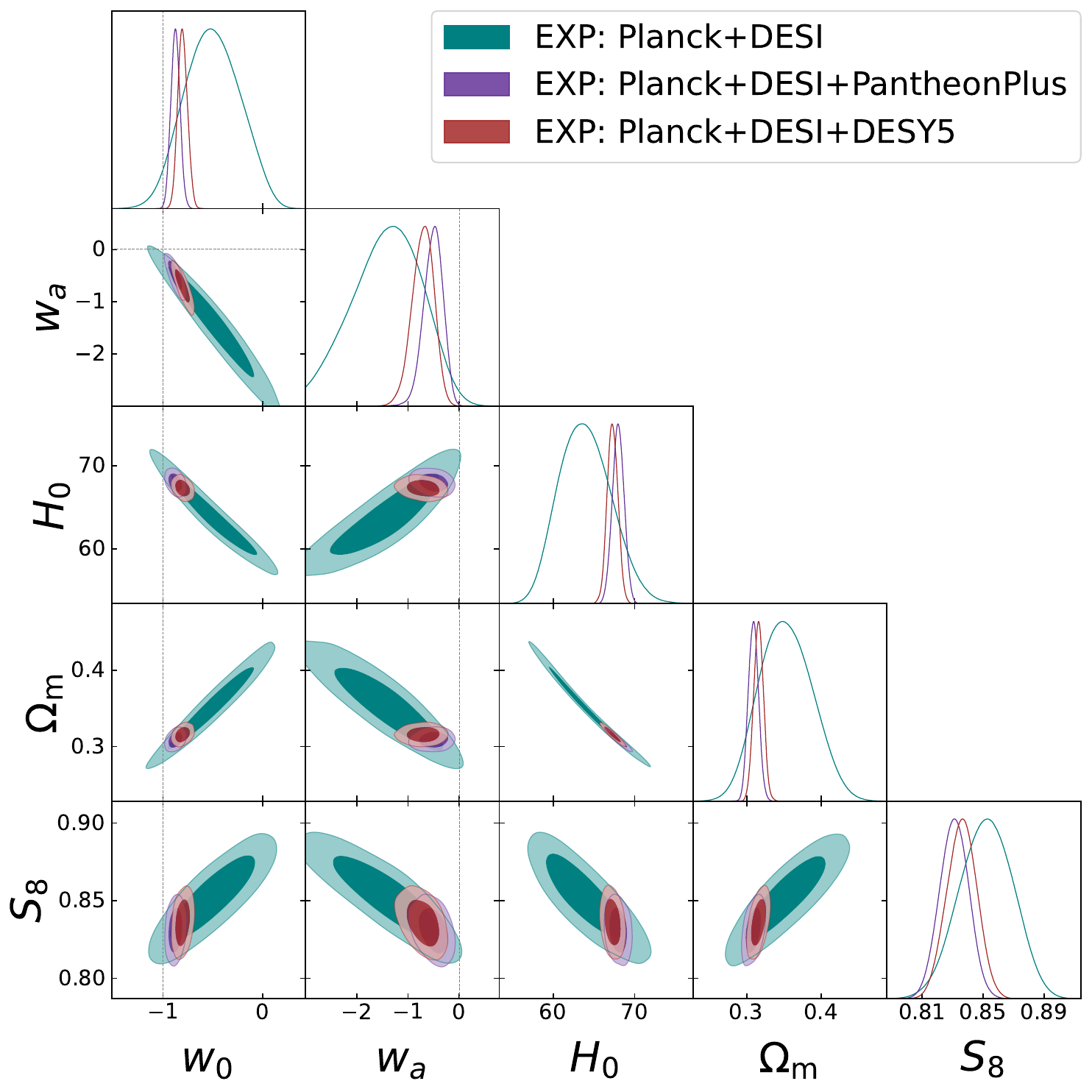}
    \caption{\textbf{Exponential parameterization \eqref{exp} --} one-dimensional posterior distributions and two-dimensional marginalized contours for the main key parameters as obtained from the Planck+DESI, Planck+DESI+PantheonPlus, and Planck+DESI+DESY5 dataset combinations.
    }
    \label{fig:exp}
\end{figure}

We present the constraints obtained by assuming an exponential parametrization for the DE EoS in Tab.~\ref{tab:exp}. In Fig.~\ref{fig:exp}, we show the one-dimensional posterior distribution functions and the two-dimensional marginalized contours for the key cosmic parameters.

As usual, we test the model against three different combinations of data involving the DESI BAO measurements. Focusing on the minimal Planck+DESI combination, we find $w_0 = -0.50 \pm 0.27$ -- significantly different from $-1$ and deep in the quintessence regime. Similarly, $w_a = -1.40^{+0.75}_{-0.62}$ is almost $2\sigma$ away from the non-dynamical $w_a = 0$ case, lending weight to the Planck+DESI preference for DDE.

The addition of PantheonPlus SN measurements reinforces this preference: the constraints on $w_0 = -0.876\pm 0.045$ shrink in the quintessence regime, deviating from $w_0 = -1$ by more than 2.5$\sigma$. Additionally, $w_a = -0.51^{+0.20}_{-0.17}$ is found to be non-zero at more than $2\sigma$.  Overall, Planck+DESI+PantheonPlus provides evidence for DDE with the present-day EoS in the quintessence regime and a dynamical evolution that crosses into the phantom regime, as clearly shown in Fig.~\ref{fig:exp}.

\begingroup
\begin{table}[h]
\centering
\renewcommand{\arraystretch}{1.2}
\resizebox{0.9\textwidth}{!}{\begin{tabular}{lcccccc}   
\toprule
 \textbf{Parameter} & \textbf{Planck+DESI} & \textbf{Planck+DESI+PantheonPlus} & \textbf{Planck+DESI+DESY5} \\
\hline \hline
$\Omega_\mathrm{c} h^2$ & $0.1201\pm 0.0010$ & $0.11962\pm 0.00098$ & $0.1198\pm 0.0010$ \\
$\Omega_\mathrm{b} h^2$ & $0.02237\pm 0.00014$ & $0.02241\pm 0.00014$ & $0.02239\pm 0.00014$ \\
$100\theta_\mathrm{MC}$ & $1.04089\pm 0.00030$ & $1.04097\pm 0.00030$ & $1.04094\pm 0.00030$ \\
$\tau$ & $0.0525\pm 0.0073$ & $0.0545\pm 0.0071$ & $0.0539\pm 0.0074$ \\
$n_\mathrm{s}$ & $0.9650\pm 0.0038$ & $0.9662\pm 0.0038$ & $0.9657\pm 0.0039$ \\
$\log(10^{10} A_\mathrm{s})$ & $3.040\pm 0.013$ & $3.044\pm 0.013$ & $3.042\pm 0.013$ \\
$w_0$ & $-0.50\pm 0.27$ & $-0.876\pm 0.045$ & $-0.804^{+0.045}_{-0.051}$ \\
$w_a$ & $-1.40^{+0.75}_{-0.62}$ & $-0.51^{+0.20}_{-0.17}$ & $-0.71^{+0.23}_{-0.19}$ \\
$\Omega_\mathrm{m}$ & $0.352\pm 0.035$ & $0.3088\pm 0.0067$ & $0.3157\pm 0.0066$ \\
$\sigma_8$ & $0.788^{+0.026}_{-0.029}$ & $0.8192\pm 0.0098$ & $0.8150\pm 0.0092$ \\
$S_8$ & $0.852\pm 0.017$ & $0.8310\pm 0.0095$ & $0.8360\pm 0.0098$ \\
$H_0 \mathrm{\,[km/s/Mpc]}$ & $64.0^{+2.9}_{-3.5}$ & $67.98\pm 0.72$ & $67.28\pm 0.65$ \\
$r_\mathrm{drag} \mathrm{\, [Mpc]}$ & $147.08\pm 0.24$ & $147.16\pm 0.23$ & $147.12\pm 0.24$ \\
$\Delta\chi^2$ & $-6.9$ & $-7.8$ & $-15.2$ \\
\bottomrule
\end{tabular} }
\caption{\textbf{Exponential Parametrization \eqref{exp} --} 68\% CL constraints on the free and derived cosmological parameters for 3 different data combinations detailed in Sec.~\ref{sec-data}. Negative values of $\Delta\chi^2 = \chi^2_{\text{exp}} - \chi^2_{\Lambda\text{CDM}}$ indicate an improvement in the fit to the data compared to $\Lambda$CDM.}
\label{tab:exp}
\end{table}
\endgroup

When we focus on the Planck+DESI+DESY5 data combination, the evidence for DDE becomes significantly more pronounced. $w_0 = -0.804^{+0.045}_{-0.051}$ remains strictly in the quintessence regime, while $w_a = -0.71^{+0.23}_{-0.19}$ is $3\sigma$ away from the non-dynamical $w_a = 0$ scenario; see again Fig.~\ref{fig:exp}.

Overall, in terms of constraints on cosmic parameters, these results are in agreement with those derived for the CPL parametrization in the previous section, underscoring the resilience of the evidence for DDE and relieving concerns about dependence on the model for these particular results.
\newpage
\subsection{Results for the JBP parametrization}
\label{sec.results.JBP}
\begin{figure}[h]
    \centering
    \includegraphics[width=0.75\linewidth]{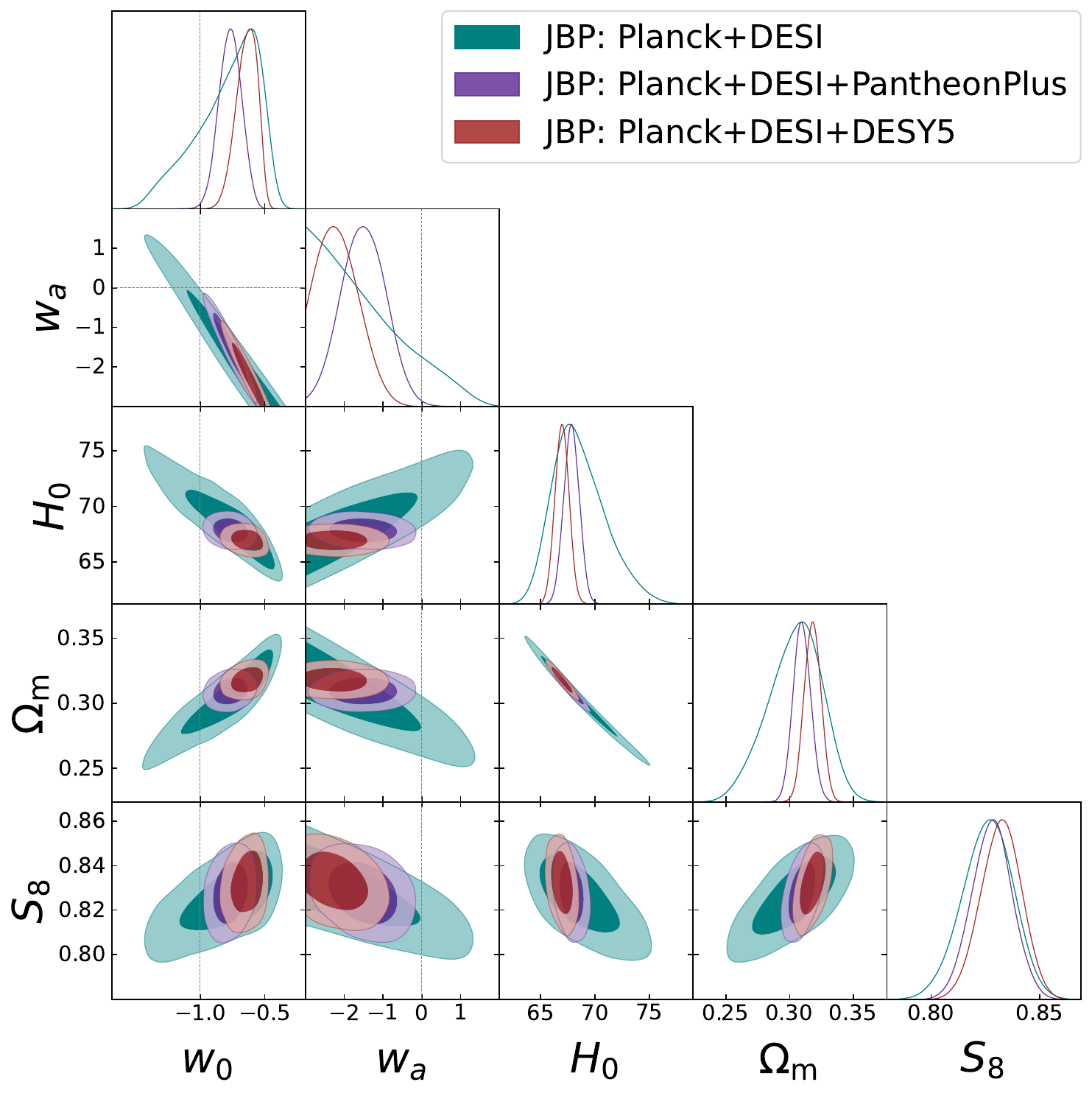}
    \caption{\textbf{JBP parametrization \eqref{jbp} --} one-dimensional posterior distributions and two-dimensional marginalized contours for the main key parameters as obtained from the Planck+DESI, Planck+DESI+PantheonPlus, and Planck+DESI+DESY5 dataset combinations.
    }
    \label{fig:jbp}
\end{figure}

The numerical constraints for the JBP parametrization are given in Tab.~\ref{tab:jbp}, while the marginalized probability contours for the usual parameters are shown in Fig.~\ref{fig:jbp}.

When considering Planck+DESI, unlike the other parametrizations described so far (e.g., CPL and exponential form), $w_a$ remains unbounded and an upper limit $w_a<0.648$ can be derived at 95\% CL. Conversely, $w_0$ remains in the quintessence regime ($w_0 = -0.79^{+0.31}_{-0.14}$), confirming the overall tendency for a present-day quintessence EoS.

When considering PantheonPlus in combination with Planck and DESI, we get $w_a = -1.50 \pm 0.57$ -- non-null at more than $2.6\sigma$. Additionally, the constraints on $w_0 = -0.767 \pm 0.086$ are narrowed down within the quintessence portion of the parameter space as seen in Fig.~\ref{fig:jbp}. Thus, effectively, evidence of DDE is confirmed for this parametrization as well.

\begin{table}[ht]
\centering
\renewcommand{\arraystretch}{1.2}
\resizebox{0.9\textwidth}{!}{\begin{tabular}{lcccccc}
\toprule
\textbf{Parameter} & \textbf{Planck+DESI} & \textbf{Planck+DESI+PantheonPlus} & \textbf{Planck+DESI+DESY5} \\
\hline \hline
$\Omega_\mathrm{c} h^2$ & $0.11940\pm 0.00098$ & $0.11934\pm 0.00096$ & $0.11947\pm 0.00094$ \\
$\Omega_\mathrm{b} h^2$ & $0.02242\pm 0.00013$ & $0.02243\pm 0.00014$ & $0.02242\pm 0.00014$ \\
$100\theta_\mathrm{MC}$ & $1.04099\pm 0.00029$ & $1.04102\pm 0.00029$ & $1.04100\pm 0.00029$ \\
$\tau$ & $0.0558\pm 0.0075$ & $0.0564\pm 0.0073$ & $0.0557\pm 0.0072$ \\
$n_\mathrm{s}$ & $0.9668\pm 0.0038$ & $0.9670\pm 0.0038$ & $0.9667\pm 0.0038$ \\
$\log(10^{10} A_\mathrm{s})$ & $3.046\pm 0.014$ & $3.048\pm 0.013$ & $3.047\pm 0.013$ \\
$w_0$ & $-0.79^{+0.31}_{-0.14}$ & $-0.767\pm 0.086$ & $-0.641^{+0.095}_{-0.067}$ \\
$w_a$ & $< 0.648$ & $-1.50\pm 0.57$ & $-2.12^{+0.38}_{-0.68}$ \\
$\Omega_\mathrm{m}$ & $0.304^{+0.023}_{-0.019}$ & $0.3096\pm 0.0068$ & $0.3180\pm 0.0065$ \\
$\sigma_8$ & $0.822^{+0.019}_{-0.025}$ & $0.8151\pm 0.0093$ & $0.8083\pm 0.0086$ \\
$S_8$ & $0.826^{+0.012}_{-0.011}$ & $0.8279\pm 0.0093$ & $0.8321\pm 0.0093$ \\
$H_0 \mathrm{\,[km/s/Mpc]}$ & $68.6^{+1.9}_{-2.8}$ & $67.83\pm 0.71$ & $66.96\pm 0.64$ \\
$r_\mathrm{drag} \mathrm{\, [Mpc]}$ & $147.20\pm 0.23$ & $147.21\pm 0.23$ & $147.19\pm 0.22$ \\
$\Delta\chi^2$ & $-5.6$ & $-6.4$ & $-16.0$ \\
\bottomrule
\end{tabular}}
\caption{\textbf{JBP parametrization \eqref{jbp} --} 68\% CL constraints and 95\% CL upper limits on the free and derived cosmological parameters for 3 different data combinations detailed in Sec.~\ref{sec-data}. Negative values of $\Delta\chi^2 = \chi^2_{\text{JBP}} - \chi^2_{\Lambda\text{CDM}}$ indicate an improvement in the fit to the data compared to $\Lambda$CDM. }
\label{tab:jbp}
\end{table}

Finally, we replace PantheonPlus with DESY5 SN measurements. In this case, we obtain $w_0 = -0.641^{+0.095}_{-0.067}$ and $w_a = -2.12^{+0.38}_{-0.68}$, confirming that the evidence of DDE becomes much more pronounced with DESY5. This evidence reaches a statistical significance $\gtrsim 3\sigma$. Having said that, comparing Fig.~\ref{fig:jbp} with the respective triangular plots of the other parameterizations, we notice that for this model, the uncertainties remain much broader, especially for the parameters describing DE EoS. This can be explained in terms of the peculiar evolution of the DE EoS obtained in this model. As discussed in detail in Appendix~\ref{app:A}, among the five models analyzed, the JBP parameterization presents a more articulated phenomenology regarding the evolution of the DE EoS. Due to its quadratic nature in the scale factor, the evolution of the EoS within the JBP parameterization crosses $w = -1$ twice. At low redshift, it behaves similarly to the other parameterizations; however, after the first quintessence-to-phantom transition, $w(z)$ approaches a minimum value around $z \sim 1$ before rising again, leading to a second phantom-to-quintessence crossing at $z \sim 4$. This behavior contrasts with other models, where the EoS remains within the phantom regime. As detailed in Appendix~\ref{app:A}, the interplay between low and high redshift behaviors results in two different pivot redshifts at low and high $z$. This interplay may contribute to tilting the 2-D probability contours in the $w_0$ and $w_a$ plane, as shown in Fig.~\ref{fig:jbp} (see also Fig.~\ref{fig:recap} for comparisons with other models). Additionally, the increased uncertainties at low redshift might suggest that this phenomenology is not ideal for consistently fitting all the data across low and high redshift. This concern is confirmed when comparing the differences between the best-fit $\chi^2$ obtained within each DDE model and the best-fit $\chi^2$ obtained within $\Lambda$CDM. Indeed, this model consistently leads to the smallest improvement over $\Lambda$CDM across all datasets and DDE models. For more details, we refer to Appendix~\ref{app:A}.

\newpage
\subsection{Results for the Logarithmic parametrization}
\label{sec.results.log}
\begin{figure}[h]
    \centering
    \includegraphics[width=0.75\linewidth]{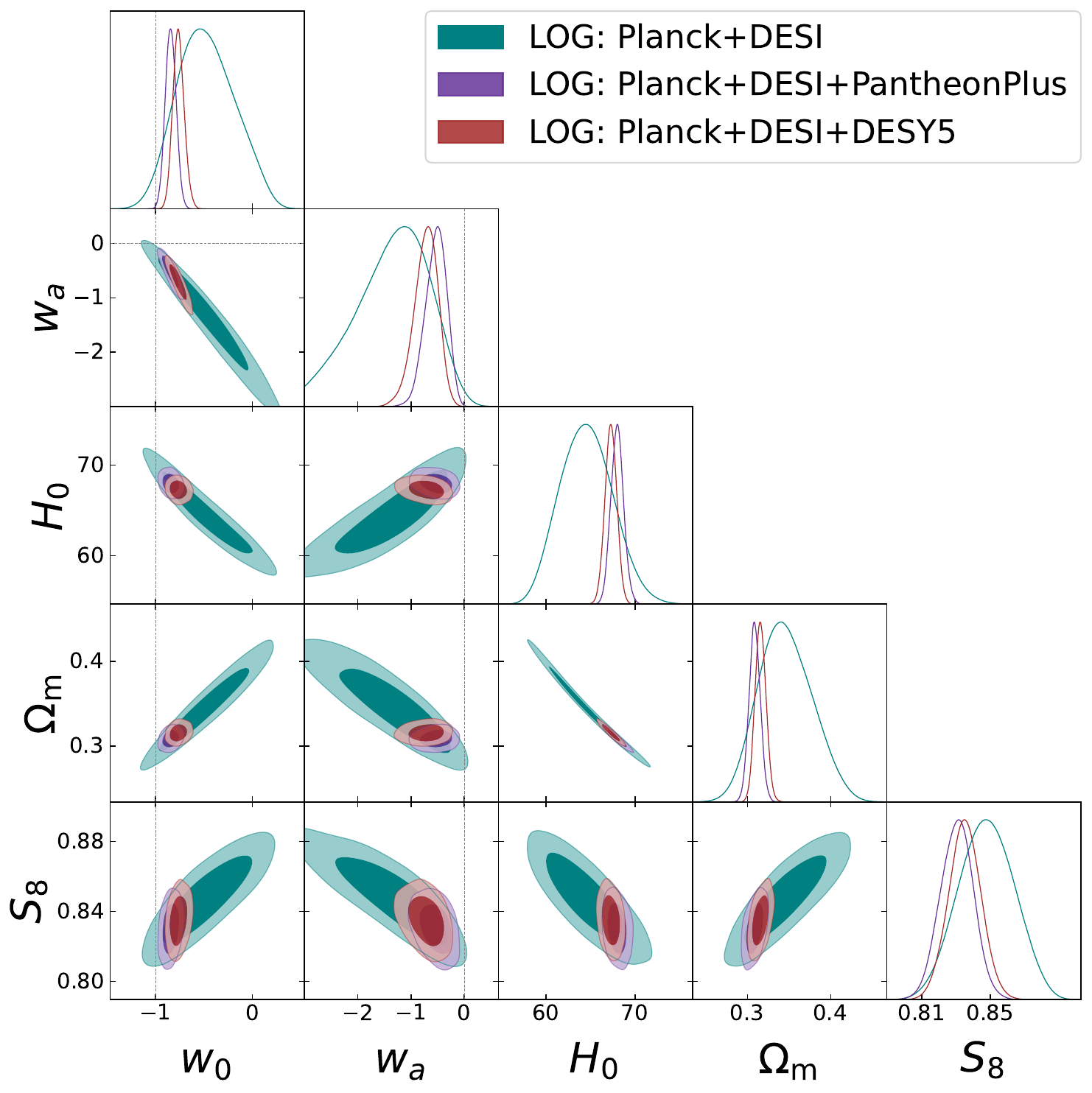}
    \caption{\textbf{Logarithmic parametrization \eqref{log} --}  one-dimensional posterior distributions and two-dimensional marginalized contours for the main key parameters as obtained from the Planck+DESI, Planck+DESI+PantheonPlus, and Planck+DESI+DESY5 dataset combinations.
    }
    \label{fig:log}
\end{figure}

Tab.~\ref{tab:log} summarizes the constraints on the model where the DE EoS is described by the logarithmic parametrization. Fig.~\ref{fig:log} displays the usual marginalized contours on relevant parameters.

\begingroup
\begin{table}[ht]
\centering
\renewcommand{\arraystretch}{1.2}
\resizebox{0.9\textwidth}{!}{\begin{tabular}{lccccccccccc}
\toprule
\textbf{Parameter} & \textbf{Planck+DESI} & \textbf{Planck+DESI+PantheonPlus} & \textbf{Planck+DESI+DESY5} \\
\hline \hline
$\Omega_\mathrm{c} h^2$ & $0.1201\pm 0.0010$ & $0.11964\pm 0.00099$ & $0.11986\pm 0.00099$\\
$\Omega_\mathrm{b} h^2$ & $0.02237\pm 0.00014$ & $0.02240\pm 0.00014$ & $0.02239\pm 0.00014$\\
$100\theta_\mathrm{MC}$ & $1.04090\pm 0.00030$ & $1.04095\pm 0.00029$ & $1.04094\pm 0.00029$\\
$\tau$ & $0.0520\pm 0.0073$ & $0.0542\pm 0.0073$  & $0.0535\pm 0.0073$ \\
$n_\mathrm{s}$ & $0.9648\pm 0.0039$ & $0.9661\pm 0.0038$ & $0.9657\pm 0.0038$\\
$\log(10^{10} A_\mathrm{s})$ & $3.039\pm 0.013$ & $3.043\pm 0.013$ & $3.042\pm 0.013$\\
$w_0$ & $-0.48^{+0.28}_{-0.33}$ & $-0.843\pm 0.055$ & $-0.763^{+0.054}_{-0.062}$\\
$w_a$ & $-1.33^{+0.79}_{-0.56}$ & $-0.53^{+0.22}_{-0.18}$ & $-0.72^{+0.25}_{-0.19}$\\
$\Omega_\mathrm{m}$ & $0.346^{+0.030}_{-0.035}$ & $0.3086\pm 0.0067$ & $0.3156\pm 0.0066$\\
$\sigma_8$ & $0.792\pm 0.025$ & $0.8189\pm 0.0097$ & $0.8142\pm 0.0092$\\
$S_8$ & $0.848\pm 0.016$ & $0.8305\pm 0.0095$ & $0.8350\pm 0.0094$\\
$H_0 \mathrm{\,[km/s/Mpc]}$ & $64.6^{+2.8}_{-3.1}$ & $68.01\pm 0.71$ & $67.29\pm 0.66$\\
$r_\mathrm{drag}\mathrm{\, [Mpc]}$ & $147.07\pm 0.23$ & $147.17\pm 0.24$ & $147.12\pm 0.23$\\
$\Delta\chi^2$ & $-6.5$ & $-9.3$ & $-14.8$\\
\bottomrule
\end{tabular}}
\caption{\textbf{Logarithmic parametrization \eqref{log} --} 68\% CL constraints on the free and derived cosmological parameters for 3 different data combinations detailed in Sec.~\ref{sec-data}. Negative values of $\Delta\chi^2 = \chi^2_{\text{log}} - \chi^2_{\Lambda\text{CDM}}$ indicate an improvement in the fit to the data compared to $\Lambda$CDM.}
\label{tab:log}
\end{table}
\endgroup

Starting with Planck+DESI, $w_0 = -0.48^{+0.28}_{-0.33}$ is confined to the quintessence regime at more than 68\% CL, while $w_a = 1.33^{+0.79}_{-0.56}$ is constrained to be different from zero at more than $2.3\sigma$ -- confirming once more the preference for DDE in Planck+DESI.

When PantheonPlus is added to Planck+DESI, we find $w_0 = -0.843 \pm 0.055$; i.e., shifted towards $w_0 = -1$ although with error bars smaller by a factor of 5. However, also in this parameterization, $w_0$ is preferred to be in the quintessence regime, excluding $w_0=-1$ at more than $2.8\sigma$. Similarly, the result on $w_a = -0.53^{+0.22}_{-0.18}$ confirms the overall preference for DDE, see Fig.~\ref{fig:log}.

Considering DESY5 in place of PantheonPlus, the constraints on $w_0$ and $w_a$ change to $w_0 = -0.763^{+0.054}_{-0.062}$ and $w_a = -0.72^{+0.25}_{-0.19}$, respectively. As a result, $w_0$ remains in the quintessence regime at more than 95\% CL, while $w_a$ is found to be non-zero at almost $3\sigma$. It is noteworthy that the evidence of DDE is consistently more pronounced in the presence of DESY5 compared to PantheonPlus, see Fig.~\ref{fig:log}.

\clearpage
\subsection{Results for the BA parametrization}
\label{sec.results.BA}

\begin{figure}[ht]
    \centering
    \includegraphics[width=0.75\linewidth]{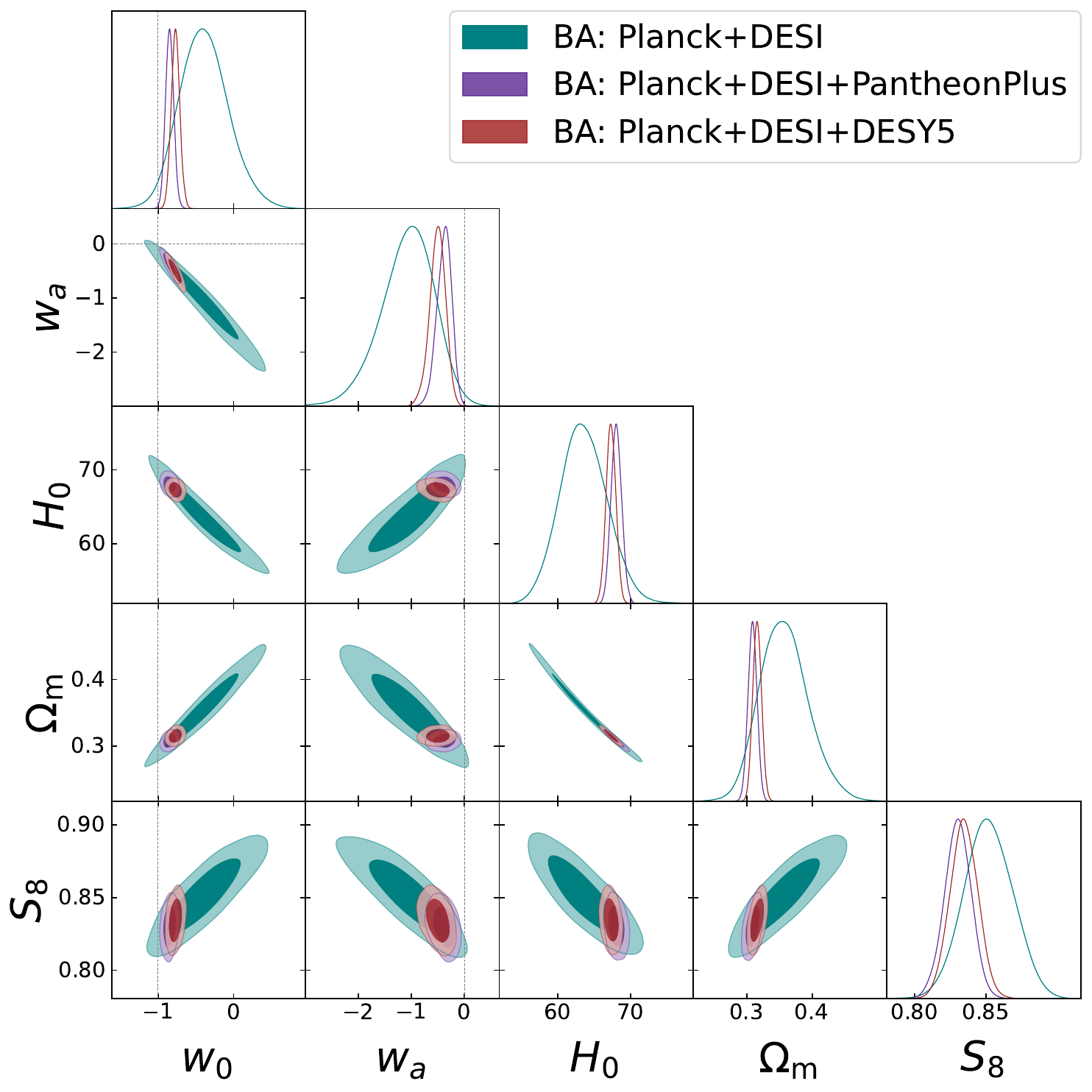}
    \caption{\textbf{BA parametrization \eqref{ba} -- } one-dimensional posterior distributions and two-dimensional marginalized contours for the main key parameters as obtained from the Planck+DESI, Planck+DESI+PantheonPlus, and Planck+DESI+DESY5 dataset combinations.
    }
    \label{fig:ba}
\end{figure}

The observational constraints for the last model analyzed in this work -- the BA parametrization -- are given in Tab.~\ref{tab:ba}. As usual, we illustrate the correlations among the key cosmic parameters in Fig.~\ref{fig:ba}.

\begingroup
\begin{table}[ht]
\centering
\renewcommand{\arraystretch}{1.2}
\resizebox{0.9\textwidth}{!}{\begin{tabular}{lccccccccccc}
\toprule
\textbf{Parameter} & \textbf{Planck+DESI} & \textbf{Planck+DESI+PantheonPlus} & \textbf{Planck+DESI+DESY5} \\
\hline \hline
$\Omega_\mathrm{c} h^2$ & $0.1201\pm 0.0010$ & $0.11963\pm 0.00099$ & $0.1198\pm 0.0010$\\
$\Omega_\mathrm{b} h^2$ & $0.02237\pm 0.00014$ & $0.02240\pm 0.00014$ & $0.02239\pm 0.00014$\\
$100\theta_\mathrm{MC}$ & $1.04090\pm 0.00029$ & $1.04097\pm 0.00030$ & $1.04095\pm 0.00029$\\
$\tau$ & $0.0523\pm 0.0073$ & $0.0544\pm 0.0074$ & $0.0539\pm 0.0074$ \\
$n_\mathrm{s}$ & $0.9649\pm 0.0039$ & $0.9663\pm 0.0039$ & $0.9658\pm 0.0038$\\
$\log(10^{10} A_\mathrm{s})$ & $3.039\pm 0.013$ & $3.044\pm 0.014$ & $3.043\pm 0.013$\\
$w_0$ & $-0.39^{+0.30}_{-0.34}$ & $-0.848\pm 0.054$ & $-0.770\pm 0.057$\\
$w_a$ & $-1.07^{+0.55}_{-0.43}$ & $-0.38^{+0.15}_{-0.13}$ & $-0.51^{+0.16}_{-0.14}$\\
$\Omega_\mathrm{m}$ & $0.357^{+0.033}_{-0.040}$ & $0.3084\pm 0.0069$ & $0.3155\pm 0.0066$\\
$\sigma_8$ & $0.783\pm 0.028$ & $0.8189\pm 0.0097$ & $0.8138\pm 0.0093$\\
$S_8$ & $0.852\pm 0.017$ & $0.8302\pm 0.0095$ & $0.8344\pm 0.0098$\\
$H_0 \mathrm{\,[km/s/Mpc]}$ & $63.6\pm 3.3$ & $68.03\pm 0.73$ & $67.30\pm 0.67$\\
$r_\mathrm{drag}\mathrm{\, [Mpc]}$ & $147.07\pm 0.24$ & $147.17\pm 0.23$ & $147.13\pm 0.24$\\
$\Delta\chi^2$ & $-8.7$ & $-9.4$ & $-16.2$\\
\bottomrule
\end{tabular}}
\caption{\textbf{BA parametrization \eqref{ba} -- } 68\% CL constraints on the free and derived cosmological parameters for 3 different data combinations detailed in Sec.~\ref{sec-data}. Negative values of $\Delta\chi^2 = \chi^2_{\text{BA}} - \chi^2_{\Lambda\text{CDM}}$ indicate an improvement in the fit to the data compared to $\Lambda$CDM.}
\label{tab:ba}
\end{table}
\endgroup

Combining Planck with DESI, we get $w_0 = -0.39^{+0.30}_{-0.34}$, approaching $-1/3$ and approximately $2\sigma$ away from $w_0 = -1$. Additionally, $w_a = -1.07^{+0.55}_{-0.43}$ is significantly different from $w_a = 0$, confirming the preference for DDE in a similar fashion to other parameterizations discussed throughout the manuscript.

The inclusion of PantheonPlus gives $w_0 = -0.848 \pm 0.054$ (deep in the quintessence regime) and $w_a = -0.38^{+0.15}_{-0.13}$ (non-zero at more than $2\sigma$). Thus, for Planck+DESI+PantheonPlus, evidence of dynamical dark energy is clearly indicated, consistent with all the other parameterizations described so far.

In the case of Planck+DESI+DESY5, $w_0 = -0.770 \pm 0.057$ shifts further away from $-1$, strengthening the preference for a quintessence EoS. Meanwhile, $w_a = -0.51^{+0.16}_{-0.14}$ is found to be non-zero at more than $3\sigma$, as visualized in Fig.~\ref{fig:ba}.

Interestingly, when comparing the difference between the best-fit $\chi^2$ obtained within each DDE model and the best-fit $\chi^2$ obtained within $\Lambda$CDM, this model consistently leads to the most significant improvement over $\Lambda$CDM across all three data combinations analyzed. It performs better than the CPL parameterization, as seen by comparing the last line in Tab.~\ref{tab:cpl} and the last line in Tab.~\ref{tab:ba}.  As discussed in Appendix~\ref{app:A}, when comparing the evolution of the EoS inferred in this model with the other cases analyzed so far, we find that at low redshift it behaves similarly to the CPL parameterization. The most notable differences emerge at $z \gtrsim 1$. In all models, the EoS moves deeply into phantom values (except for the JBP model, where it is compelled to rise back towards quintessence-like values). In contrast, within the BA model, $w(z)$ does not trend towards very negative values at $z \gtrsim 1$. While it remains phantom, it stabilizes on a distinctive, nearly flat plateau.

\section{Discussions and Conclusions}
\label{sec-summary}

The recent DESI BAO measurements, when combined with CMB data from Planck and two samples of Type Ia supernovae (Pantheon-Plus and DESY5), reveal a preference for a present-day quintessence-like equation of state that crossed into the phantom regime in the past. The statistical significance of this preference for dynamic dark energy ranges between 2.5$\sigma$ and 3.9$\sigma$, depending on the specific data combinations analyzed. A core \textit{ansatz} for this result is the use of the Chevallier-Polarski-Linder (CPL) parameterization to describe the redshift evolution of the equation of state. Despite its several advantages -- such as capturing the effective behavior of a wide range of models with up to 0.1\% accuracy -- the CPL parameterization forces the evolution of the equation of state to be linear in the scale factor.

In this paper, we tested whether and to what extent the preference for a present-day quintessence equation of state that evolves towards the phantom regime depends on the parameterization adopted to describe its dynamical behavior. To avoid broadening uncertainties in cosmological parameters and facilitate direct comparison with the baseline CPL case, we focused on some well-known alternative models: the exponential, Jassal-Bagla-Padmanabhan, logarithmic, and Barboza-Alcaniz parameterizations for the equation of state. Like the CPL model, all these parameterizations consist of only two free parameters: the present-day value of the equation of state ($w_0$) and a parameter quantifying its dynamical evolution ($w_a$). However, they allow for deviations from linear behavior at both high and low redshifts. Therefore, given the same pair of values ($w_0$, $w_a$), different late-time expansion histories are obtained within the four models, thereby affecting cosmological observables differently.

To assess whether the preference for a dynamical dark energy component characterized by $w_0 > -1$ and $w_a < 0$ remains a robust prediction of the data, we tested these models against the most recent high and low redshift observations: the Planck 2018 CMB measurements, DESI BAO, as well as PantheonPlus and DESY5 SN measurements. For all the dataset combinations explored -- i.e., Planck+DESI, Planck+DESI+PantheonPlus, and Planck+DESI+DESY5 -- we find that $w_0$ consistently remains in the quintessence regime. Additionally, the constraints on $w_a$ consistently indicate a preference for a dynamical evolution that crossed into the phantom regime ($w_a < 0$). Therefore, our findings confirm the DESI results, regardless of the parameterization adopted to describe the dynamics of the dark energy sector.

\begin{figure}[hp!]
    \centering
    \includegraphics[width=0.4\textwidth]{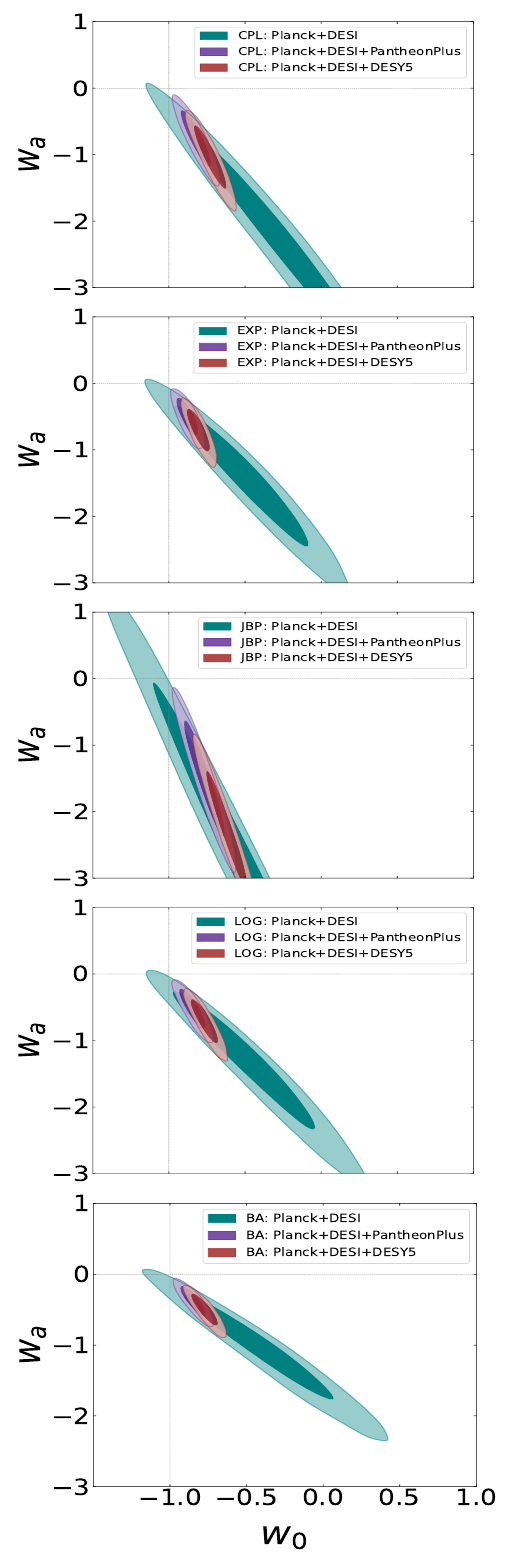}
    \caption{\textbf{Summary Plot --} two-dimensional marginalized contours in the ($w_0$, $w_a$) plane for all models and datasets analyzed in this study.}
    \label{fig:recap}
\end{figure}

Notably, convincing hints of a dynamical evolution of the equation of state are found even with just Planck+DESI. As clearly seen in Fig.~\ref{fig:recap} -- which summarizes the results for the different models -- the pair $w_0 = -1$ and $w_a = 0$ (corresponding to the standard cosmological constant model of structure formation, $\Lambda$CDM) always falls outside the 95\% confidence level contour.

However, the real step forward in terms of preference for dynamical dark energy comes when we consider Type Ia supernovae. Including distance moduli measurements gathered from the PantheonPlus catalog, the error bars on $w_0$ and $w_a$ tighten by a factor of 5 compared to Planck+DESI alone. The contours on $w_0$ significantly shrink within the quintessence portion of the parameter space $w_0 > -1$, while the contours on $w_a$ significantly reduce within the $w_a < 0$ region. Replacing PantheonPlus data with DESY5 SN measurements, the preference for dynamical dark energy becomes substantially more significant, to the point where it is not an exaggeration to refer to it as \textit{evidence} rather than mere preference. This is again clearly illustrated in Fig.~\ref{fig:recap}: for all models, the constraints shift further away from a cosmological constant, which always falls well outside the 95\% marginalized probability contours.

At first glance, Fig.~\ref{fig:recap} also reveals that the contours in the $w_0$-$w_a$ plane show similar trends for all four parameterizations (including the baseline CPL case), especially when SN measurements are included in the analysis. This simultaneously underscores the intrinsic robustness of the preference for dynamical dark energy as reported by the DESI BAO and SN measurements and its resilience against different parameterizations. Given these results, there is solid ground to conclude that the choice of parameterization has a minimal impact.

Last but not least, we examined statistical metrics to quantify the extent to which the different parameterizations analyzed in this study are successful in explaining observations. Specifically, for each model and data combination, we report the difference between the best-fit $\chi^2$ obtained within each dynamical dark energy model and the best-fit $\chi^2$ obtained within $\Lambda$CDM.
\newpage

Once more, all models exhibit similar trends: for Planck+DESI, we consistently observe an improvement in the fit over $\Lambda$CDM, with $\Delta \chi^2$ ranging from $-5.6$ to $-8.7$, depending on the specific model. This improvement in the fit is further enhanced when PantheonPlus SN measurements are included ($\Delta \chi^2$ ranges from $-6.4$ to $-9.4$) and is substantially increased -- by up to a factor of $\sim 2$ -- when adopting DESY5 SN data (in this case, $\Delta \chi^2$ ranges from $-14.8$ to $-16.2$). This trend follows the overall preference for dynamical dark energy discussed thus far. Interestingly, the linear CPL parameterization is never the best-fitting model. The equation of state proposed by Barboza-Alcaniz, given by Eq.~\eqref{ba}, consistently leads to the most significant improvement in $\Delta \chi^2$ over $\Lambda$CDM across all three data combinations analyzed. Conversely, the Jassal-Bagla-Padmanabhan parameterization, given by Eq.~\eqref{jbp}, shows the smallest improvement in fit compared to $\Lambda$CDM among the models considered. The only exception is for Planck+DESI+DESY5, where $\Delta \chi^2 = -16$ indicates a better fit to this dataset compared to the CPL, logarithmic, and exponential parameterizations, although it is still smaller compared to the Barboza-Alcaniz model. For further discussion and physical interpretation of the different phenomenological behaviors of the models analyzed so far, we refer to Appendix~\ref{app:A}. Specifically, Fig.~\ref{fig:7} presents constraints on the evolution of the equation of state with respect to redshift, as inferred from various datasets across all models. Tab.~\ref{tab.Results.z} provides constraints on other important properties, such as the pivot redshift, the corresponding values (and uncertainties) of the equation of state, and the epoch of the quintessence-to-phantom transition. Overall, these results support the main conclusions regarding the resilience of the DESI and SN preference for evolving dark energy, while suggesting that current data are approaching a precision that could enhance our understanding of its physical nature, should future surveys and data releases confirm these findings.

\acknowledgments 
The authors thank the referee for many important comments that improved the manuscript. W.G.\ acknowledges support from the Lancaster–Sheffield Consortium for Fundamental Physics through the Science and Technology Facilities Council (STFC) grant ST/X000621/1.  S.P.\ acknowledges the financial support from the Department of Science and Technology (DST), Govt. of India under the Scheme   ``Fund for Improvement of S\&T Infrastructure (FIST)'' (File No. SR/FST/MS-I/2019/41).  E.D.V\ acknowledges support from the Royal Society through a Royal Society Dorothy Hodgkin Research Fellowship. This article is based upon work from the COST Action CA21136 ``Addressing observational tensions in cosmology with systematics and fundamental physics'' (CosmoVerse), supported by COST (European Cooperation in Science and Technology). 

\appendix
\section{Equation of State, pivot redshift, and phantom crossing across the different models}
\label{app:A}

\begin{figure}[htp!]
    \centering
    \includegraphics[width=0.6\textwidth]{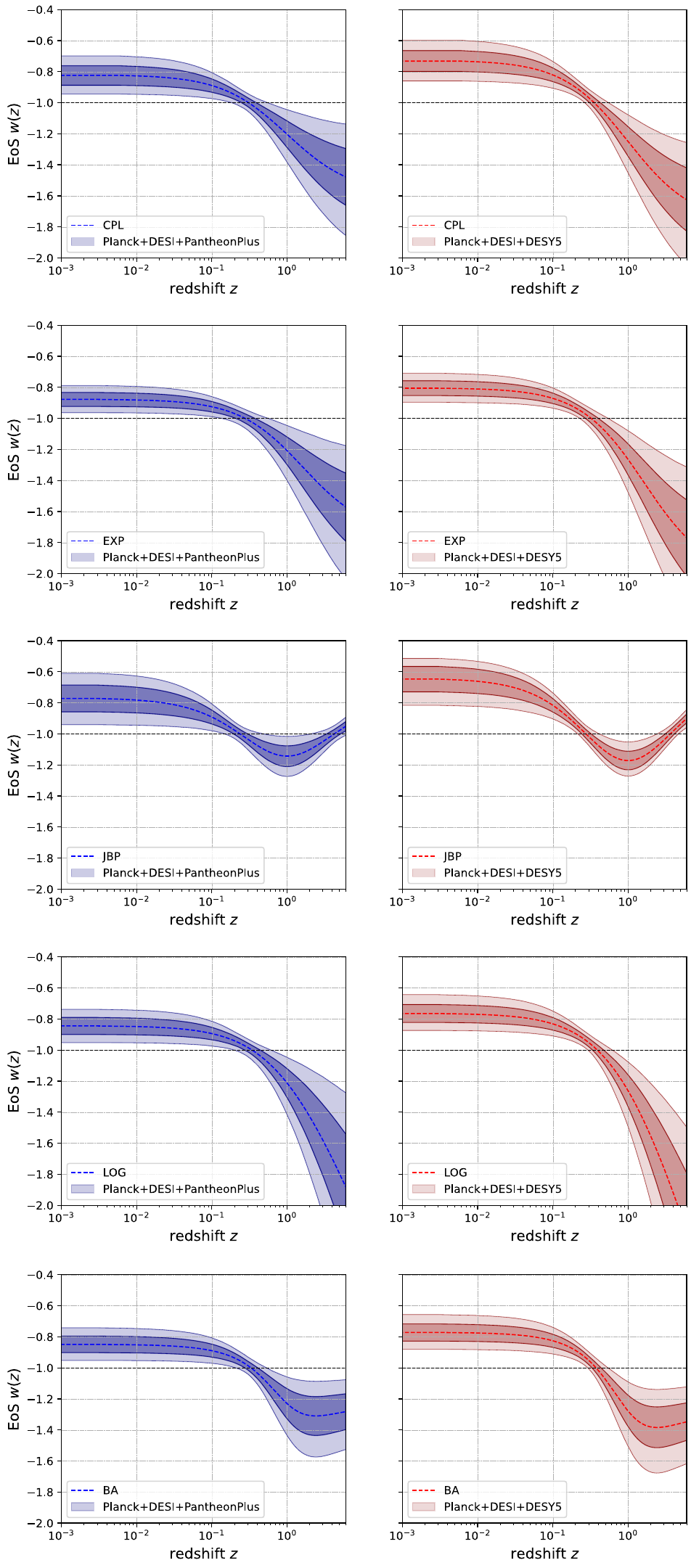}
    \caption{Evolution of $w(z)$ for $0 < z < 6$ across all DDE models, inferred from CMB+DESI+PantheonPlus (left panels) and CMB+DESI+DESY5 (right panels). The dashed lines represent the mean values, while the dark and light shaded regions indicate the $1\sigma$ and $2\sigma$ uncertainties, respectively.}
    \label{fig:7}
\end{figure}

\begin{table}[htb]
\centering
\renewcommand{\arraystretch}{1.5}
\resizebox{0.9\textwidth}{!}{\begin{tabular}{l @{\hspace{1 cm}} l @{\hspace{1 cm}} c @{\hspace{1 cm}} c @{\hspace{1 cm}} c}
\toprule
\textbf{Model} & \textbf{Dataset} & \boldmath{$z_p$} & \boldmath{$w(z_p)$} & \boldmath{$z_c$} \\
\hline\hline

CPL & Planck+DESI+PantheonPlus   & $0.27$  & $-0.982\pm  0.028$  & $0.31^{+0.08}_{-0.06}$ \\
& Planck+DESI+DESY5 & $0.25$ & $-0.937 \pm 0.026$ & $0.35^{+0.07}_{-0.05}$ \\
\hline\hline
Exponential & Planck+DESI+PantheonPlus   & $0.21$ & $-0.974\pm  0.028$ &$0.27^{+0.10}_{-0.07}$ \\
& Planck+DESI+DESY5 & $0.22$ & $-0.942\pm  0.026$ & $0.32^{+0.08}_{-0.05}$ \\
\hline\hline
JBP & Planck+DESI+PantheonPlus   & $0.21$   & $-0.985 \pm 0.027$  &$0.24^{+0.06}_{-0.04}$ \\
&& $4.6$&$-0.988\pm0.027$&$4.2\pm0.9$\\
& Planck+DESI+DESY5 & $0.21$ & $-0.945 \pm 0.026$ & $0.27^{+0.05}_{-0.03}$ \\
&&$4.7$&$-0.946\pm 0.026$&$3.6\pm0.5$\\
\hline\hline
Logarithmic & Planck+DESI+PantheonPlus   & $0.29$  & $-0.979 \pm 0.028$  &$0.34^{+0.10}_{-0.07}$ \\
& Planck+DESI+DESY5 & $0.26$ & $-0.930^{+0.027}_{-0.026}$ & $0.39^{+0.08}_{-0.05}$ \\
\hline\hline
BA & Planck+DESI+PantheonPlus   & $0.28$  & $-0.974^{+0.027}_{-0.028}$  &$0.33^{+0.08}_{-0.06}$ \\
& Planck+DESI+DESY5 & $0.28$ & $-0.937^{+0.026}_{-0.027}$ & $0.37^{+0.06}_{-0.04}$ \\
\hline
\bottomrule
\end{tabular}}
\caption{Constraints at 68\% CL on the pivot redshift $z_p$, the corresponding value of the EoS $w(z_p)$, and the redshift $z_c$ where the EoS crosses the phantom divide, for Planck+DESI+PantheonPlus and Planck+DESI+DESY5.}
\label{tab.Results.z}
\end{table}
In this article, we have emphasized the resilience of the results recently delivered by the DESI collaboration, showing that DESI BAO data, in combination with Planck CMB observations and two different catalogs of SN distance moduli measurements (i.e., PantheonPlus and DESY5), consistently indicate a preference for DDE across various parameterizations of the EoS. While the primary goal was to confirm that this preference remains stable regardless of the specific DDE model, minor differences have emerged across the five cases analyzed, warranting further investigation. In this appendix, we explore these differences in more detail, aiming to provide a stronger physical interpretation of the results presented in the manuscript. For all five models, we reconstruct the evolution of the EoS with redshift, $w(z)$, based on the constraints on $w_0$ and $w_a$ inferred from the Planck+DESI+PantheonPlus and Planck+DESI+DESY5 datasets. In Fig.~\ref{fig:7}, we present the mean value of $w(z)$ (dashed lines), along with its uncertainties at the 68\% (dark regions) and 95\% (light regions) CL, across the redshift range $0 \lesssim z \lesssim 6$ for all models and both data combinations. From the reconstructed shape of $w(z)$, we extract crucial information that helps compare the different models and clarify the results presented in the manuscript. Specifically, in Tab.~\ref{tab.Results.z}, we present the results for:
\begin{itemize}
    \item[\textit{(i)}] the pivot redshift $z_p$ and the corresponding value of the EoS, $w(z_p)$, which indicate the redshift and the EoS value at which $w(z)$ is best constrained by the two datasets across the five models;\footnote{See, e.g., Refs~\cite{Huterer:2000mj,Albrecht:2006um} for discussions on the importance of the pivot redshift.}
    \item [\textit{(ii)}] the redshift $z_c$ when the EoS crosses the phantom divide (i.e., $w(z_c) = -1$), informing us of when the phantom crossing occurs, along with their respective uncertainties. 
\end{itemize}
Starting from the baseline CPL model as the reference case,\footnote{Note that the features presented in this appendix for the CPL model have been discussed in detail by the DESI collaboration -- see, e.g., Sec. 5.2 of Ref.~\cite{DESI:2024mwx}. As we essentially recover all of the DESI results, we omit further discussion of the CPL model here.} we summarize the main features and differences across the various models.

\clearpage

\begin{itemize}[leftmargin=*]
    \item \textbf{Exponential:} As seen when comparing the top panels with those in the second row from the top in Fig.~\ref{fig:7}, from a phenomenological perspective, the exponential parameterization closely resembles the CPL model. This similarity was already highlighted in the main text when comparing the improvement in $\Delta \chi^2$ over $\Lambda$CDM (which is quite similar for both scenarios). The agreement in predictions is further supported by Tab.~\ref{tab.Results.z}. The only noticeable difference between the models is a slightly smaller pivot redshift $z_p$ in the exponential parameterization. However, the EoS at this pivot is constrained with comparable precision. Additionally, the predictions regarding the redshift of the phantom crossing $z_c$ agree within one standard deviation for both Planck+DESI+PantheonPlus and Planck+DESI+DESY5.
    
    \item \textbf{JBP:} Among the five models analyzed, the JBP parameterization presents a more nuanced phenomenology regarding the evolution of the DE EoS. As shown in the third panels from the top in Fig.~\ref{fig:7}, due to its quadratic nature in the scale factor, the evolution of the EoS within the JBP parameterization crosses $w = -1$ twice. At very low redshift, it behaves similarly to the other parameterizations, remaining within the quintessence region $w(z) > -1$, albeit with larger uncertainties compared to the other models. The first quintessence-to-phantom transition is estimated to occur at $z = 0.24^{+0.06}_{-0.04}$ ($z = 0.27^{+0.05}_{-0.03}$) for Planck+DESI+PantheonPlus (Planck+DESI+DESY5) at 68\% CL. After this transition, $w(z)$ approaches a minimum value around $z = 1$ before rising towards less negative values. Eventually, a second phantom-to-quintessence crossing occurs at $z = 4.2 \pm 0.9$ ($z = 3.6 \pm 0.5$) for Planck+DESI+PantheonPlus (Planck+DESI+DESY5), both at 95\% CL. This behavior contrasts with other models, where the EoS remains within the phantom regime, often trending towards more negative $w(z)$ values at high redshift. In contrast, within the JBP model, the EoS cannot move towards (more) phantom values but is compelled to transition back towards less negative values at $z > 1$. The double crossing of regimes achieved within this parameterization is also reflected in the pivot scale $z_p$, at which the EoS is well constrained by data. In Tab.~\ref{tab.Results.z}, we distinguish between two different regimes: the redshift range $0 < z < 1$ (capturing the first quintessence-to-phantom crossing) and the range $z > 1$ (covering the second phantom-to-quintessence crossing). In these two regions, we identify two distinct pivot redshifts: the first at $z_p \sim 0.21$ and the second at $z_p \sim 4.6$. In both cases (and for both datasets), the EoS is constrained within the same minimal error. This confirms that, within this model, due to the functional form of the EoS, constraints at low redshift (i.e., around $z \sim 0.21$) also dictate the behavior of the parameterization at higher redshifts. The interplay between low and high redshift behaviors, as highlighted by the two pivot redshifts, could contribute to the increased uncertainties at low redshift and the tilting of the probability contours seen in Fig.~\ref{fig:recap}. As discussed in the main manuscript, this model offers relatively modest improvements in the fit compared to $\Lambda$CDM, particularly in datasets covering $z \gtrsim 1$, where the model’s deviations from the others become more pronounced.
    
    \item \textbf{Logarithmic:} When it comes to the logarithmic parameterization, the behavior of $w(z)$ for $z \lesssim 1$, shown in the fourth panel of Fig.~\ref{fig:7}, is similar to that of the CPL and exponential models. This is also reflected in the values we inferred for $z_p$, $w(z_p)$, and $z_c$, all summarized in Tab.~\ref{tab.Results.z} and consistent with those models. However, we observe that at $z \gtrsim 1$, the EoS is forced down into deep phantom values, and the descent towards these very negative values is steeper than in the CPL and exponential cases. This is due to the fact that at $z \gtrsim 1$, the scale factor $a$ approaches small values (moving towards $a \to 0$), causing $\log(a)$ to decrease to negative values quite rapidly. This sudden decline in $w(z)$ for $z \gtrsim 1$ can lead to changes in the fit to data spanning $1 \lesssim z \lesssim 3$, which is covered by BAO and SN observations, resulting in the differences in the $\chi^2$ of the fit discussed in the manuscript.
    
    \item \textbf{BA:} Last but not least, the evolution of $w(z)$ obtained for the BA model is presented in the bottom panel of Fig.~\ref{fig:7}. As we argued in the manuscript, this model provides the most significant improvement in the fit over $\Lambda$CDM across all datasets analyzed in this study. Therefore, it is interesting to examine what is different in the evolution of the EoS compared to the other models. Looking at the low-redshift part of the EoS, we see that the model behaves very similarly to CPL (and its relatives). However, for the pivot redshift, we obtain $z_p = 0.28$, slightly larger than in any other model, while $w(z_p)$ takes values consistent with the other cases. We estimate the quintessence-to-phantom transition to occur at $z_c = 0.33^{+0.08}_{-0.06}$ ($z_c = 0.37^{+0.06}_{-0.04}$) for Planck+DESI+PantheonPlus (Planck+DESI+DESY5) at 68\% CL. The most noticeable difference in the EoS occurs at $z \gtrsim 1$. In all other models studied so far, the EoS either moved deeply into phantom values (characterized by more or less steep functional forms of $w(z)$) or was compelled to increase back towards quintessence-like values in the JBP model. Referring to the bottom panel of Fig.~\ref{fig:7}, we observe that for $z \gtrsim 1$, the evolution of $w(z)$ in the BA model remains phantom but does not trend towards very negative values. Instead, $w(z)$ stabilizes on a sort of second plateau that is distinctive of the BA model.
\end{itemize}

\bibliographystyle{JHEP}
\bibliography{biblio}

\providecommand{\href}[2]{#2}\begingroup\raggedright\begin{thebibliography}{100}

\bibitem{SupernovaSearchTeam:1998fmf}
{\scshape Supernova Search Team} collaboration, \emph{{Observational evidence from supernovae for an accelerating universe and a cosmological constant}}, \href{https://doi.org/10.1086/300499}{\emph{Astron. J.} {\bfseries 116} (1998) 1009} [\href{https://arxiv.org/abs/astro-ph/9805201}{{\ttfamily astro-ph/9805201}}].

\bibitem{SupernovaCosmologyProject:1998vns}
{\scshape Supernova Cosmology Project} collaboration, \emph{{Measurements of $\Omega$ and $\Lambda$ from 42 High Redshift Supernovae}}, \href{https://doi.org/10.1086/307221}{\emph{Astrophys. J.} {\bfseries 517} (1999) 565} [\href{https://arxiv.org/abs/astro-ph/9812133}{{\ttfamily astro-ph/9812133}}].

\bibitem{SupernovaSearchTeam:2001qse}
{\scshape Supernova Search Team} collaboration, \emph{{The farthest known supernova: support for an accelerating universe and a glimpse of the epoch of deceleration}}, \href{https://doi.org/10.1086/322348}{\emph{Astrophys. J.} {\bfseries 560} (2001) 49} [\href{https://arxiv.org/abs/astro-ph/0104455}{{\ttfamily astro-ph/0104455}}].

\bibitem{SDSS:2003eyi}
{\scshape SDSS} collaboration, \emph{{Cosmological parameters from SDSS and WMAP}}, \href{https://doi.org/10.1103/PhysRevD.69.103501}{\emph{Phys. Rev. D} {\bfseries 69} (2004) 103501} [\href{https://arxiv.org/abs/astro-ph/0310723}{{\ttfamily astro-ph/0310723}}].

\bibitem{SDSS:2003lnz}
{\scshape SDSS} collaboration, \emph{{Physical evidence for dark energy}},  \href{https://arxiv.org/abs/astro-ph/0307335}{{\ttfamily astro-ph/0307335}}.

\bibitem{SupernovaSearchTeam:2003cyd}
{\scshape Supernova Search Team} collaboration, \emph{{Cosmological results from high-z supernovae}}, \href{https://doi.org/10.1086/376865}{\emph{Astrophys. J.} {\bfseries 594} (2003) 1} [\href{https://arxiv.org/abs/astro-ph/0305008}{{\ttfamily astro-ph/0305008}}].

\bibitem{SupernovaCosmologyProject:2003dcn}
{\scshape Supernova Cosmology Project} collaboration, \emph{{New constraints on Omega(M), Omega(lambda), and w from an independent set of eleven high-redshift supernovae observed with HST}}, \href{https://doi.org/10.1086/378560}{\emph{Astrophys. J.} {\bfseries 598} (2003) 102} [\href{https://arxiv.org/abs/astro-ph/0309368}{{\ttfamily astro-ph/0309368}}].

\bibitem{SDSS:2004kqt}
{\scshape SDSS} collaboration, \emph{{Cosmological parameter analysis including SDSS Ly-alpha forest and galaxy bias: Constraints on the primordial spectrum of fluctuations, neutrino mass, and dark energy}}, \href{https://doi.org/10.1103/PhysRevD.71.103515}{\emph{Phys. Rev. D} {\bfseries 71} (2005) 103515} [\href{https://arxiv.org/abs/astro-ph/0407372}{{\ttfamily astro-ph/0407372}}].

\bibitem{Feng:2004ad}
B.~Feng, X.-L.~Wang and X.-M.~Zhang, \emph{{Dark energy constraints from the cosmic age and supernova}}, \href{https://doi.org/10.1016/j.physletb.2004.12.071}{\emph{Phys. Lett. B} {\bfseries 607} (2005) 35} [\href{https://arxiv.org/abs/astro-ph/0404224}{{\ttfamily astro-ph/0404224}}].

\bibitem{SupernovaSearchTeam:2004lze}
{\scshape Supernova Search Team} collaboration, \emph{{Type Ia supernova discoveries at z \ensuremath{>} 1 from the Hubble Space Telescope: Evidence for past deceleration and constraints on dark energy evolution}}, \href{https://doi.org/10.1086/383612}{\emph{Astrophys. J.} {\bfseries 607} (2004) 665} [\href{https://arxiv.org/abs/astro-ph/0402512}{{\ttfamily astro-ph/0402512}}].

\bibitem{SNLS:2005qlf}
{\scshape SNLS} collaboration, \emph{{The Supernova Legacy Survey: Measurement of $\Omega_M$, $\Omega_\Lambda$ and ${\cal w}$ from the first year data set}}, \href{https://doi.org/10.1051/0004-6361:20054185}{\emph{Astron. Astrophys.} {\bfseries 447} (2006) 31} [\href{https://arxiv.org/abs/astro-ph/0510447}{{\ttfamily astro-ph/0510447}}].

\bibitem{SDSS:2005xqv}
{\scshape SDSS} collaboration, \emph{{Detection of the Baryon Acoustic Peak in the Large-Scale Correlation Function of SDSS Luminous Red Galaxies}}, \href{https://doi.org/10.1086/466512}{\emph{Astrophys. J.} {\bfseries 633} (2005) 560} [\href{https://arxiv.org/abs/astro-ph/0501171}{{\ttfamily astro-ph/0501171}}].

\bibitem{Eisenstein:2006nk}
D.J.~Eisenstein, H.-j.~Seo, E.~Sirko and D.~Spergel, \emph{{Improving Cosmological Distance Measurements by Reconstruction of the Baryon Acoustic Peak}}, \href{https://doi.org/10.1086/518712}{\emph{Astrophys. J.} {\bfseries 664} (2007) 675} [\href{https://arxiv.org/abs/astro-ph/0604362}{{\ttfamily astro-ph/0604362}}].

\bibitem{SDSS:2006lmn}
{\scshape SDSS} collaboration, \emph{{Cosmological Constraints from the SDSS Luminous Red Galaxies}}, \href{https://doi.org/10.1103/PhysRevD.74.123507}{\emph{Phys. Rev. D} {\bfseries 74} (2006) 123507} [\href{https://arxiv.org/abs/astro-ph/0608632}{{\ttfamily astro-ph/0608632}}].

\bibitem{Sahni:2006pa}
V.~Sahni and A.~Starobinsky, \emph{{Reconstructing Dark Energy}}, \href{https://doi.org/10.1142/S0218271806009704}{\emph{Int. J. Mod. Phys. D} {\bfseries 15} (2006) 2105} [\href{https://arxiv.org/abs/astro-ph/0610026}{{\ttfamily astro-ph/0610026}}].

\bibitem{ESSENCE:2007acn}
{\scshape ESSENCE} collaboration, \emph{{Observational Constraints on the Nature of the Dark Energy: First Cosmological Results from the ESSENCE Supernova Survey}}, \href{https://doi.org/10.1086/518642}{\emph{Astrophys. J.} {\bfseries 666} (2007) 694} [\href{https://arxiv.org/abs/astro-ph/0701041}{{\ttfamily astro-ph/0701041}}].

\bibitem{Vikhlinin:2008ym}
A.~Vikhlinin et~al., \emph{{Chandra Cluster Cosmology Project III: Cosmological Parameter Constraints}}, \href{https://doi.org/10.1088/0004-637X/692/2/1060}{\emph{Astrophys. J.} {\bfseries 692} (2009) 1060} [\href{https://arxiv.org/abs/0812.2720}{{\ttfamily 0812.2720}}].

\bibitem{Stern:2009ep}
D.~Stern, R.~Jimenez, L.~Verde, M.~Kamionkowski and S.A.~Stanford, \emph{{Cosmic Chronometers: Constraining the Equation of State of Dark Energy. I: H(z) Measurements}}, \href{https://doi.org/10.1088/1475-7516/2010/02/008}{\emph{JCAP} {\bfseries 02} (2010) 008} [\href{https://arxiv.org/abs/0907.3149}{{\ttfamily 0907.3149}}].

\bibitem{Sherwin:2011gv}
B.D.~Sherwin et~al., \emph{{Evidence for dark energy from the cosmic microwave background alone using the Atacama Cosmology Telescope lensing measurements}}, \href{https://doi.org/10.1103/PhysRevLett.107.021302}{\emph{Phys. Rev. Lett.} {\bfseries 107} (2011) 021302} [\href{https://arxiv.org/abs/1105.0419}{{\ttfamily 1105.0419}}].

\bibitem{WMAP:2012fli}
{\scshape WMAP} collaboration, \emph{{Nine-Year Wilkinson Microwave Anisotropy Probe (WMAP) Observations: Final Maps and Results}}, \href{https://doi.org/10.1088/0067-0049/208/2/20}{\emph{Astrophys. J. Suppl.} {\bfseries 208} (2013) 20} [\href{https://arxiv.org/abs/1212.5225}{{\ttfamily 1212.5225}}].

\bibitem{WMAP:2012nax}
{\scshape WMAP} collaboration, \emph{{Nine-Year Wilkinson Microwave Anisotropy Probe (WMAP) Observations: Cosmological Parameter Results}}, \href{https://doi.org/10.1088/0067-0049/208/2/19}{\emph{Astrophys. J. Suppl.} {\bfseries 208} (2013) 19} [\href{https://arxiv.org/abs/1212.5226}{{\ttfamily 1212.5226}}].

\bibitem{BOSS:2012dmf}
{\scshape BOSS} collaboration, \emph{{The Baryon Oscillation Spectroscopic Survey of SDSS-III}}, \href{https://doi.org/10.1088/0004-6256/145/1/10}{\emph{Astron. J.} {\bfseries 145} (2013) 10} [\href{https://arxiv.org/abs/1208.0022}{{\ttfamily 1208.0022}}].

\bibitem{deJong:2012zb}
{\scshape Astro-WISE, KiDS} collaboration, \emph{{The Kilo-Degree Survey}}, \href{https://doi.org/10.1007/s10686-012-9306-1}{\emph{Exper. Astron.} {\bfseries 35} (2013) 25} [\href{https://arxiv.org/abs/1206.1254}{{\ttfamily 1206.1254}}].

\bibitem{BOSS:2013rlg}
{\scshape BOSS} collaboration, \emph{{The clustering of galaxies in the SDSS-III Baryon Oscillation Spectroscopic Survey: baryon acoustic oscillations in the Data Releases 10 and 11 Galaxy samples}}, \href{https://doi.org/10.1093/mnras/stu523}{\emph{Mon. Not. Roy. Astron. Soc.} {\bfseries 441} (2014) 24} [\href{https://arxiv.org/abs/1312.4877}{{\ttfamily 1312.4877}}].

\bibitem{Weinberg:2013agg}
D.H.~Weinberg, M.J.~Mortonson, D.J.~Eisenstein, C.~Hirata, A.G.~Riess and E.~Rozo, \emph{{Observational Probes of Cosmic Acceleration}}, \href{https://doi.org/10.1016/j.physrep.2013.05.001}{\emph{Phys. Rept.} {\bfseries 530} (2013) 87} [\href{https://arxiv.org/abs/1201.2434}{{\ttfamily 1201.2434}}].

\bibitem{BOSS:2013uda}
{\scshape BOSS} collaboration, \emph{{The clustering of galaxies in the SDSS-III Baryon Oscillation Spectroscopic Survey: Testing gravity with redshift-space distortions using the power spectrum multipoles}}, \href{https://doi.org/10.1093/mnras/stu1051}{\emph{Mon. Not. Roy. Astron. Soc.} {\bfseries 443} (2014) 1065} [\href{https://arxiv.org/abs/1312.4611}{{\ttfamily 1312.4611}}].

\bibitem{BOSS:2014hwf}
{\scshape BOSS} collaboration, \emph{{Baryon acoustic oscillations in the Ly\ensuremath{\alpha} forest of BOSS DR11 quasars}}, \href{https://doi.org/10.1051/0004-6361/201423969}{\emph{Astron. Astrophys.} {\bfseries 574} (2015) A59} [\href{https://arxiv.org/abs/1404.1801}{{\ttfamily 1404.1801}}].

\bibitem{SDSS:2014iwm}
{\scshape SDSS} collaboration, \emph{{Improved cosmological constraints from a joint analysis of the SDSS-II and SNLS supernova samples}}, \href{https://doi.org/10.1051/0004-6361/201423413}{\emph{Astron. Astrophys.} {\bfseries 568} (2014) A22} [\href{https://arxiv.org/abs/1401.4064}{{\ttfamily 1401.4064}}].

\bibitem{BOSS:2014hhw}
{\scshape BOSS} collaboration, \emph{{Cosmological implications of baryon acoustic oscillation measurements}}, \href{https://doi.org/10.1103/PhysRevD.92.123516}{\emph{Phys. Rev. D} {\bfseries 92} (2015) 123516} [\href{https://arxiv.org/abs/1411.1074}{{\ttfamily 1411.1074}}].

\bibitem{Ross:2014qpa}
A.J.~Ross, L.~Samushia, C.~Howlett, W.J.~Percival, A.~Burden and M.~Manera, \emph{{The clustering of the SDSS DR7 main Galaxy sample \textendash{} I. A 4 per cent distance measure at $z = 0.15$}}, \href{https://doi.org/10.1093/mnras/stv154}{\emph{Mon. Not. Roy. Astron. Soc.} {\bfseries 449} (2015) 835} [\href{https://arxiv.org/abs/1409.3242}{{\ttfamily 1409.3242}}].

\bibitem{Moresco:2016mzx}
M.~Moresco, L.~Pozzetti, A.~Cimatti, R.~Jimenez, C.~Maraston, L.~Verde et~al., \emph{{A 6\% measurement of the Hubble parameter at $z\sim0.45$: direct evidence of the epoch of cosmic re-acceleration}}, \href{https://doi.org/10.1088/1475-7516/2016/05/014}{\emph{JCAP} {\bfseries 05} (2016) 014} [\href{https://arxiv.org/abs/1601.01701}{{\ttfamily 1601.01701}}].

\bibitem{Moresco:2016nqq}
M.~Moresco, R.~Jimenez, L.~Verde, A.~Cimatti, L.~Pozzetti, C.~Maraston et~al., \emph{{Constraining the time evolution of dark energy, curvature and neutrino properties with cosmic chronometers}}, \href{https://doi.org/10.1088/1475-7516/2016/12/039}{\emph{JCAP} {\bfseries 12} (2016) 039} [\href{https://arxiv.org/abs/1604.00183}{{\ttfamily 1604.00183}}].

\bibitem{Rubin:2016iqe}
D.~Rubin and B.~Hayden, \emph{{Is the expansion of the universe accelerating? All signs point to yes}}, \href{https://doi.org/10.3847/2041-8213/833/2/L30}{\emph{Astrophys. J. Lett.} {\bfseries 833} (2016) L30} [\href{https://arxiv.org/abs/1610.08972}{{\ttfamily 1610.08972}}].

\bibitem{BOSS:2016wmc}
{\scshape BOSS} collaboration, \emph{{The clustering of galaxies in the completed SDSS-III Baryon Oscillation Spectroscopic Survey: cosmological analysis of the DR12 galaxy sample}}, \href{https://doi.org/10.1093/mnras/stx721}{\emph{Mon. Not. Roy. Astron. Soc.} {\bfseries 470} (2017) 2617} [\href{https://arxiv.org/abs/1607.03155}{{\ttfamily 1607.03155}}].

\bibitem{DES:2016jjg}
{\scshape DES} collaboration, \emph{{The Dark Energy Survey: more than dark energy \textendash{} an overview}}, \href{https://doi.org/10.1093/mnras/stw641}{\emph{Mon. Not. Roy. Astron. Soc.} {\bfseries 460} (2016) 1270} [\href{https://arxiv.org/abs/1601.00329}{{\ttfamily 1601.00329}}].

\bibitem{Haridasu:2017lma}
B.S.~Haridasu, V.V.~Lukovi\'c, R.~D'Agostino and N.~Vittorio, \emph{{Strong evidence for an accelerating universe}}, \href{https://doi.org/10.1051/0004-6361/201730469}{\emph{Astron. Astrophys.} {\bfseries 600} (2017) L1} [\href{https://arxiv.org/abs/1702.08244}{{\ttfamily 1702.08244}}].

\bibitem{DES:2017qwj}
{\scshape DES} collaboration, \emph{{Dark Energy Survey Year 1 results: Cosmological constraints from cosmic shear}}, \href{https://doi.org/10.1103/PhysRevD.98.043528}{\emph{Phys. Rev. D} {\bfseries 98} (2018) 043528} [\href{https://arxiv.org/abs/1708.01538}{{\ttfamily 1708.01538}}].

\bibitem{Pan-STARRS1:2017jku}
{\scshape Pan-STARRS1} collaboration, \emph{{The Complete Light-curve Sample of Spectroscopically Confirmed SNe Ia from Pan-STARRS1 and Cosmological Constraints from the Combined Pantheon Sample}}, \href{https://doi.org/10.3847/1538-4357/aab9bb}{\emph{Astrophys. J.} {\bfseries 859} (2018) 101} [\href{https://arxiv.org/abs/1710.00845}{{\ttfamily 1710.00845}}].

\bibitem{Planck:2018nkj}
{\scshape Planck} collaboration, \emph{{Planck 2018 results. I. Overview and the cosmological legacy of Planck}}, \href{https://doi.org/10.1051/0004-6361/201833880}{\emph{Astron. Astrophys.} {\bfseries 641} (2020) A1} [\href{https://arxiv.org/abs/1807.06205}{{\ttfamily 1807.06205}}].

\bibitem{Planck:2018vyg}
{\scshape Planck} collaboration, \emph{{Planck 2018 results. VI. Cosmological parameters}}, \href{https://doi.org/10.1051/0004-6361/201833910}{\emph{Astron. Astrophys.} {\bfseries 641} (2020) A6} [\href{https://arxiv.org/abs/1807.06209}{{\ttfamily 1807.06209}}].

\bibitem{Gomez-Valent:2018gvm}
A.~G\'omez-Valent, \emph{{Quantifying the evidence for the current speed-up of the Universe with low and intermediate-redshift data. A more model-independent approach}}, \href{https://doi.org/10.1088/1475-7516/2019/05/026}{\emph{JCAP} {\bfseries 05} (2019) 026} [\href{https://arxiv.org/abs/1810.02278}{{\ttfamily 1810.02278}}].

\bibitem{Yang:2019fjt}
Y.~Yang and Y.~Gong, \emph{{The evidence of cosmic acceleration and observational constraints}}, \href{https://doi.org/10.1088/1475-7516/2020/06/059}{\emph{JCAP} {\bfseries 06} (2020) 059} [\href{https://arxiv.org/abs/1912.07375}{{\ttfamily 1912.07375}}].

\bibitem{ACT:2020frw}
{\scshape ACT} collaboration, \emph{{The Atacama Cosmology Telescope: a measurement of the Cosmic Microwave Background power spectra at 98 and 150 GHz}}, \href{https://doi.org/10.1088/1475-7516/2020/12/045}{\emph{JCAP} {\bfseries 12} (2020) 045} [\href{https://arxiv.org/abs/2007.07289}{{\ttfamily 2007.07289}}].

\bibitem{ACT:2020gnv}
{\scshape ACT} collaboration, \emph{{The Atacama Cosmology Telescope: DR4 Maps and Cosmological Parameters}}, \href{https://doi.org/10.1088/1475-7516/2020/12/047}{\emph{JCAP} {\bfseries 12} (2020) 047} [\href{https://arxiv.org/abs/2007.07288}{{\ttfamily 2007.07288}}].

\bibitem{eBOSS:2020yzd}
{\scshape eBOSS} collaboration, \emph{{Completed SDSS-IV extended Baryon Oscillation Spectroscopic Survey: Cosmological implications from two decades of spectroscopic surveys at the Apache Point Observatory}}, \href{https://doi.org/10.1103/PhysRevD.103.083533}{\emph{Phys. Rev. D} {\bfseries 103} (2021) 083533} [\href{https://arxiv.org/abs/2007.08991}{{\ttfamily 2007.08991}}].

\bibitem{Nadathur:2020kvq}
S.~Nadathur, W.J.~Percival, F.~Beutler and H.~Winther, \emph{{Testing Low-Redshift Cosmic Acceleration with Large-Scale Structure}}, \href{https://doi.org/10.1103/PhysRevLett.124.221301}{\emph{Phys. Rev. Lett.} {\bfseries 124} (2020) 221301} [\href{https://arxiv.org/abs/2001.11044}{{\ttfamily 2001.11044}}].

\bibitem{Rose:2020shp}
B.M.~Rose, D.~Rubin, A.~Cikota, S.E.~Deustua, S.~Dixon, A.~Fruchter et~al., \emph{{Evidence for Cosmic Acceleration is Robust to Observed Correlations Between Type Ia Supernova Luminosity and Stellar Age}}, \href{https://doi.org/10.3847/2041-8213/ab94ad}{\emph{Astrophys. J. Lett.} {\bfseries 896} (2020) L4} [\href{https://arxiv.org/abs/2002.12382}{{\ttfamily 2002.12382}}].

\bibitem{DiValentino:2020evt}
E.~Di~Valentino, S.~Gariazzo, O.~Mena and S.~Vagnozzi, \emph{{Soundness of Dark Energy properties}}, \href{https://doi.org/10.1088/1475-7516/2020/07/045}{\emph{JCAP} {\bfseries 07} (2020) 045} [\href{https://arxiv.org/abs/2005.02062}{{\ttfamily 2005.02062}}].

\bibitem{KiDS:2020suj}
{\scshape KiDS} collaboration, \emph{{KiDS-1000 Cosmology: Cosmic shear constraints and comparison between two point statistics}}, \href{https://doi.org/10.1051/0004-6361/202039070}{\emph{Astron. Astrophys.} {\bfseries 645} (2021) A104} [\href{https://arxiv.org/abs/2007.15633}{{\ttfamily 2007.15633}}].

\bibitem{KiDS:2020ghu}
{\scshape KiDS} collaboration, \emph{{KiDS-1000 Cosmology: Constraints beyond flat \ensuremath{\Lambda}CDM}}, \href{https://doi.org/10.1051/0004-6361/202039805}{\emph{Astron. Astrophys.} {\bfseries 649} (2021) A88} [\href{https://arxiv.org/abs/2010.16416}{{\ttfamily 2010.16416}}].

\bibitem{SPT-3G:2021eoc}
{\scshape SPT-3G} collaboration, \emph{{Measurements of the E-mode polarization and temperature-E-mode correlation of the CMB from SPT-3G 2018 data}}, \href{https://doi.org/10.1103/PhysRevD.104.022003}{\emph{Phys. Rev. D} {\bfseries 104} (2021) 022003} [\href{https://arxiv.org/abs/2101.01684}{{\ttfamily 2101.01684}}].

\bibitem{DES:2021wwk}
{\scshape DES} collaboration, \emph{{Dark Energy Survey Year 3 results: Cosmological constraints from galaxy clustering and weak lensing}}, \href{https://doi.org/10.1103/PhysRevD.105.023520}{\emph{Phys. Rev. D} {\bfseries 105} (2022) 023520} [\href{https://arxiv.org/abs/2105.13549}{{\ttfamily 2105.13549}}].

\bibitem{Moresco:2022phi}
M.~Moresco et~al., \emph{{Unveiling the Universe with emerging cosmological probes}}, \href{https://doi.org/10.1007/s41114-022-00040-z}{\emph{Living Rev. Rel.} {\bfseries 25} (2022) 6} [\href{https://arxiv.org/abs/2201.07241}{{\ttfamily 2201.07241}}].

\bibitem{DES:2022ccp}
{\scshape DES} collaboration, \emph{{Dark Energy Survey Year 3 results: Constraints on extensions to \ensuremath{\Lambda}CDM with weak lensing and galaxy clustering}}, \href{https://doi.org/10.1103/PhysRevD.107.083504}{\emph{Phys. Rev. D} {\bfseries 107} (2023) 083504} [\href{https://arxiv.org/abs/2207.05766}{{\ttfamily 2207.05766}}].

\bibitem{Brout:2022vxf}
D.~Brout et~al., \emph{{The Pantheon+ Analysis: Cosmological Constraints}}, \href{https://doi.org/10.3847/1538-4357/ac8e04}{\emph{Astrophys. J.} {\bfseries 938} (2022) 110} [\href{https://arxiv.org/abs/2202.04077}{{\ttfamily 2202.04077}}].

\bibitem{ACT:2023kun}
{\scshape ACT} collaboration, \emph{{The Atacama Cosmology Telescope: DR6 Gravitational Lensing Map and Cosmological Parameters}}, \href{https://doi.org/10.3847/1538-4357/acff5f}{\emph{Astrophys. J.} {\bfseries 962} (2024) 113} [\href{https://arxiv.org/abs/2304.05203}{{\ttfamily 2304.05203}}].

\bibitem{Kilo-DegreeSurvey:2023gfr}
{\scshape Kilo-Degree Survey, Dark Energy Survey} collaboration, \emph{{DES Y3 + KiDS-1000: Consistent cosmology combining cosmic shear surveys}}, \href{https://doi.org/10.21105/astro.2305.17173}{\emph{Open J. Astrophys.} {\bfseries 6} (2023) 2305.17173} [\href{https://arxiv.org/abs/2305.17173}{{\ttfamily 2305.17173}}].

\bibitem{DESI:2024uvr}
{\scshape DESI} collaboration, \emph{{DESI 2024 III: Baryon Acoustic Oscillations from Galaxies and Quasars}},  \href{https://arxiv.org/abs/2404.03000}{{\ttfamily 2404.03000}}.

\bibitem{DESI:2024kob}
{\scshape DESI} collaboration, \emph{{DESI 2024: Constraints on Physics-Focused Aspects of Dark Energy using DESI DR1 BAO Data}},  \href{https://arxiv.org/abs/2405.13588}{{\ttfamily 2405.13588}}.

\bibitem{DES:2024tys}
{\scshape DES} collaboration, \emph{{The Dark Energy Survey: Cosmology Results With \textasciitilde{}1500 New High-redshift Type Ia Supernovae Using The Full 5-year Dataset}},  \href{https://arxiv.org/abs/2401.02929}{{\ttfamily 2401.02929}}.

\bibitem{DES:2024upw}
{\scshape DES} collaboration, \emph{{The Dark Energy Survey Supernova Program: Light curves and 5-Year data release}},  \href{https://arxiv.org/abs/2406.05046}{{\ttfamily 2406.05046}}.

\bibitem{DES:2024hip}
{\scshape DES} collaboration, \emph{{The Dark Energy Survey Supernova Program: Cosmological Analysis and Systematic Uncertainties}},  \href{https://arxiv.org/abs/2401.02945}{{\ttfamily 2401.02945}}.

\bibitem{Buchert:1999er}
T.~Buchert, \emph{{On average properties of inhomogeneous fluids in general relativity. 1. Dust cosmologies}}, \href{https://doi.org/10.1023/A:1001800617177}{\emph{Gen. Rel. Grav.} {\bfseries 32} (2000) 105} [\href{https://arxiv.org/abs/gr-qc/9906015}{{\ttfamily gr-qc/9906015}}].

\bibitem{Buchert:2001sa}
T.~Buchert, \emph{{On average properties of inhomogeneous fluids in general relativity: Perfect fluid cosmologies}}, \href{https://doi.org/10.1023/A:1012061725841}{\emph{Gen. Rel. Grav.} {\bfseries 33} (2001) 1381} [\href{https://arxiv.org/abs/gr-qc/0102049}{{\ttfamily gr-qc/0102049}}].

\bibitem{Buchert:2007ik}
T.~Buchert, \emph{{Dark Energy from Structure: A Status Report}}, \href{https://doi.org/10.1007/s10714-007-0554-8}{\emph{Gen. Rel. Grav.} {\bfseries 40} (2008) 467} [\href{https://arxiv.org/abs/0707.2153}{{\ttfamily 0707.2153}}].

\bibitem{Hunt:2008wp}
P.~Hunt and S.~Sarkar, \emph{{Constraints on large scale inhomogeneities from WMAP-5 and SDSS: confrontation with recent observations}}, \href{https://doi.org/10.1111/j.1365-2966.2009.15670.x}{\emph{Mon. Not. Roy. Astron. Soc.} {\bfseries 401} (2010) 547} [\href{https://arxiv.org/abs/0807.4508}{{\ttfamily 0807.4508}}].

\bibitem{Nielsen:2015pga}
J.T.~Nielsen, A.~Guffanti and S.~Sarkar, \emph{{Marginal evidence for cosmic acceleration from Type Ia supernovae}}, \href{https://doi.org/10.1038/srep35596}{\emph{Sci. Rep.} {\bfseries 6} (2016) 35596} [\href{https://arxiv.org/abs/1506.01354}{{\ttfamily 1506.01354}}].

\bibitem{Tutusaus:2017ibk}
I.~Tutusaus, B.~Lamine, A.~Dupays and A.~Blanchard, \emph{{Is cosmic acceleration proven by local cosmological probes?}}, \href{https://doi.org/10.1051/0004-6361/201630289}{\emph{Astron. Astrophys.} {\bfseries 602} (2017) A73} [\href{https://arxiv.org/abs/1706.05036}{{\ttfamily 1706.05036}}].

\bibitem{Dam:2017xqs}
L.H.~Dam, A.~Heinesen and D.L.~Wiltshire, \emph{{Apparent cosmic acceleration from type Ia supernovae}}, \href{https://doi.org/10.1093/mnras/stx1858}{\emph{Mon. Not. Roy. Astron. Soc.} {\bfseries 472} (2017) 835} [\href{https://arxiv.org/abs/1706.07236}{{\ttfamily 1706.07236}}].

\bibitem{Colin:2019opb}
J.~Colin, R.~Mohayaee, M.~Rameez and S.~Sarkar, \emph{{Evidence for anisotropy of cosmic acceleration}}, \href{https://doi.org/10.1051/0004-6361/201936373}{\emph{Astron. Astrophys.} {\bfseries 631} (2019) L13} [\href{https://arxiv.org/abs/1808.04597}{{\ttfamily 1808.04597}}].

\bibitem{Desgrange:2019npu}
C.~Desgrange, A.~Heinesen and T.~Buchert, \emph{{Dynamical spatial curvature as a fit to type Ia supernovae}}, \href{https://doi.org/10.1142/S0218271819501438}{\emph{Int. J. Mod. Phys. D} {\bfseries 28} (2019) 1950143} [\href{https://arxiv.org/abs/1902.07915}{{\ttfamily 1902.07915}}].

\bibitem{Koksbang:2019cen}
S.M.~Koksbang, \emph{{Towards statistically homogeneous and isotropic perfect fluid universes with cosmic backreaction}}, \href{https://doi.org/10.1088/1361-6382/ab376c}{\emph{Class. Quant. Grav.} {\bfseries 36} (2019) 185004} [\href{https://arxiv.org/abs/1907.08681}{{\ttfamily 1907.08681}}].

\bibitem{Koksbang:2019glb}
S.M.~Koksbang, \emph{{Another look at redshift drift and the backreaction conjecture}}, \href{https://doi.org/10.1088/1475-7516/2019/10/036}{\emph{JCAP} {\bfseries 10} (2019) 036} [\href{https://arxiv.org/abs/1909.13489}{{\ttfamily 1909.13489}}].

\bibitem{Heinesen:2022lqs}
A.~Heinesen, \emph{{Reconciling a decelerating Universe with cosmological observations}}, \href{https://doi.org/10.1103/PhysRevD.107.L101301}{\emph{Phys. Rev. D} {\bfseries 107} (2023) L101301} [\href{https://arxiv.org/abs/2212.05568}{{\ttfamily 2212.05568}}].

\bibitem{Sahni:1999gb}
V.~Sahni and A.A.~Starobinsky, \emph{{The Case for a positive cosmological Lambda term}}, \href{https://doi.org/10.1142/S0218271800000542}{\emph{Int. J. Mod. Phys. D} {\bfseries 9} (2000) 373} [\href{https://arxiv.org/abs/astro-ph/9904398}{{\ttfamily astro-ph/9904398}}].

\bibitem{Carroll:2000fy}
S.M.~Carroll, \emph{{The Cosmological constant}}, \href{https://doi.org/10.12942/lrr-2001-1}{\emph{Living Rev. Rel.} {\bfseries 4} (2001) 1} [\href{https://arxiv.org/abs/astro-ph/0004075}{{\ttfamily astro-ph/0004075}}].

\bibitem{Peebles:2002gy}
P.J.E.~Peebles and B.~Ratra, \emph{{The Cosmological Constant and Dark Energy}}, \href{https://doi.org/10.1103/RevModPhys.75.559}{\emph{Rev. Mod. Phys.} {\bfseries 75} (2003) 559} [\href{https://arxiv.org/abs/astro-ph/0207347}{{\ttfamily astro-ph/0207347}}].

\bibitem{Padmanabhan:2002ji}
T.~Padmanabhan, \emph{{Cosmological constant: The Weight of the vacuum}}, \href{https://doi.org/10.1016/S0370-1573(03)00120-0}{\emph{Phys. Rept.} {\bfseries 380} (2003) 235} [\href{https://arxiv.org/abs/hep-th/0212290}{{\ttfamily hep-th/0212290}}].

\bibitem{Copeland:2006wr}
E.J.~Copeland, M.~Sami and S.~Tsujikawa, \emph{{Dynamics of dark energy}}, \href{https://doi.org/10.1142/S021827180600942X}{\emph{Int. J. Mod. Phys. D} {\bfseries 15} (2006) 1753} [\href{https://arxiv.org/abs/hep-th/0603057}{{\ttfamily hep-th/0603057}}].

\bibitem{Caldwell:2009ix}
R.R.~Caldwell and M.~Kamionkowski, \emph{{The Physics of Cosmic Acceleration}}, \href{https://doi.org/10.1146/annurev-nucl-010709-151330}{\emph{Ann. Rev. Nucl. Part. Sci.} {\bfseries 59} (2009) 397} [\href{https://arxiv.org/abs/0903.0866}{{\ttfamily 0903.0866}}].

\bibitem{Li:2011sd}
M.~Li, X.-D.~Li, S.~Wang and Y.~Wang, \emph{{Dark Energy}}, \href{https://doi.org/10.1088/0253-6102/56/3/24}{\emph{Commun. Theor. Phys.} {\bfseries 56} (2011) 525} [\href{https://arxiv.org/abs/1103.5870}{{\ttfamily 1103.5870}}].

\bibitem{Martin:2012bt}
J.~Martin, \emph{{Everything You Always Wanted To Know About The Cosmological Constant Problem (But Were Afraid To Ask)}}, \href{https://doi.org/10.1016/j.crhy.2012.04.008}{\emph{Comptes Rendus Physique} {\bfseries 13} (2012) 566} [\href{https://arxiv.org/abs/1205.3365}{{\ttfamily 1205.3365}}].

\bibitem{Weinberg:1988cp}
S.~Weinberg, \emph{{The Cosmological Constant Problem}}, \href{https://doi.org/10.1103/RevModPhys.61.1}{\emph{Rev. Mod. Phys.} {\bfseries 61} (1989) 1}.

\bibitem{Krauss:1995yb}
L.M.~Krauss and M.S.~Turner, \emph{{The Cosmological constant is back}}, \href{https://doi.org/10.1007/BF02108229}{\emph{Gen. Rel. Grav.} {\bfseries 27} (1995) 1137} [\href{https://arxiv.org/abs/astro-ph/9504003}{{\ttfamily astro-ph/9504003}}].

\bibitem{Weinberg:2000yb}
S.~Weinberg, \emph{{The Cosmological constant problems}},  2, 2000.

\bibitem{Sahni:2002kh}
V.~Sahni, \emph{{The Cosmological constant problem and quintessence}}, \href{https://doi.org/10.1088/0264-9381/19/13/304}{\emph{Class. Quant. Grav.} {\bfseries 19} (2002) 3435} [\href{https://arxiv.org/abs/astro-ph/0202076}{{\ttfamily astro-ph/0202076}}].

\bibitem{Yokoyama:2003ii}
J.~Yokoyama, \emph{{Issues on the cosmological constant}},  5, 2003.

\bibitem{Nobbenhuis:2004wn}
S.~Nobbenhuis, \emph{{Categorizing different approaches to the cosmological constant problem}}, \href{https://doi.org/10.1007/s10701-005-9042-8}{\emph{Found. Phys.} {\bfseries 36} (2006) 613} [\href{https://arxiv.org/abs/gr-qc/0411093}{{\ttfamily gr-qc/0411093}}].

\bibitem{Burgess:2013ara}
C.P.~Burgess, \emph{{The Cosmological Constant Problem: Why it's hard to get Dark Energy from Micro-physics}},  2015.
\newblock 10.1093/acprof:oso/9780198728856.003.0004.

\bibitem{Joyce:2014kja}
A.~Joyce, B.~Jain, J.~Khoury and M.~Trodden, \emph{{Beyond the Cosmological Standard Model}}, \href{https://doi.org/10.1016/j.physrep.2014.12.002}{\emph{Phys. Rept.} {\bfseries 568} (2015) 1} [\href{https://arxiv.org/abs/1407.0059}{{\ttfamily 1407.0059}}].

\bibitem{Bull:2015stt}
P.~Bull et~al., \emph{{Beyond $\Lambda$CDM: Problems, solutions, and the road ahead}}, \href{https://doi.org/10.1016/j.dark.2016.02.001}{\emph{Phys. Dark Univ.} {\bfseries 12} (2016) 56} [\href{https://arxiv.org/abs/1512.05356}{{\ttfamily 1512.05356}}].

\bibitem{Wang:2016lxa}
B.~Wang, E.~Abdalla, F.~Atrio-Barandela and D.~Pavon, \emph{{Dark Matter and Dark Energy Interactions: Theoretical Challenges, Cosmological Implications and Observational Signatures}}, \href{https://doi.org/10.1088/0034-4885/79/9/096901}{\emph{Rept. Prog. Phys.} {\bfseries 79} (2016) 096901} [\href{https://arxiv.org/abs/1603.08299}{{\ttfamily 1603.08299}}].

\bibitem{Brustein:1992nk}
R.~Brustein and P.J.~Steinhardt, \emph{{Challenges for superstring cosmology}}, \href{https://doi.org/10.1016/0370-2693(93)90384-T}{\emph{Phys. Lett. B} {\bfseries 302} (1993) 196} [\href{https://arxiv.org/abs/hep-th/9212049}{{\ttfamily hep-th/9212049}}].

\bibitem{Witten:2000zk}
E.~Witten, \emph{{The Cosmological constant from the viewpoint of string theory}},  3, 2000.

\bibitem{Kachru:2003aw}
S.~Kachru, R.~Kallosh, A.D.~Linde and S.P.~Trivedi, \emph{{De Sitter vacua in string theory}}, \href{https://doi.org/10.1103/PhysRevD.68.046005}{\emph{Phys. Rev. D} {\bfseries 68} (2003) 046005} [\href{https://arxiv.org/abs/hep-th/0301240}{{\ttfamily hep-th/0301240}}].

\bibitem{Polchinski:2006gy}
J.~Polchinski, \emph{{The Cosmological Constant and the String Landscape}},  3, 2006.

\bibitem{Danielsson:2018ztv}
U.H.~Danielsson and T.~Van~Riet, \emph{{What if string theory has no de Sitter vacua?}}, \href{https://doi.org/10.1142/S0218271818300070}{\emph{Int. J. Mod. Phys. D} {\bfseries 27} (2018) 1830007} [\href{https://arxiv.org/abs/1804.01120}{{\ttfamily 1804.01120}}].

\bibitem{Zlatev:1998tr}
I.~Zlatev, L.-M.~Wang and P.J.~Steinhardt, \emph{{Quintessence, cosmic coincidence, and the cosmological constant}}, \href{https://doi.org/10.1103/PhysRevLett.82.896}{\emph{Phys. Rev. Lett.} {\bfseries 82} (1999) 896} [\href{https://arxiv.org/abs/astro-ph/9807002}{{\ttfamily astro-ph/9807002}}].

\bibitem{Pavon:2005yx}
D.~Pavon and W.~Zimdahl, \emph{{Holographic dark energy and cosmic coincidence}}, \href{https://doi.org/10.1016/j.physletb.2005.08.134}{\emph{Phys. Lett. B} {\bfseries 628} (2005) 206} [\href{https://arxiv.org/abs/gr-qc/0505020}{{\ttfamily gr-qc/0505020}}].

\bibitem{Velten:2014nra}
H.E.S.~Velten, R.F.~vom Marttens and W.~Zimdahl, \emph{{Aspects of the cosmological \textquotedblleft{}coincidence problem\textquotedblright{}}}, \href{https://doi.org/10.1140/epjc/s10052-014-3160-4}{\emph{Eur. Phys. J. C} {\bfseries 74} (2014) 3160} [\href{https://arxiv.org/abs/1410.2509}{{\ttfamily 1410.2509}}].

\bibitem{Dolgov:1997za}
A.D.~Dolgov, \emph{{The Problem of vacuum energy and cosmology}},  6, 1997.

\bibitem{Straumann:1999ia}
N.~Straumann, \emph{{The Mystery of the cosmic vacuum energy density and the accelerated expansion of the universe}}, \href{https://doi.org/10.1088/0143-0807/20/6/307}{\emph{Eur. J. Phys.} {\bfseries 20} (1999) 419} [\href{https://arxiv.org/abs/astro-ph/9908342}{{\ttfamily astro-ph/9908342}}].

\bibitem{Sola:2013gha}
J.~Sola, \emph{{Cosmological constant and vacuum energy: old and new ideas}}, \href{https://doi.org/10.1088/1742-6596/453/1/012015}{\emph{J. Phys. Conf. Ser.} {\bfseries 453} (2013) 012015} [\href{https://arxiv.org/abs/1306.1527}{{\ttfamily 1306.1527}}].

\bibitem{Amendola:1999er}
L.~Amendola, \emph{{Coupled quintessence}}, \href{https://doi.org/10.1103/PhysRevD.62.043511}{\emph{Phys. Rev. D} {\bfseries 62} (2000) 043511} [\href{https://arxiv.org/abs/astro-ph/9908023}{{\ttfamily astro-ph/9908023}}].

\bibitem{Kamenshchik:2001cp}
A.Y.~Kamenshchik, U.~Moschella and V.~Pasquier, \emph{{An Alternative to quintessence}}, \href{https://doi.org/10.1016/S0370-2693(01)00571-8}{\emph{Phys. Lett. B} {\bfseries 511} (2001) 265} [\href{https://arxiv.org/abs/gr-qc/0103004}{{\ttfamily gr-qc/0103004}}].

\bibitem{Capozziello:2002rd}
S.~Capozziello, \emph{{Curvature quintessence}}, \href{https://doi.org/10.1142/S0218271802002025}{\emph{Int. J. Mod. Phys. D} {\bfseries 11} (2002) 483} [\href{https://arxiv.org/abs/gr-qc/0201033}{{\ttfamily gr-qc/0201033}}].

\bibitem{Bento:2002ps}
M.C.~Bento, O.~Bertolami and A.A.~Sen, \emph{{Generalized Chaplygin gas, accelerated expansion and dark energy matter unification}}, \href{https://doi.org/10.1103/PhysRevD.66.043507}{\emph{Phys. Rev. D} {\bfseries 66} (2002) 043507} [\href{https://arxiv.org/abs/gr-qc/0202064}{{\ttfamily gr-qc/0202064}}].

\bibitem{Mangano:2002gg}
G.~Mangano, G.~Miele and V.~Pettorino, \emph{{Coupled quintessence and the coincidence problem}}, \href{https://doi.org/10.1142/S0217732303009940}{\emph{Mod. Phys. Lett. A} {\bfseries 18} (2003) 831} [\href{https://arxiv.org/abs/astro-ph/0212518}{{\ttfamily astro-ph/0212518}}].

\bibitem{Farrar:2003uw}
G.R.~Farrar and P.J.E.~Peebles, \emph{{Interacting dark matter and dark energy}}, \href{https://doi.org/10.1086/381728}{\emph{Astrophys. J.} {\bfseries 604} (2004) 1} [\href{https://arxiv.org/abs/astro-ph/0307316}{{\ttfamily astro-ph/0307316}}].

\bibitem{Khoury:2003aq}
J.~Khoury and A.~Weltman, \emph{{Chameleon fields: Awaiting surprises for tests of gravity in space}}, \href{https://doi.org/10.1103/PhysRevLett.93.171104}{\emph{Phys. Rev. Lett.} {\bfseries 93} (2004) 171104} [\href{https://arxiv.org/abs/astro-ph/0309300}{{\ttfamily astro-ph/0309300}}].

\bibitem{Li:2004rb}
M.~Li, \emph{{A Model of holographic dark energy}}, \href{https://doi.org/10.1016/j.physletb.2004.10.014}{\emph{Phys. Lett. B} {\bfseries 603} (2004) 1} [\href{https://arxiv.org/abs/hep-th/0403127}{{\ttfamily hep-th/0403127}}].

\bibitem{Amendola:2006we}
L.~Amendola, R.~Gannouji, D.~Polarski and S.~Tsujikawa, \emph{{Conditions for the cosmological viability of f(R) dark energy models}}, \href{https://doi.org/10.1103/PhysRevD.75.083504}{\emph{Phys. Rev. D} {\bfseries 75} (2007) 083504} [\href{https://arxiv.org/abs/gr-qc/0612180}{{\ttfamily gr-qc/0612180}}].

\bibitem{Hu:2007nk}
W.~Hu and I.~Sawicki, \emph{{Models of f(R) Cosmic Acceleration that Evade Solar-System Tests}}, \href{https://doi.org/10.1103/PhysRevD.76.064004}{\emph{Phys. Rev. D} {\bfseries 76} (2007) 064004} [\href{https://arxiv.org/abs/0705.1158}{{\ttfamily 0705.1158}}].

\bibitem{Cognola:2007zu}
G.~Cognola, E.~Elizalde, S.~Nojiri, S.D.~Odintsov, L.~Sebastiani and S.~Zerbini, \emph{{A Class of viable modified f(R) gravities describing inflation and the onset of accelerated expansion}}, \href{https://doi.org/10.1103/PhysRevD.77.046009}{\emph{Phys. Rev. D} {\bfseries 77} (2008) 046009} [\href{https://arxiv.org/abs/0712.4017}{{\ttfamily 0712.4017}}].

\bibitem{Nojiri:2010pw}
S.~Nojiri, S.D.~Odintsov, M.~Sasaki and Y.-l.~Zhang, \emph{{Screening of cosmological constant in non-local gravity}}, \href{https://doi.org/10.1016/j.physletb.2010.12.035}{\emph{Phys. Lett. B} {\bfseries 696} (2011) 278} [\href{https://arxiv.org/abs/1010.5375}{{\ttfamily 1010.5375}}].

\bibitem{Zhang:2011uv}
Y.-l.~Zhang and M.~Sasaki, \emph{{Screening of cosmological constant in non-local cosmology}}, \href{https://doi.org/10.1142/S021827181250006X}{\emph{Int. J. Mod. Phys. D} {\bfseries 21} (2012) 1250006} [\href{https://arxiv.org/abs/1108.2112}{{\ttfamily 1108.2112}}].

\bibitem{Rinaldi:2014yta}
M.~Rinaldi, \emph{{Higgs Dark Energy}}, \href{https://doi.org/10.1088/0264-9381/32/4/045002}{\emph{Class. Quant. Grav.} {\bfseries 32} (2015) 045002} [\href{https://arxiv.org/abs/1404.0532}{{\ttfamily 1404.0532}}].

\bibitem{Luongo:2014nld}
O.~Luongo and H.~Quevedo, \emph{{A Unified Dark Energy Model from a Vanishing Speed of Sound with Emergent Cosmological Constant}}, \href{https://doi.org/10.1142/S0218271814500126}{\emph{Int. J. Mod. Phys. D} {\bfseries 23} (2014) 1450012}.

\bibitem{Rinaldi:2015iza}
M.~Rinaldi, \emph{{Dark energy as a fixed point of the Einstein Yang-Mills Higgs Equations}}, \href{https://doi.org/10.1088/1475-7516/2015/10/023}{\emph{JCAP} {\bfseries 10} (2015) 023} [\href{https://arxiv.org/abs/1508.04576}{{\ttfamily 1508.04576}}].

\bibitem{DeFelice:2016yws}
A.~De~Felice, L.~Heisenberg, R.~Kase, S.~Mukohyama, S.~Tsujikawa and Y.-l.~Zhang, \emph{{Cosmology in generalized Proca theories}}, \href{https://doi.org/10.1088/1475-7516/2016/06/048}{\emph{JCAP} {\bfseries 06} (2016) 048} [\href{https://arxiv.org/abs/1603.05806}{{\ttfamily 1603.05806}}].

\bibitem{Josset:2016vrq}
T.~Josset, A.~Perez and D.~Sudarsky, \emph{{Dark Energy from Violation of Energy Conservation}}, \href{https://doi.org/10.1103/PhysRevLett.118.021102}{\emph{Phys. Rev. Lett.} {\bfseries 118} (2017) 021102} [\href{https://arxiv.org/abs/1604.04183}{{\ttfamily 1604.04183}}].

\bibitem{Burrage:2016bwy}
C.~Burrage and J.~Sakstein, \emph{{A Compendium of Chameleon Constraints}}, \href{https://doi.org/10.1088/1475-7516/2016/11/045}{\emph{JCAP} {\bfseries 11} (2016) 045} [\href{https://arxiv.org/abs/1609.01192}{{\ttfamily 1609.01192}}].

\bibitem{Sebastiani:2016ras}
L.~Sebastiani, S.~Vagnozzi and R.~Myrzakulov, \emph{{Mimetic gravity: a review of recent developments and applications to cosmology and astrophysics}}, \href{https://doi.org/10.1155/2017/3156915}{\emph{Adv. High Energy Phys.} {\bfseries 2017} (2017) 3156915} [\href{https://arxiv.org/abs/1612.08661}{{\ttfamily 1612.08661}}].

\bibitem{Nojiri:2017ncd}
S.~Nojiri, S.D.~Odintsov and V.K.~Oikonomou, \emph{{Modified Gravity Theories on a Nutshell: Inflation, Bounce and Late-time Evolution}}, \href{https://doi.org/10.1016/j.physrep.2017.06.001}{\emph{Phys. Rept.} {\bfseries 692} (2017) 1} [\href{https://arxiv.org/abs/1705.11098}{{\ttfamily 1705.11098}}].

\bibitem{Burrage:2017qrf}
C.~Burrage and J.~Sakstein, \emph{{Tests of Chameleon Gravity}}, \href{https://doi.org/10.1007/s41114-018-0011-x}{\emph{Living Rev. Rel.} {\bfseries 21} (2018) 1} [\href{https://arxiv.org/abs/1709.09071}{{\ttfamily 1709.09071}}].

\bibitem{Capozziello:2017buj}
S.~Capozziello, R.~D'Agostino and O.~Luongo, \emph{{Cosmic acceleration from a single fluid description}}, \href{https://doi.org/10.1016/j.dark.2018.02.002}{\emph{Phys. Dark Univ.} {\bfseries 20} (2018) 1} [\href{https://arxiv.org/abs/1712.04317}{{\ttfamily 1712.04317}}].

\bibitem{Benisty:2018qed}
D.~Benisty and E.I.~Guendelman, \emph{{Unified dark energy and dark matter from dynamical spacetime}}, \href{https://doi.org/10.1103/PhysRevD.98.023506}{\emph{Phys. Rev. D} {\bfseries 98} (2018) 023506} [\href{https://arxiv.org/abs/1802.07981}{{\ttfamily 1802.07981}}].

\bibitem{Casalino:2018tcd}
A.~Casalino, M.~Rinaldi, L.~Sebastiani and S.~Vagnozzi, \emph{{Mimicking dark matter and dark energy in a mimetic model compatible with GW170817}}, \href{https://doi.org/10.1016/j.dark.2018.10.001}{\emph{Phys. Dark Univ.} {\bfseries 22} (2018) 108} [\href{https://arxiv.org/abs/1803.02620}{{\ttfamily 1803.02620}}].

\bibitem{Yang:2018euj}
W.~Yang, S.~Pan, E.~Di~Valentino, R.C.~Nunes, S.~Vagnozzi and D.F.~Mota, \emph{{Tale of stable interacting dark energy, observational signatures, and the $H_0$ tension}}, \href{https://doi.org/10.1088/1475-7516/2018/09/019}{\emph{JCAP} {\bfseries 09} (2018) 019} [\href{https://arxiv.org/abs/1805.08252}{{\ttfamily 1805.08252}}].

\bibitem{Saridakis:2018unr}
E.N.~Saridakis, K.~Bamba, R.~Myrzakulov and F.K.~Anagnostopoulos, \emph{{Holographic dark energy through Tsallis entropy}}, \href{https://doi.org/10.1088/1475-7516/2018/12/012}{\emph{JCAP} {\bfseries 12} (2018) 012} [\href{https://arxiv.org/abs/1806.01301}{{\ttfamily 1806.01301}}].

\bibitem{Visinelli:2018utg}
L.~Visinelli and S.~Vagnozzi, \emph{{Cosmological window onto the string axiverse and the supersymmetry breaking scale}}, \href{https://doi.org/10.1103/PhysRevD.99.063517}{\emph{Phys. Rev. D} {\bfseries 99} (2019) 063517} [\href{https://arxiv.org/abs/1809.06382}{{\ttfamily 1809.06382}}].

\bibitem{Langlois:2018dxi}
D.~Langlois, \emph{{Dark energy and modified gravity in degenerate higher-order scalar\textendash{}tensor (DHOST) theories: A review}}, \href{https://doi.org/10.1142/S0218271819420069}{\emph{Int. J. Mod. Phys. D} {\bfseries 28} (2019) 1942006} [\href{https://arxiv.org/abs/1811.06271}{{\ttfamily 1811.06271}}].

\bibitem{Benisty:2018oyy}
D.~Benisty, E.~Guendelman and Z.~Haba, \emph{{Unification of dark energy and dark matter from diffusive cosmology}}, \href{https://doi.org/10.1103/PhysRevD.99.123521}{\emph{Phys. Rev. D} {\bfseries 99} (2019) 123521} [\href{https://arxiv.org/abs/1812.06151}{{\ttfamily 1812.06151}}].

\bibitem{Boshkayev:2019qcx}
K.~Boshkayev, R.~D'Agostino and O.~Luongo, \emph{{Extended logotropic fluids as unified dark energy models}}, \href{https://doi.org/10.1140/epjc/s10052-019-6854-9}{\emph{Eur. Phys. J. C} {\bfseries 79} (2019) 332} [\href{https://arxiv.org/abs/1901.01031}{{\ttfamily 1901.01031}}].

\bibitem{Heckman:2019dsj}
J.J.~Heckman, C.~Lawrie, L.~Lin, J.~Sakstein and G.~Zoccarato, \emph{{Pixelated Dark Energy}}, \href{https://doi.org/10.1002/prop.201900071}{\emph{Fortsch. Phys.} {\bfseries 67} (2019) 1900071} [\href{https://arxiv.org/abs/1901.10489}{{\ttfamily 1901.10489}}].

\bibitem{DAgostino:2019wko}
R.~D'Agostino, \emph{{Holographic dark energy from nonadditive entropy: cosmological perturbations and observational constraints}}, \href{https://doi.org/10.1103/PhysRevD.99.103524}{\emph{Phys. Rev. D} {\bfseries 99} (2019) 103524} [\href{https://arxiv.org/abs/1903.03836}{{\ttfamily 1903.03836}}].

\bibitem{Mukhopadhyay:2019wrw}
U.~Mukhopadhyay, D.~Majumdar and D.~Adak, \emph{{Evolution of Dark Energy Perturbations for Slotheon Field and Power Spectrum}}, \href{https://doi.org/10.1140/epjc/s10052-020-8165-6}{\emph{Eur. Phys. J. C} {\bfseries 80} (2020) 593} [\href{https://arxiv.org/abs/1903.08650}{{\ttfamily 1903.08650}}].

\bibitem{Mukhopadhyay:2019cai}
U.~Mukhopadhyay and D.~Majumdar, \emph{{Swampland criteria in the slotheon field dark energy}}, \href{https://doi.org/10.1103/PhysRevD.100.024006}{\emph{Phys. Rev. D} {\bfseries 100} (2019) 024006} [\href{https://arxiv.org/abs/1904.01455}{{\ttfamily 1904.01455}}].

\bibitem{Mukhopadhyay:2019jla}
U.~Mukhopadhyay, A.~Paul and D.~Majumdar, \emph{{Addressing the high-$f$ problem in pseudo-Nambu\textendash{}Goldstone boson dark energy models with dark matter\textendash{}dark energy interaction}}, \href{https://doi.org/10.1140/epjc/s10052-020-08457-y}{\emph{Eur. Phys. J. C} {\bfseries 80} (2020) 904} [\href{https://arxiv.org/abs/1909.03925}{{\ttfamily 1909.03925}}].

\bibitem{Vagnozzi:2019kvw}
S.~Vagnozzi, L.~Visinelli, O.~Mena and D.F.~Mota, \emph{{Do we have any hope of detecting scattering between dark energy and baryons through cosmology?}}, \href{https://doi.org/10.1093/mnras/staa311}{\emph{Mon. Not. Roy. Astron. Soc.} {\bfseries 493} (2020) 1139} [\href{https://arxiv.org/abs/1911.12374}{{\ttfamily 1911.12374}}].

\bibitem{Akarsu:2019hmw}
O.~Akarsu, J.D.~Barrow, L.A.~Escamilla and J.A.~Vazquez, \emph{{Graduated dark energy: Observational hints of a spontaneous sign switch in the cosmological constant}}, \href{https://doi.org/10.1103/PhysRevD.101.063528}{\emph{Phys. Rev. D} {\bfseries 101} (2020) 063528} [\href{https://arxiv.org/abs/1912.08751}{{\ttfamily 1912.08751}}].

\bibitem{Saridakis:2020zol}
E.N.~Saridakis, \emph{{Barrow holographic dark energy}}, \href{https://doi.org/10.1103/PhysRevD.102.123525}{\emph{Phys. Rev. D} {\bfseries 102} (2020) 123525} [\href{https://arxiv.org/abs/2005.04115}{{\ttfamily 2005.04115}}].

\bibitem{Ruchika:2020avj}
Ruchika, S.A.~Adil, K.~Dutta, A.~Mukherjee and A.A.~Sen, \emph{{Observational constraints on axion(s) dark energy with a cosmological constant}}, \href{https://doi.org/10.1016/j.dark.2023.101199}{\emph{Phys. Dark Univ.} {\bfseries 40} (2023) 101199} [\href{https://arxiv.org/abs/2005.08813}{{\ttfamily 2005.08813}}].

\bibitem{Odintsov:2020zct}
S.D.~Odintsov, V.K.~Oikonomou and T.~Paul, \emph{{From a Bounce to the Dark Energy Era with $F(R)$ Gravity}}, \href{https://doi.org/10.1088/1361-6382/abbc47}{\emph{Class. Quant. Grav.} {\bfseries 37} (2020) 235005} [\href{https://arxiv.org/abs/2009.09947}{{\ttfamily 2009.09947}}].

\bibitem{Odintsov:2020vjb}
S.D.~Odintsov, V.K.~Oikonomou, F.P.~Fronimos and K.V.~Fasoulakos, \emph{{Unification of a Bounce with a Viable Dark Energy Era in Gauss-Bonnet Gravity}}, \href{https://doi.org/10.1103/PhysRevD.102.104042}{\emph{Phys. Rev. D} {\bfseries 102} (2020) 104042} [\href{https://arxiv.org/abs/2010.13580}{{\ttfamily 2010.13580}}].

\bibitem{Oikonomou:2020qah}
V.K.~Oikonomou, \emph{{Unifying inflation with early and late dark energy epochs in axion $F(R)$ gravity}}, \href{https://doi.org/10.1103/PhysRevD.103.044036}{\emph{Phys. Rev. D} {\bfseries 103} (2021) 044036} [\href{https://arxiv.org/abs/2012.00586}{{\ttfamily 2012.00586}}].

\bibitem{Oikonomou:2020oex}
V.K.~Oikonomou, \emph{{Rescaled Einstein-Hilbert Gravity from $f(R)$ Gravity: Inflation, Dark Energy and the Swampland Criteria}}, \href{https://doi.org/10.1103/PhysRevD.103.124028}{\emph{Phys. Rev. D} {\bfseries 103} (2021) 124028} [\href{https://arxiv.org/abs/2012.01312}{{\ttfamily 2012.01312}}].

\bibitem{Vagnozzi:2021quy}
S.~Vagnozzi, L.~Visinelli, P.~Brax, A.-C.~Davis and J.~Sakstein, \emph{{Direct detection of dark energy: The XENON1T excess and future prospects}}, \href{https://doi.org/10.1103/PhysRevD.104.063023}{\emph{Phys. Rev. D} {\bfseries 104} (2021) 063023} [\href{https://arxiv.org/abs/2103.15834}{{\ttfamily 2103.15834}}].

\bibitem{Solanki:2021qni}
R.~Solanki, S.K.J.~Pacif, A.~Parida and P.K.~Sahoo, \emph{{Cosmic acceleration with bulk viscosity in modified f(Q) gravity}}, \href{https://doi.org/10.1016/j.dark.2021.100820}{\emph{Phys. Dark Univ.} {\bfseries 32} (2021) 100820} [\href{https://arxiv.org/abs/2105.00876}{{\ttfamily 2105.00876}}].

\bibitem{Saridakis:2021qxb}
E.N.~Saridakis, \emph{{Do we need soft cosmology?}}, \href{https://doi.org/10.1016/j.physletb.2021.136649}{\emph{Phys. Lett. B} {\bfseries 822} (2021) 136649} [\href{https://arxiv.org/abs/2105.08646}{{\ttfamily 2105.08646}}].

\bibitem{Arora:2021tuh}
S.~Arora, S.K.J.~Pacif, A.~Parida and P.K.~Sahoo, \emph{{Bulk viscous matter and the cosmic acceleration of the universe in f(Q,T) gravity}}, \href{https://doi.org/10.1016/j.jheap.2021.10.001}{\emph{JHEAp} {\bfseries 33} (2022) 1} [\href{https://arxiv.org/abs/2106.00491}{{\ttfamily 2106.00491}}].

\bibitem{Capozziello:2022jbw}
S.~Capozziello, R.~D'Agostino and O.~Luongo, \emph{{Thermodynamic parametrization of dark energy}}, \href{https://doi.org/10.1016/j.dark.2022.101045}{\emph{Phys. Dark Univ.} {\bfseries 36} (2022) 101045} [\href{https://arxiv.org/abs/2202.03300}{{\ttfamily 2202.03300}}].

\bibitem{Narawade:2022jeg}
S.A.~Narawade, L.~Pati, B.~Mishra and S.K.~Tripathy, \emph{{Dynamical system analysis for accelerating models in non-metricity f(Q) gravity}}, \href{https://doi.org/10.1016/j.dark.2022.101020}{\emph{Phys. Dark Univ.} {\bfseries 36} (2022) 101020} [\href{https://arxiv.org/abs/2203.14121}{{\ttfamily 2203.14121}}].

\bibitem{DAgostino:2022fcx}
R.~D'Agostino, O.~Luongo and M.~Muccino, \emph{{Healing the cosmological constant problem during inflation through a unified quasi-quintessence matter field}}, \href{https://doi.org/10.1088/1361-6382/ac8af2}{\emph{Class. Quant. Grav.} {\bfseries 39} (2022) 195014} [\href{https://arxiv.org/abs/2204.02190}{{\ttfamily 2204.02190}}].

\bibitem{Oikonomou:2022wuk}
V.K.~Oikonomou and I.~Giannakoudi, \emph{{A panorama of viable F(R) gravity dark energy models}}, \href{https://doi.org/10.1142/S0218271822500754}{\emph{Int. J. Mod. Phys. D} {\bfseries 31} (2022) 2250075} [\href{https://arxiv.org/abs/2205.08599}{{\ttfamily 2205.08599}}].

\bibitem{Belfiglio:2022egm}
A.~Belfiglio, R.~Giamb\`o and O.~Luongo, \emph{{Alleviating the cosmological constant problem from particle production}}, \href{https://doi.org/10.1088/1361-6382/accc00}{\emph{Class. Quant. Grav.} {\bfseries 40} (2023) 105004} [\href{https://arxiv.org/abs/2206.14158}{{\ttfamily 2206.14158}}].

\bibitem{Luciano:2022hhy}
G.G.~Luciano and J.~Gin\'e, \emph{{Generalized interacting Barrow Holographic Dark Energy: Cosmological predictions and thermodynamic considerations}}, \href{https://doi.org/10.1016/j.dark.2023.101256}{\emph{Phys. Dark Univ.} {\bfseries 41} (2023) 101256} [\href{https://arxiv.org/abs/2210.09755}{{\ttfamily 2210.09755}}].

\bibitem{Kadam:2022yrj}
S.A.~Kadam, J.~Levi~Said and B.~Mishra, \emph{{Accelerating cosmological models in f(T,B) gravitational theory}}, \href{https://doi.org/10.1142/S0219887823500834}{\emph{Int. J. Geom. Meth. Mod. Phys.} {\bfseries 20} (2023) 2350083} [\href{https://arxiv.org/abs/2210.17075}{{\ttfamily 2210.17075}}].

\bibitem{Ong:2022wrs}
Y.C.~Ong, \emph{{An Effective Sign Switching Dark Energy: Lotka\textendash{}Volterra Model of Two Interacting Fluids}}, \href{https://doi.org/10.3390/universe9100437}{\emph{Universe} {\bfseries 9} (2023) 437} [\href{https://arxiv.org/abs/2212.04429}{{\ttfamily 2212.04429}}].

\bibitem{Bernui:2023byc}
A.~Bernui, E.~Di~Valentino, W.~Giar\`e, S.~Kumar and R.C.~Nunes, \emph{{Exploring the H0 tension and the evidence for dark sector interactions from 2D BAO measurements}}, \href{https://doi.org/10.1103/PhysRevD.107.103531}{\emph{Phys. Rev. D} {\bfseries 107} (2023) 103531} [\href{https://arxiv.org/abs/2301.06097}{{\ttfamily 2301.06097}}].

\bibitem{Luciano:2023wtx}
G.G.~Luciano, \emph{{Saez\textendash{}Ballester gravity in Kantowski\textendash{}Sachs Universe: A new reconstruction paradigm for Barrow Holographic Dark Energy}}, \href{https://doi.org/10.1016/j.dark.2023.101237}{\emph{Phys. Dark Univ.} {\bfseries 41} (2023) 101237} [\href{https://arxiv.org/abs/2301.12488}{{\ttfamily 2301.12488}}].

\bibitem{Giani:2023tai}
L.~Giani and O.F.~Piattella, \emph{{Induced non-local cosmology}}, \href{https://doi.org/10.1016/j.dark.2023.101219}{\emph{Phys. Dark Univ.} {\bfseries 40} (2023) 101219} [\href{https://arxiv.org/abs/2302.06762}{{\ttfamily 2302.06762}}].

\bibitem{Belfiglio:2023rxb}
A.~Belfiglio, Y.~Carloni and O.~Luongo, \emph{{Particle production from non-minimal coupling in a symmetry breaking potential transporting vacuum energy}}, \href{https://doi.org/10.1016/j.dark.2024.101458}{\emph{Phys. Dark Univ.} {\bfseries 44} (2024) 101458} [\href{https://arxiv.org/abs/2307.04739}{{\ttfamily 2307.04739}}].

\bibitem{Frion:2023xwq}
E.~Frion, D.~Camarena, L.~Giani, T.~Miranda, D.~Bertacca, V.~Marra et~al., \emph{{Bayesian analysis of a Unified Dark Matter model with transition: can it alleviate the $H_{0}$ tension?}},  \href{https://arxiv.org/abs/2307.06320}{{\ttfamily 2307.06320}}.

\bibitem{Adil:2023ara}
S.A.~Adil, U.~Mukhopadhyay, A.A.~Sen and S.~Vagnozzi, \emph{{Dark energy in light of the early JWST observations: case for a negative cosmological constant?}}, \href{https://doi.org/10.1088/1475-7516/2023/10/072}{\emph{JCAP} {\bfseries 10} (2023) 072} [\href{https://arxiv.org/abs/2307.12763}{{\ttfamily 2307.12763}}].

\bibitem{Halder:2024gag}
S.~Halder, S.~Pan, P.M.~S\'a and T.~Saha, \emph{{Coupled phantom cosmological model motivated by the warm inflationary paradigm}},  \href{https://arxiv.org/abs/2407.15804}{{\ttfamily 2407.15804}}.

\bibitem{Fischer:2024eic}
H.~Fischer, C.~K\"ading and M.~Pitschmann, \emph{{Screened Scalar Fields in the Laboratory and the Solar System}}, \href{https://doi.org/10.3390/universe10070297}{\emph{Universe} {\bfseries 10} (2024) 297} [\href{https://arxiv.org/abs/2405.14638}{{\ttfamily 2405.14638}}].

\bibitem{Cooray:1999da}
A.R.~Cooray and D.~Huterer, \emph{{Gravitational lensing as a probe of quintessence}}, \href{https://doi.org/10.1086/311927}{\emph{Astrophys. J. Lett.} {\bfseries 513} (1999) L95} [\href{https://arxiv.org/abs/astro-ph/9901097}{{\ttfamily astro-ph/9901097}}].

\bibitem{Efstathiou:1999tm}
G.~Efstathiou, \emph{{Constraining the equation of state of the universe from distant type Ia supernovae and cosmic microwave background anisotropies}}, \href{https://doi.org/10.1046/j.1365-8711.1999.02997.x}{\emph{Mon. Not. Roy. Astron. Soc.} {\bfseries 310} (1999) 842} [\href{https://arxiv.org/abs/astro-ph/9904356}{{\ttfamily astro-ph/9904356}}].

\bibitem{Chevallier:2000qy}
M.~Chevallier and D.~Polarski, \emph{{Accelerating universes with scaling dark matter}}, \href{https://doi.org/10.1142/S0218271801000822}{\emph{Int. J. Mod. Phys. D} {\bfseries 10} (2001) 213} [\href{https://arxiv.org/abs/gr-qc/0009008}{{\ttfamily gr-qc/0009008}}].

\bibitem{Melchiorri:2002ux}
A.~Melchiorri, L.~Mersini-Houghton, C.J.~Odman and M.~Trodden, \emph{{The State of the dark energy equation of state}}, \href{https://doi.org/10.1103/PhysRevD.68.043509}{\emph{Phys. Rev. D} {\bfseries 68} (2003) 043509} [\href{https://arxiv.org/abs/astro-ph/0211522}{{\ttfamily astro-ph/0211522}}].

\bibitem{Linder:2002et}
E.V.~Linder, \emph{{Exploring the expansion history of the universe}}, \href{https://doi.org/10.1103/PhysRevLett.90.091301}{\emph{Phys. Rev. Lett.} {\bfseries 90} (2003) 091301} [\href{https://arxiv.org/abs/astro-ph/0208512}{{\ttfamily astro-ph/0208512}}].

\bibitem{Wetterich:2004pv}
C.~Wetterich, \emph{{Phenomenological parameterization of quintessence}}, \href{https://doi.org/10.1016/j.physletb.2004.05.008}{\emph{Phys. Lett. B} {\bfseries 594} (2004) 17} [\href{https://arxiv.org/abs/astro-ph/0403289}{{\ttfamily astro-ph/0403289}}].

\bibitem{Feng:2004ff}
B.~Feng, M.~Li, Y.-S.~Piao and X.~Zhang, \emph{{Oscillating quintom and the recurrent universe}}, \href{https://doi.org/10.1016/j.physletb.2006.01.066}{\emph{Phys. Lett. B} {\bfseries 634} (2006) 101} [\href{https://arxiv.org/abs/astro-ph/0407432}{{\ttfamily astro-ph/0407432}}].

\bibitem{Xia:2004rw}
J.-Q.~Xia, B.~Feng and X.-M.~Zhang, \emph{{Constraints on oscillating quintom from supernova, microwave background and galaxy clustering}}, \href{https://doi.org/10.1142/S0217732305017445}{\emph{Mod. Phys. Lett. A} {\bfseries 20} (2005) 2409} [\href{https://arxiv.org/abs/astro-ph/0411501}{{\ttfamily astro-ph/0411501}}].

\bibitem{Hannestad:2004cb}
S.~Hannestad and E.~Mortsell, \emph{{Cosmological constraints on the dark energy equation of state and its evolution}}, \href{https://doi.org/10.1088/1475-7516/2004/09/001}{\emph{JCAP} {\bfseries 09} (2004) 001} [\href{https://arxiv.org/abs/astro-ph/0407259}{{\ttfamily astro-ph/0407259}}].

\bibitem{Gong:2005de}
Y.-g.~Gong and Y.-Z.~Zhang, \emph{{Probing the curvature and dark energy}}, \href{https://doi.org/10.1103/PhysRevD.72.043518}{\emph{Phys. Rev. D} {\bfseries 72} (2005) 043518} [\href{https://arxiv.org/abs/astro-ph/0502262}{{\ttfamily astro-ph/0502262}}].

\bibitem{Jassal:2005qc}
H.K.~Jassal, J.S.~Bagla and T.~Padmanabhan, \emph{{Observational constraints on low redshift evolution of dark energy: How consistent are different observations?}}, \href{https://doi.org/10.1103/PhysRevD.72.103503}{\emph{Phys. Rev. D} {\bfseries 72} (2005) 103503} [\href{https://arxiv.org/abs/astro-ph/0506748}{{\ttfamily astro-ph/0506748}}].

\bibitem{Nesseris:2005ur}
S.~Nesseris and L.~Perivolaropoulos, \emph{{Comparison of the legacy and gold snia dataset constraints on dark energy models}}, \href{https://doi.org/10.1103/PhysRevD.72.123519}{\emph{Phys. Rev. D} {\bfseries 72} (2005) 123519} [\href{https://arxiv.org/abs/astro-ph/0511040}{{\ttfamily astro-ph/0511040}}].

\bibitem{Liu:2008vy}
D.-J.~Liu, X.-Z.~Li, J.~Hao and X.-H.~Jin, \emph{{Revisiting the parametrization of Equation of State of Dark Energy via SNIa Data}}, \href{https://doi.org/10.1111/j.1365-2966.2008.13380.x}{\emph{Mon. Not. Roy. Astron. Soc.} {\bfseries 388} (2008) 275} [\href{https://arxiv.org/abs/0804.3829}{{\ttfamily 0804.3829}}].

\bibitem{Barboza:2008rh}
E.M.~Barboza, Jr. and J.S.~Alcaniz, \emph{{A parametric model for dark energy}}, \href{https://doi.org/10.1016/j.physletb.2008.08.012}{\emph{Phys. Lett. B} {\bfseries 666} (2008) 415} [\href{https://arxiv.org/abs/0805.1713}{{\ttfamily 0805.1713}}].

\bibitem{Ma:2011nc}
J.-Z.~Ma and X.~Zhang, \emph{{Probing the dynamics of dark energy with novel parametrizations}}, \href{https://doi.org/10.1016/j.physletb.2011.04.013}{\emph{Phys. Lett. B} {\bfseries 699} (2011) 233} [\href{https://arxiv.org/abs/1102.2671}{{\ttfamily 1102.2671}}].

\bibitem{Sendra:2011pt}
I.~Sendra and R.~Lazkoz, \emph{{SN and BAO constraints on (new) polynomial dark energy parametrizations: current results and forecasts}}, \href{https://doi.org/10.1111/j.1365-2966.2012.20661.x}{\emph{Mon. Not. Roy. Astron. Soc.} {\bfseries 422} (2012) 776} [\href{https://arxiv.org/abs/1105.4943}{{\ttfamily 1105.4943}}].

\bibitem{DeFelice:2012vd}
A.~De~Felice, S.~Nesseris and S.~Tsujikawa, \emph{{Observational constraints on dark energy with a fast varying equation of state}}, \href{https://doi.org/10.1088/1475-7516/2012/05/029}{\emph{JCAP} {\bfseries 05} (2012) 029} [\href{https://arxiv.org/abs/1203.6760}{{\ttfamily 1203.6760}}].

\bibitem{Li:2012vn}
H.~Li and X.~Zhang, \emph{{Constraining dynamical dark energy with a divergence-free parametrization in the presence of spatial curvature and massive neutrinos}}, \href{https://doi.org/10.1016/j.physletb.2012.06.030}{\emph{Phys. Lett. B} {\bfseries 713} (2012) 160} [\href{https://arxiv.org/abs/1202.4071}{{\ttfamily 1202.4071}}].

\bibitem{Feng:2012gf}
C.-J.~Feng, X.-Y.~Shen, P.~Li and X.-Z.~Li, \emph{{A New Class of Parametrization for Dark Energy without Divergence}}, \href{https://doi.org/10.1088/1475-7516/2012/09/023}{\emph{JCAP} {\bfseries 09} (2012) 023} [\href{https://arxiv.org/abs/1206.0063}{{\ttfamily 1206.0063}}].

\bibitem{Magana:2014voa}
J.~Maga\~na, V.H.~C\'ardenas and V.~Motta, \emph{{Cosmic slowing down of acceleration for several dark energy parametrizations}}, \href{https://doi.org/10.1088/1475-7516/2014/10/017}{\emph{JCAP} {\bfseries 10} (2014) 017} [\href{https://arxiv.org/abs/1407.1632}{{\ttfamily 1407.1632}}].

\bibitem{Pantazis:2016nky}
G.~Pantazis, S.~Nesseris and L.~Perivolaropoulos, \emph{{Comparison of thawing and freezing dark energy parametrizations}}, \href{https://doi.org/10.1103/PhysRevD.93.103503}{\emph{Phys. Rev. D} {\bfseries 93} (2016) 103503} [\href{https://arxiv.org/abs/1603.02164}{{\ttfamily 1603.02164}}].

\bibitem{DiValentino:2016hlg}
E.~Di~Valentino, A.~Melchiorri and J.~Silk, \emph{{Reconciling Planck with the local value of $H_0$ in extended parameter space}}, \href{https://doi.org/10.1016/j.physletb.2016.08.043}{\emph{Phys. Lett. B} {\bfseries 761} (2016) 242} [\href{https://arxiv.org/abs/1606.00634}{{\ttfamily 1606.00634}}].

\bibitem{Yang:2017alx}
W.~Yang, S.~Pan and A.~Paliathanasis, \emph{{Latest astronomical constraints on some non-linear parametric dark energy models}}, \href{https://doi.org/10.1093/mnras/sty019}{\emph{Mon. Not. Roy. Astron. Soc.} {\bfseries 475} (2018) 2605} [\href{https://arxiv.org/abs/1708.01717}{{\ttfamily 1708.01717}}].

\bibitem{Pan:2017zoh}
S.~Pan, E.N.~Saridakis and W.~Yang, \emph{{Observational Constraints on Oscillating Dark-Energy Parametrizations}}, \href{https://doi.org/10.1103/PhysRevD.98.063510}{\emph{Phys. Rev. D} {\bfseries 98} (2018) 063510} [\href{https://arxiv.org/abs/1712.05746}{{\ttfamily 1712.05746}}].

\bibitem{Mortsell:2018mfj}
E.~M\"ortsell and S.~Dhawan, \emph{{Does the Hubble constant tension call for new physics?}}, \href{https://doi.org/10.1088/1475-7516/2018/09/025}{\emph{JCAP} {\bfseries 09} (2018) 025} [\href{https://arxiv.org/abs/1801.07260}{{\ttfamily 1801.07260}}].

\bibitem{Dutta:2018vmq}
K.~Dutta, Ruchika, A.~Roy, A.A.~Sen and M.M.~Sheikh-Jabbari, \emph{{Beyond $\Lambda $CDM with low and high redshift data: implications for dark energy}}, \href{https://doi.org/10.1007/s10714-020-2665-4}{\emph{Gen. Rel. Grav.} {\bfseries 52} (2020) 15} [\href{https://arxiv.org/abs/1808.06623}{{\ttfamily 1808.06623}}].

\bibitem{Yang:2018prh}
W.~Yang, S.~Pan, E.~Di~Valentino and E.N.~Saridakis, \emph{{Observational constraints on dynamical dark energy with pivoting redshift}}, \href{https://doi.org/10.3390/universe5110219}{\emph{Universe} {\bfseries 5} (2019) 219} [\href{https://arxiv.org/abs/1811.06932}{{\ttfamily 1811.06932}}].

\bibitem{Yang:2018qmz}
W.~Yang, S.~Pan, E.~Di~Valentino, E.N.~Saridakis and S.~Chakraborty, \emph{{Observational constraints on one-parameter dynamical dark-energy parametrizations and the $H_0$ tension}}, \href{https://doi.org/10.1103/PhysRevD.99.043543}{\emph{Phys. Rev. D} {\bfseries 99} (2019) 043543} [\href{https://arxiv.org/abs/1810.05141}{{\ttfamily 1810.05141}}].

\bibitem{Li:2019yem}
X.~Li and A.~Shafieloo, \emph{{A Simple Phenomenological Emergent Dark Energy Model can Resolve the Hubble Tension}}, \href{https://doi.org/10.3847/2041-8213/ab3e09}{\emph{Astrophys. J. Lett.} {\bfseries 883} (2019) L3} [\href{https://arxiv.org/abs/1906.08275}{{\ttfamily 1906.08275}}].

\bibitem{Vagnozzi:2019ezj}
S.~Vagnozzi, \emph{{New physics in light of the $H_0$ tension: An alternative view}}, \href{https://doi.org/10.1103/PhysRevD.102.023518}{\emph{Phys. Rev. D} {\bfseries 102} (2020) 023518} [\href{https://arxiv.org/abs/1907.07569}{{\ttfamily 1907.07569}}].

\bibitem{Visinelli:2019qqu}
L.~Visinelli, S.~Vagnozzi and U.~Danielsson, \emph{{Revisiting a negative cosmological constant from low-redshift data}}, \href{https://doi.org/10.3390/sym11081035}{\emph{Symmetry} {\bfseries 11} (2019) 1035} [\href{https://arxiv.org/abs/1907.07953}{{\ttfamily 1907.07953}}].

\bibitem{DiValentino:2019ffd}
E.~Di~Valentino, A.~Melchiorri, O.~Mena and S.~Vagnozzi, \emph{{Interacting dark energy in the early 2020s: A promising solution to the $H_0$ and cosmic shear tensions}}, \href{https://doi.org/10.1016/j.dark.2020.100666}{\emph{Phys. Dark Univ.} {\bfseries 30} (2020) 100666} [\href{https://arxiv.org/abs/1908.04281}{{\ttfamily 1908.04281}}].

\bibitem{Dutta:2019pio}
K.~Dutta, A.~Roy, Ruchika, A.A.~Sen and M.M.~Sheikh-Jabbari, \emph{{Cosmology with low-redshift observations: No signal for new physics}}, \href{https://doi.org/10.1103/PhysRevD.100.103501}{\emph{Phys. Rev. D} {\bfseries 100} (2019) 103501} [\href{https://arxiv.org/abs/1908.07267}{{\ttfamily 1908.07267}}].

\bibitem{DiValentino:2019jae}
E.~Di~Valentino, A.~Melchiorri, O.~Mena and S.~Vagnozzi, \emph{{Nonminimal dark sector physics and cosmological tensions}}, \href{https://doi.org/10.1103/PhysRevD.101.063502}{\emph{Phys. Rev. D} {\bfseries 101} (2020) 063502} [\href{https://arxiv.org/abs/1910.09853}{{\ttfamily 1910.09853}}].

\bibitem{Pan:2019hac}
S.~Pan, W.~Yang, E.~Di~Valentino, A.~Shafieloo and S.~Chakraborty, \emph{{Reconciling $H_0$ tension in a six parameter space?}}, \href{https://doi.org/10.1088/1475-7516/2020/06/062}{\emph{JCAP} {\bfseries 06} (2020) 062} [\href{https://arxiv.org/abs/1907.12551}{{\ttfamily 1907.12551}}].

\bibitem{Teixeira:2019hil}
E.M.~Teixeira, A.~Nunes and N.J.~Nunes, \emph{{Disformally Coupled Quintessence}}, \href{https://doi.org/10.1103/PhysRevD.101.083506}{\emph{Phys. Rev. D} {\bfseries 101} (2020) 083506} [\href{https://arxiv.org/abs/1912.13348}{{\ttfamily 1912.13348}}].

\bibitem{Teixeira:2019tfi}
E.M.~Teixeira, A.~Nunes and N.J.~Nunes, \emph{{Conformally Coupled Tachyonic Dark Energy}}, \href{https://doi.org/10.1103/PhysRevD.100.043539}{\emph{Phys. Rev. D} {\bfseries 100} (2019) 043539} [\href{https://arxiv.org/abs/1903.06028}{{\ttfamily 1903.06028}}].

\bibitem{Martinelli:2019dau}
M.~Martinelli, N.B.~Hogg, S.~Peirone, M.~Bruni and D.~Wands, \emph{{Constraints on the interacting vacuum\textendash{}geodesic CDM scenario}}, \href{https://doi.org/10.1093/mnras/stz1915}{\emph{Mon. Not. Roy. Astron. Soc.} {\bfseries 488} (2019) 3423} [\href{https://arxiv.org/abs/1902.10694}{{\ttfamily 1902.10694}}].

\bibitem{Zumalacarregui:2020cjh}
M.~Zumalacarregui, \emph{{Gravity in the Era of Equality: Towards solutions to the Hubble problem without fine-tuned initial conditions}}, \href{https://doi.org/10.1103/PhysRevD.102.023523}{\emph{Phys. Rev. D} {\bfseries 102} (2020) 023523} [\href{https://arxiv.org/abs/2003.06396}{{\ttfamily 2003.06396}}].

\bibitem{Hogg:2020rdp}
N.B.~Hogg, M.~Bruni, R.~Crittenden, M.~Martinelli and S.~Peirone, \emph{{Latest evidence for a late time vacuum\textendash{}geodesic CDM interaction}}, \href{https://doi.org/10.1016/j.dark.2020.100583}{\emph{Phys. Dark Univ.} {\bfseries 29} (2020) 100583} [\href{https://arxiv.org/abs/2002.10449}{{\ttfamily 2002.10449}}].

\bibitem{Alestas:2020mvb}
G.~Alestas, L.~Kazantzidis and L.~Perivolaropoulos, \emph{{$H_0$ tension, phantom dark energy, and cosmological parameter degeneracies}}, \href{https://doi.org/10.1103/PhysRevD.101.123516}{\emph{Phys. Rev. D} {\bfseries 101} (2020) 123516} [\href{https://arxiv.org/abs/2004.08363}{{\ttfamily 2004.08363}}].

\bibitem{DiValentino:2020naf}
E.~Di~Valentino, A.~Mukherjee and A.A.~Sen, \emph{{Dark Energy with Phantom Crossing and the $H_0$ Tension}}, \href{https://doi.org/10.3390/e23040404}{\emph{Entropy} {\bfseries 23} (2021) 404} [\href{https://arxiv.org/abs/2005.12587}{{\ttfamily 2005.12587}}].

\bibitem{Alestas:2020zol}
G.~Alestas, L.~Kazantzidis and L.~Perivolaropoulos, \emph{{$w-M$ phantom transition at $z_t$ \ensuremath{<}0.1 as a resolution of the Hubble tension}}, \href{https://doi.org/10.1103/PhysRevD.103.083517}{\emph{Phys. Rev. D} {\bfseries 103} (2021) 083517} [\href{https://arxiv.org/abs/2012.13932}{{\ttfamily 2012.13932}}].

\bibitem{Rezaei:2020mrj}
M.~Rezaei, T.~Naderi, M.~Malekjani and A.~Mehrabi, \emph{{A Bayesian comparison between $\Lambda$CDM and phenomenologically emergent dark energy models}}, \href{https://doi.org/10.1140/epjc/s10052-020-7942-6}{\emph{Eur. Phys. J. C} {\bfseries 80} (2020) 374} [\href{https://arxiv.org/abs/2004.08168}{{\ttfamily 2004.08168}}].

\bibitem{Perkovic:2020mph}
D.~Perkovic and H.~Stefancic, \emph{{Barotropic fluid compatible parametrizations of dark energy}}, \href{https://doi.org/10.1140/epjc/s10052-020-8199-9}{\emph{Eur. Phys. J. C} {\bfseries 80} (2020) 629} [\href{https://arxiv.org/abs/2004.05342}{{\ttfamily 2004.05342}}].

\bibitem{Benaoum:2020qsi}
H.B.~Benaoum, W.~Yang, S.~Pan and E.~Di~Valentino, \emph{{Modified emergent dark energy and its astronomical constraints}}, \href{https://doi.org/10.1142/S0218271822500158}{\emph{Int. J. Mod. Phys. D} {\bfseries 31} (2022) 2250015} [\href{https://arxiv.org/abs/2008.09098}{{\ttfamily 2008.09098}}].

\bibitem{Kumar:2021eev}
S.~Kumar, \emph{{Remedy of some cosmological tensions via effective phantom-like behavior of interacting vacuum energy}}, \href{https://doi.org/10.1016/j.dark.2021.100862}{\emph{Phys. Dark Univ.} {\bfseries 33} (2021) 100862} [\href{https://arxiv.org/abs/2102.12902}{{\ttfamily 2102.12902}}].

\bibitem{Vagnozzi:2021tjv}
S.~Vagnozzi, F.~Pacucci and A.~Loeb, \emph{{Implications for the Hubble tension from the ages of the oldest astrophysical objects}}, \href{https://doi.org/10.1016/j.jheap.2022.07.004}{\emph{JHEAp} {\bfseries 36} (2022) 27} [\href{https://arxiv.org/abs/2105.10421}{{\ttfamily 2105.10421}}].

\bibitem{Escamilla:2021uoj}
L.A.~Escamilla and J.A.~Vazquez, \emph{{Model selection applied to reconstructions of the Dark Energy}}, \href{https://doi.org/10.1140/epjc/s10052-023-11404-2}{\emph{Eur. Phys. J. C} {\bfseries 83} (2023) 251} [\href{https://arxiv.org/abs/2111.10457}{{\ttfamily 2111.10457}}].

\bibitem{Bag:2021cqm}
S.~Bag, V.~Sahni, A.~Shafieloo and Y.~Shtanov, \emph{{Phantom Braneworld and the Hubble Tension}}, \href{https://doi.org/10.3847/1538-4357/ac307e}{\emph{Astrophys. J.} {\bfseries 923} (2021) 212} [\href{https://arxiv.org/abs/2107.03271}{{\ttfamily 2107.03271}}].

\bibitem{Theodoropoulos:2021hkk}
A.~Theodoropoulos and L.~Perivolaropoulos, \emph{{The Hubble Tension, the M Crisis of Late Time H(z) Deformation Models and the Reconstruction of Quintessence Lagrangians}}, \href{https://doi.org/10.3390/universe7080300}{\emph{Universe} {\bfseries 7} (2021) 300} [\href{https://arxiv.org/abs/2109.06256}{{\ttfamily 2109.06256}}].

\bibitem{Alestas:2021luu}
G.~Alestas, D.~Camarena, E.~Di~Valentino, L.~Kazantzidis, V.~Marra, S.~Nesseris et~al., \emph{{Late-transition versus smooth $H(z)$-deformation models for the resolution of the Hubble crisis}}, \href{https://doi.org/10.1103/PhysRevD.105.063538}{\emph{Phys. Rev. D} {\bfseries 105} (2022) 063538} [\href{https://arxiv.org/abs/2110.04336}{{\ttfamily 2110.04336}}].

\bibitem{Sen:2021wld}
A.A.~Sen, S.A.~Adil and S.~Sen, \emph{{Do cosmological observations allow a negative \ensuremath{\Lambda}?}}, \href{https://doi.org/10.1093/mnras/stac2796}{\emph{Mon. Not. Roy. Astron. Soc.} {\bfseries 518} (2022) 1098} [\href{https://arxiv.org/abs/2112.10641}{{\ttfamily 2112.10641}}].

\bibitem{Yang:2021eud}
W.~Yang, E.~Di~Valentino, S.~Pan, A.~Shafieloo and X.~Li, \emph{{Generalized emergent dark energy model and the Hubble constant tension}}, \href{https://doi.org/10.1103/PhysRevD.104.063521}{\emph{Phys. Rev. D} {\bfseries 104} (2021) 063521} [\href{https://arxiv.org/abs/2103.03815}{{\ttfamily 2103.03815}}].

\bibitem{Hogg:2021yiz}
N.B.~Hogg and M.~Bruni, \emph{{Shan\textendash{}Chen interacting vacuum cosmology}}, \href{https://doi.org/10.1093/mnras/stac324}{\emph{Mon. Not. Roy. Astron. Soc.} {\bfseries 511} (2022) 4430} [\href{https://arxiv.org/abs/2109.08676}{{\ttfamily 2109.08676}}].

\bibitem{Roy:2022fif}
N.~Roy, S.~Goswami and S.~Das, \emph{{Quintessence or phantom: Study of scalar field dark energy models through a general parametrization of the Hubble parameter}}, \href{https://doi.org/10.1016/j.dark.2022.101037}{\emph{Phys. Dark Univ.} {\bfseries 36} (2022) 101037} [\href{https://arxiv.org/abs/2201.09306}{{\ttfamily 2201.09306}}].

\bibitem{Heisenberg:2022lob}
L.~Heisenberg, H.~Villarrubia-Rojo and J.~Zosso, \emph{{Simultaneously solving the H0 and \ensuremath{\sigma}8 tensions with late dark energy}}, \href{https://doi.org/10.1016/j.dark.2022.101163}{\emph{Phys. Dark Univ.} {\bfseries 39} (2023) 101163} [\href{https://arxiv.org/abs/2201.11623}{{\ttfamily 2201.11623}}].

\bibitem{Chudaykin:2022rnl}
A.~Chudaykin, D.~Gorbunov and N.~Nedelko, \emph{{Exploring ${\Lambda}$CDM extensions with SPT-3G and Planck data: 4$\sigma$ evidence for neutrino masses and implications of extended dark energy models for cosmological tensions}},  \href{https://arxiv.org/abs/2203.03666}{{\ttfamily 2203.03666}}.

\bibitem{Akarsu:2022typ}
O.~Akarsu, S.~Kumar, E.~\"Oz\"ulker, J.A.~Vazquez and A.~Yadav, \emph{{Relaxing cosmological tensions with a sign switching cosmological constant: Improved results with Planck, BAO, and Pantheon data}}, \href{https://doi.org/10.1103/PhysRevD.108.023513}{\emph{Phys. Rev. D} {\bfseries 108} (2023) 023513} [\href{https://arxiv.org/abs/2211.05742}{{\ttfamily 2211.05742}}].

\bibitem{Santos:2022atq}
F.B.M.d.~Santos, \emph{{Updating constraints on phantom crossing f(T) gravity}}, \href{https://doi.org/10.1088/1475-7516/2023/06/039}{\emph{JCAP} {\bfseries 06} (2023) 039} [\href{https://arxiv.org/abs/2211.16370}{{\ttfamily 2211.16370}}].

\bibitem{Schiavone:2022wvq}
T.~Schiavone, G.~Montani and F.~Bombacigno, \emph{{f(R) gravity in the Jordan frame as a paradigm for the Hubble tension}}, \href{https://doi.org/10.1093/mnrasl/slad041}{\emph{Mon. Not. Roy. Astron. Soc.} {\bfseries 522} (2023) L72} [\href{https://arxiv.org/abs/2211.16737}{{\ttfamily 2211.16737}}].

\bibitem{vandeBruck:2022xbk}
C.~van~de Bruck, G.~Poulot and E.M.~Teixeira, \emph{{Scalar field dark matter and dark energy: a hybrid model for the dark sector}}, \href{https://doi.org/10.1088/1475-7516/2023/07/019}{\emph{JCAP} {\bfseries 07} (2023) 019} [\href{https://arxiv.org/abs/2211.13653}{{\ttfamily 2211.13653}}].

\bibitem{Ozulker:2022slu}
E.~Ozulker, \emph{{Is the dark energy equation of state parameter singular?}}, \href{https://doi.org/10.1103/PhysRevD.106.063509}{\emph{Phys. Rev. D} {\bfseries 106} (2022) 063509} [\href{https://arxiv.org/abs/2203.04167}{{\ttfamily 2203.04167}}].

\bibitem{Teixeira:2022sjr}
E.M.~Teixeira, B.J.~Barros, V.M.C.~Ferreira and N.~Frusciante, \emph{{Dissecting kinetically coupled quintessence: phenomenology and observational tests}}, \href{https://doi.org/10.1088/1475-7516/2022/11/059}{\emph{JCAP} {\bfseries 11} (2022) 059} [\href{https://arxiv.org/abs/2207.13682}{{\ttfamily 2207.13682}}].

\bibitem{Ben-Dayan:2023rgt}
I.~Ben-Dayan and U.~Kumar, \emph{{Emergent Unparticles Dark Energy can restore cosmological concordance}}, \href{https://doi.org/10.1088/1475-7516/2023/12/047}{\emph{JCAP} {\bfseries 12} (2023) 047} [\href{https://arxiv.org/abs/2302.00067}{{\ttfamily 2302.00067}}].

\bibitem{Ballardini:2023mzm}
M.~Ballardini, A.G.~Ferrari and F.~Finelli, \emph{{Phantom scalar-tensor models and cosmological tensions}}, \href{https://doi.org/10.1088/1475-7516/2023/04/029}{\emph{JCAP} {\bfseries 04} (2023) 029} [\href{https://arxiv.org/abs/2302.05291}{{\ttfamily 2302.05291}}].

\bibitem{Yang:2022kho}
W.~Yang, W.~Giar\`e, S.~Pan, E.~Di~Valentino, A.~Melchiorri and J.~Silk, \emph{{Revealing the effects of curvature on the cosmological models}}, \href{https://doi.org/10.1103/PhysRevD.107.063509}{\emph{Phys. Rev. D} {\bfseries 107} (2023) 063509} [\href{https://arxiv.org/abs/2210.09865}{{\ttfamily 2210.09865}}].

\bibitem{deCruzPerez:2023wzd}
J.~de~Cruz~Perez and J.~Sola~Peracaula, \emph{{Running vacuum in Brans \& Dicke theory: A possible cure for the \ensuremath{\sigma}8 and H0 tensions}}, \href{https://doi.org/10.1016/j.dark.2023.101406}{\emph{Phys. Dark Univ.} {\bfseries 43} (2024) 101406} [\href{https://arxiv.org/abs/2302.04807}{{\ttfamily 2302.04807}}].

\bibitem{Patil:2023rqy}
T.~Patil, Ruchika and S.~Panda, \emph{{Coupled quintessence scalar field model in light of observational datasets}}, \href{https://doi.org/10.1088/1475-7516/2024/05/033}{\emph{JCAP} {\bfseries 05} (2024) 033} [\href{https://arxiv.org/abs/2307.03740}{{\ttfamily 2307.03740}}].

\bibitem{Zhai:2023yny}
Y.~Zhai, W.~Giar\`e, C.~van~de Bruck, E.~Di~Valentino, O.~Mena and R.C.~Nunes, \emph{{A consistent view of interacting dark energy from multiple CMB probes}}, \href{https://doi.org/10.1088/1475-7516/2023/07/032}{\emph{JCAP} {\bfseries 07} (2023) 032} [\href{https://arxiv.org/abs/2303.08201}{{\ttfamily 2303.08201}}].

\bibitem{Adil:2023exv}
S.A.~Adil, O.~Akarsu, E.~Di~Valentino, R.C.~Nunes, E.~\"Oz\"ulker, A.A.~Sen et~al., \emph{{Omnipotent dark energy: A phenomenological answer to the Hubble tension}}, \href{https://doi.org/10.1103/PhysRevD.109.023527}{\emph{Phys. Rev. D} {\bfseries 109} (2024) 023527} [\href{https://arxiv.org/abs/2306.08046}{{\ttfamily 2306.08046}}].

\bibitem{Montani:2023xpd}
G.~Montani, M.~De~Angelis, F.~Bombacigno and N.~Carlevaro, \emph{{Metric f(R) gravity with dynamical dark energy as a scenario for the Hubble tension}}, \href{https://doi.org/10.1093/mnrasl/slad159}{\emph{Mon. Not. Roy. Astron. Soc.} {\bfseries 527} (2023) L156} [\href{https://arxiv.org/abs/2306.11101}{{\ttfamily 2306.11101}}].

\bibitem{Akarsu:2023mfb}
O.~Akarsu, E.~Di~Valentino, S.~Kumar, R.C.~Nunes, J.A.~Vazquez and A.~Yadav, \emph{{$\Lambda_{\rm s}$CDM model: A promising scenario for alleviation of cosmological tensions}},  \href{https://arxiv.org/abs/2307.10899}{{\ttfamily 2307.10899}}.

\bibitem{Vagnozzi:2023nrq}
S.~Vagnozzi, \emph{{Seven Hints That Early-Time New Physics Alone Is Not Sufficient to Solve the Hubble Tension}}, \href{https://doi.org/10.3390/universe9090393}{\emph{Universe} {\bfseries 9} (2023) 393} [\href{https://arxiv.org/abs/2308.16628}{{\ttfamily 2308.16628}}].

\bibitem{Avsajanishvili:2023jcl}
O.~Avsajanishvili, G.Y.~Chitov, T.~Kahniashvili, S.~Mandal and L.~Samushia, \emph{{Observational Constraints on Dynamical Dark Energy Models}}, \href{https://doi.org/10.3390/universe10030122}{\emph{Universe} {\bfseries 10} (2024) 122} [\href{https://arxiv.org/abs/2310.16911}{{\ttfamily 2310.16911}}].

\bibitem{Giani:2023aor}
L.~Giani, C.~Howlett, K.~Said, T.~Davis and S.~Vagnozzi, \emph{{An effective description of Laniakea: impact on cosmology and the local determination of the Hubble constant}}, \href{https://doi.org/10.1088/1475-7516/2024/01/071}{\emph{JCAP} {\bfseries 01} (2024) 071} [\href{https://arxiv.org/abs/2311.00215}{{\ttfamily 2311.00215}}].

\bibitem{Lazkoz:2023oqc}
R.~Lazkoz, V.~Salzano, L.~Fernandez-Jambrina and M.~Bouhmadi-L\'opez, \emph{{Ripped \ensuremath{\Lambda}CDM: An observational contender to the consensus cosmological model}}, \href{https://doi.org/10.1016/j.dark.2024.101511}{\emph{Phys. Dark Univ.} {\bfseries 45} (2024) 101511} [\href{https://arxiv.org/abs/2311.10526}{{\ttfamily 2311.10526}}].

\bibitem{Escamilla:2023oce}
L.A.~Escamilla, W.~Giar\`e, E.~Di~Valentino, R.C.~Nunes and S.~Vagnozzi, \emph{{The state of the dark energy equation of state circa 2023}}, \href{https://doi.org/10.1088/1475-7516/2024/05/091}{\emph{JCAP} {\bfseries 05} (2024) 091} [\href{https://arxiv.org/abs/2307.14802}{{\ttfamily 2307.14802}}].

\bibitem{Escamilla:2023shf}
L.A.~Escamilla, O.~Akarsu, E.~Di~Valentino and J.A.~Vazquez, \emph{{Model-independent reconstruction of the interacting dark energy kernel: Binned and Gaussian process}}, \href{https://doi.org/10.1088/1475-7516/2023/11/051}{\emph{JCAP} {\bfseries 11} (2023) 051} [\href{https://arxiv.org/abs/2305.16290}{{\ttfamily 2305.16290}}].

\bibitem{Rezaei:2023xkj}
M.~Rezaei, S.~Pan, W.~Yang and D.F.~Mota, \emph{{Evidence of dynamical dark energy in a non-flat universe: current and future observations}}, \href{https://doi.org/10.1088/1475-7516/2024/01/052}{\emph{JCAP} {\bfseries 01} (2024) 052} [\href{https://arxiv.org/abs/2305.18544}{{\ttfamily 2305.18544}}].

\bibitem{Teixeira:2023zjt}
E.M.~Teixeira, R.~Daniel, N.~Frusciante and C.~van~de Bruck, \emph{{Forecasts on interacting dark energy with standard sirens}}, \href{https://doi.org/10.1103/PhysRevD.108.084070}{\emph{Phys. Rev. D} {\bfseries 108} (2023) 084070} [\href{https://arxiv.org/abs/2309.06544}{{\ttfamily 2309.06544}}].

\bibitem{Forconi:2023hsj}
M.~Forconi, W.~Giar\`e, O.~Mena, Ruchika, E.~Di~Valentino, A.~Melchiorri et~al., \emph{{A double take on early and interacting dark energy from JWST}}, \href{https://doi.org/10.1088/1475-7516/2024/05/097}{\emph{JCAP} {\bfseries 05} (2024) 097} [\href{https://arxiv.org/abs/2312.11074}{{\ttfamily 2312.11074}}].

\bibitem{Sebastianutti:2023dbt}
M.~Sebastianutti, N.B.~Hogg and M.~Bruni, \emph{{The interacting vacuum and tensions: A comparison of theoretical models}}, \href{https://doi.org/10.1016/j.dark.2024.101546}{\emph{Phys. Dark Univ.} {\bfseries 46} (2024) 101546} [\href{https://arxiv.org/abs/2312.14123}{{\ttfamily 2312.14123}}].

\bibitem{Wolf:2023uno}
W.J.~Wolf and P.G.~Ferreira, \emph{{Underdetermination of dark energy}}, \href{https://doi.org/10.1103/PhysRevD.108.103519}{\emph{Phys. Rev. D} {\bfseries 108} (2023) 103519} [\href{https://arxiv.org/abs/2310.07482}{{\ttfamily 2310.07482}}].

\bibitem{Giare:2024sdl}
W.~Giar\`e, E.~Di~Valentino, E.V.~Linder and E.~Specogna, \emph{{Testing $\alpha$-attractor quintessential inflation against CMB and low-redshift data}},  \href{https://arxiv.org/abs/2402.01560}{{\ttfamily 2402.01560}}.

\bibitem{Giare:2024smz}
W.~Giar\`e, M.A.~Sabogal, R.C.~Nunes and E.~Di~Valentino, \emph{{Interacting Dark Energy after DESI Baryon Acoustic Oscillation measurements}},  \href{https://arxiv.org/abs/2404.15232}{{\ttfamily 2404.15232}}.

\bibitem{Giare:2024ytc}
W.~Giar\`e, Y.~Zhai, S.~Pan, E.~Di~Valentino, R.C.~Nunes and C.~van~de Bruck, \emph{{Tightening the reins on non-minimal dark sector physics: Interacting Dark Energy with dynamical and non-dynamical equation of state}},  \href{https://arxiv.org/abs/2404.02110}{{\ttfamily 2404.02110}}.

\bibitem{Menci:2024rbq}
N.~Menci, S.A.~Adil, U.~Mukhopadhyay, A.A.~Sen and S.~Vagnozzi, \emph{{Negative cosmological constant in the dark energy sector: tests from JWST photometric and spectroscopic observations of high-redshift galaxies}},  \href{https://arxiv.org/abs/2401.12659}{{\ttfamily 2401.12659}}.

\bibitem{Akarsu:2024qiq}
O.~Akarsu, E.O.~Colg\'ain, A.A.~Sen and M.M.~Sheikh-Jabbari, \emph{{$\Lambda$CDM Tensions: Localising Missing Physics through Consistency Checks}},  \href{https://arxiv.org/abs/2402.04767}{{\ttfamily 2402.04767}}.

\bibitem{Teixeira:2024wsw}
E.M.~Teixeira, \emph{{Illuminating the Dark Sector: Searching for new interactions between dark matter and dark energy}},  1, 2024.

\bibitem{Benisty:2024lmj}
D.~Benisty, S.~Pan, D.~Staicova, E.~Di~Valentino and R.C.~Nunes, \emph{{Late-Time constraints on Interacting Dark Energy: Analysis independent of $H_0$, $r_d$ and $M_B$}},  \href{https://arxiv.org/abs/2403.00056}{{\ttfamily 2403.00056}}.

\bibitem{Najafi:2024qzm}
M.~Najafi, S.~Pan, E.~Di~Valentino and J.T.~Firouzjaee, \emph{{Dynamical dark energy confronted with multiple CMB missions}}, \href{https://doi.org/10.1016/j.dark.2024.101539}{\emph{Phys. Dark Univ.} {\bfseries 45} (2024) 101539}.

\bibitem{Moshafi:2024guo}
H.~Moshafi, A.~Talebian, E.~Yusofi and E.~Di~Valentino, \emph{{Observational constraints on the dark energy with a quadratic equation of state}}, \href{https://doi.org/10.1016/j.dark.2024.101524}{\emph{Phys. Dark Univ.} {\bfseries 45} (2024) 101524} [\href{https://arxiv.org/abs/2403.02000}{{\ttfamily 2403.02000}}].

\bibitem{Silva:2024ift}
E.~Silva, U.~Z\'u\~niga Bola\~no, R.C.~Nunes and E.~Di~Valentino, \emph{{Non-Linear Matter Power Spectrum Modeling in Interacting Dark Energy Cosmologies}},  \href{https://arxiv.org/abs/2403.19590}{{\ttfamily 2403.19590}}.

\bibitem{Reyhani:2024cnr}
M.~Reyhani, M.~Najafi, J.T.~Firouzjaee and E.~Di~Valentino, \emph{{Structure formation in various dynamical dark energy scenarios}}, \href{https://doi.org/10.1016/j.dark.2024.101477}{\emph{Phys. Dark Univ.} {\bfseries 44} (2024) 101477} [\href{https://arxiv.org/abs/2403.15202}{{\ttfamily 2403.15202}}].

\bibitem{Escamilla:2024olw}
L.A.~Escamilla, S.~Pan, E.~Di~Valentino, A.~Paliathanasis, J.A.~V\'azquez and W.~Yang, \emph{{Oscillations in the Dark?}},  \href{https://arxiv.org/abs/2404.00181}{{\ttfamily 2404.00181}}.

\bibitem{Wang:2024sgo}
H.~Wang, G.~Ye and Y.-S.~Piao, \emph{{Impact of evolving dark energy on the search for primordial gravitational waves}},  \href{https://arxiv.org/abs/2407.11263}{{\ttfamily 2407.11263}}.

\bibitem{Montani:2024xys}
G.~Montani, N.~Carlevaro and M.~De~Angelis, \emph{{Modified gravity in the presence of matter creation in the late Universe: alleviation of the Hubble tension}},  \href{https://arxiv.org/abs/2407.12409}{{\ttfamily 2407.12409}}.

\bibitem{Li:2024qso}
T.-N.~Li, P.-J.~Wu, G.-H.~Du, S.-J.~Jin, H.-L.~Li, J.-F.~Zhang et~al., \emph{{Constraints on interacting dark energy models from the DESI BAO and DES supernovae data}},  \href{https://arxiv.org/abs/2407.14934}{{\ttfamily 2407.14934}}.

\bibitem{Yang:2024kdo}
Y.~Yang, X.~Ren, Q.~Wang, Z.~Lu, D.~Zhang, Y.-F.~Cai et~al., \emph{{Quintom cosmology and modified gravity after DESI 2024}},  \href{https://arxiv.org/abs/2404.19437}{{\ttfamily 2404.19437}}.

\bibitem{Dwivedi:2024okk}
S.~Dwivedi and M.~H\"og\r{a}s, \emph{{2D BAO vs 3D BAO: solving the Hubble tension with alternative cosmological models}},  \href{https://arxiv.org/abs/2407.04322}{{\ttfamily 2407.04322}}.

\bibitem{ACT:2023dou}
{\scshape ACT} collaboration, \emph{{The Atacama Cosmology Telescope: A Measurement of the DR6 CMB Lensing Power Spectrum and Its Implications for Structure Growth}}, \href{https://doi.org/10.3847/1538-4357/acfe06}{\emph{Astrophys. J.} {\bfseries 962} (2024) 112} [\href{https://arxiv.org/abs/2304.05202}{{\ttfamily 2304.05202}}].

\bibitem{SPT-3G:2014dbx}
{\scshape SPT-3G} collaboration, \emph{{SPT-3G: A Next-Generation Cosmic Microwave Background Polarization Experiment on the South Pole Telescope}}, \href{https://doi.org/10.1117/12.2057305}{\emph{Proc. SPIE Int. Soc. Opt. Eng.} {\bfseries 9153} (2014) 91531P} [\href{https://arxiv.org/abs/1407.2973}{{\ttfamily 1407.2973}}].

\bibitem{SPT-3G:2022hvq}
{\scshape SPT-3G} collaboration, \emph{{Measurement of the CMB temperature power spectrum and constraints on cosmology from the SPT-3G 2018 TT, TE, and EE dataset}}, \href{https://doi.org/10.1103/PhysRevD.108.023510}{\emph{Phys. Rev. D} {\bfseries 108} (2023) 023510} [\href{https://arxiv.org/abs/2212.05642}{{\ttfamily 2212.05642}}].

\bibitem{Lewis:2006fu}
A.~Lewis and A.~Challinor, \emph{{Weak gravitational lensing of the CMB}}, \href{https://doi.org/10.1016/j.physrep.2006.03.002}{\emph{Phys. Rept.} {\bfseries 429} (2006) 1} [\href{https://arxiv.org/abs/astro-ph/0601594}{{\ttfamily astro-ph/0601594}}].

\bibitem{Giare:2020vzo}
W.~Giar\`e, E.~Di~Valentino, A.~Melchiorri and O.~Mena, \emph{{New cosmological bounds on hot relics: axions and neutrinos}}, \href{https://doi.org/10.1093/mnras/stab1442}{\emph{Mon. Not. Roy. Astron. Soc.} {\bfseries 505} (2021) 2703} [\href{https://arxiv.org/abs/2011.14704}{{\ttfamily 2011.14704}}].

\bibitem{DiValentino:2021imh}
E.~Di~Valentino and A.~Melchiorri, \emph{{Neutrino Mass Bounds in the Era of Tension Cosmology}}, \href{https://doi.org/10.3847/2041-8213/ac6ef5}{\emph{Astrophys. J. Lett.} {\bfseries 931} (2022) L18} [\href{https://arxiv.org/abs/2112.02993}{{\ttfamily 2112.02993}}].

\bibitem{DEramo:2022nvb}
F.~D'Eramo, E.~Di~Valentino, W.~Giar\`e, F.~Hajkarim, A.~Melchiorri, O.~Mena et~al., \emph{{Cosmological bound on the QCD axion mass, redux}}, \href{https://doi.org/10.1088/1475-7516/2022/09/022}{\emph{JCAP} {\bfseries 09} (2022) 022} [\href{https://arxiv.org/abs/2205.07849}{{\ttfamily 2205.07849}}].

\bibitem{DiValentino:2022edq}
E.~Di~Valentino, S.~Gariazzo, W.~Giar\`e, A.~Melchiorri, O.~Mena and F.~Renzi, \emph{{Novel model-marginalized cosmological bound on the QCD axion mass}}, \href{https://doi.org/10.1103/PhysRevD.107.103528}{\emph{Phys. Rev. D} {\bfseries 107} (2023) 103528} [\href{https://arxiv.org/abs/2212.11926}{{\ttfamily 2212.11926}}].

\bibitem{Giare:2023aix}
W.~Giar\`e, O.~Mena and E.~Di~Valentino, \emph{{Lensing impact on cosmic relics and tensions}}, \href{https://doi.org/10.1103/PhysRevD.108.103539}{\emph{Phys. Rev. D} {\bfseries 108} (2023) 103539} [\href{https://arxiv.org/abs/2307.14204}{{\ttfamily 2307.14204}}].

\bibitem{Planck:2018lbu}
{\scshape Planck} collaboration, \emph{{Planck 2018 results. VIII. Gravitational lensing}}, \href{https://doi.org/10.1051/0004-6361/201833886}{\emph{Astron. Astrophys.} {\bfseries 641} (2020) A8} [\href{https://arxiv.org/abs/1807.06210}{{\ttfamily 1807.06210}}].

\bibitem{Ye:2023zel}
G.~Ye, J.-Q.~Jiang and Y.-S.~Piao, \emph{{Shape of CMB lensing in the early dark energy cosmology}}, \href{https://doi.org/10.1103/PhysRevD.108.063512}{\emph{Phys. Rev. D} {\bfseries 108} (2023) 063512} [\href{https://arxiv.org/abs/2305.18873}{{\ttfamily 2305.18873}}].

\bibitem{ACT:2023skz}
{\scshape ACT, DES} collaboration, \emph{{Cosmology from cross-correlation of ACT-DR4 CMB lensing and DES-Y3 cosmic shear}}, \href{https://doi.org/10.1093/mnras/stad3987}{\emph{Mon. Not. Roy. Astron. Soc.} {\bfseries 528} (2024) 2112} [\href{https://arxiv.org/abs/2309.04412}{{\ttfamily 2309.04412}}].

\bibitem{ACT:2023ipp}
{\scshape ACT, DES} collaboration, \emph{{Cosmological constraints from the tomography of DES-Y3 galaxies with CMB lensing from ACT DR4}}, \href{https://doi.org/10.1088/1475-7516/2024/01/033}{\emph{JCAP} {\bfseries 01} (2024) 033} [\href{https://arxiv.org/abs/2306.17268}{{\ttfamily 2306.17268}}].

\bibitem{ACT:2024okh}
{\scshape ACT, DESI} collaboration, \emph{{The Atacama Cosmology Telescope DR6 and DESI: Structure formation over cosmic time with a measurement of the cross-correlation of CMB Lensing and Luminous Red Galaxies}},  \href{https://arxiv.org/abs/2407.04606}{{\ttfamily 2407.04606}}.

\bibitem{Sailer:2024coh}
N.~Sailer et~al., \emph{{Cosmological constraints from the cross-correlation of DESI Luminous Red Galaxies with CMB lensing from Planck PR4 and ACT DR6}},  \href{https://arxiv.org/abs/2407.04607}{{\ttfamily 2407.04607}}.

\bibitem{ACT:2024npz}
{\scshape ACT} collaboration, \emph{{The Atacama Cosmology Telescope: DR6 Gravitational Lensing and SDSS BOSS cross-correlation measurement and constraints on gravity with the $E_G$ statistic}},  \href{https://arxiv.org/abs/2405.12795}{{\ttfamily 2405.12795}}.

\bibitem{EUCLID:2011zbd}
{\scshape EUCLID} collaboration, \emph{{Euclid Definition Study Report}},  \href{https://arxiv.org/abs/1110.3193}{{\ttfamily 1110.3193}}.

\bibitem{LSST:2008ijt}
{\scshape LSST} collaboration, \emph{{LSST: from Science Drivers to Reference Design and Anticipated Data Products}}, \href{https://doi.org/10.3847/1538-4357/ab042c}{\emph{Astrophys. J.} {\bfseries 873} (2019) 111} [\href{https://arxiv.org/abs/0805.2366}{{\ttfamily 0805.2366}}].

\bibitem{Spergel:2013tha}
D.~Spergel et~al., \emph{{Wide-Field InfraRed Survey Telescope-Astrophysics Focused Telescope Assets WFIRST-AFTA Final Report}},  \href{https://arxiv.org/abs/1305.5422}{{\ttfamily 1305.5422}}.

\bibitem{SKA:2018ckk}
{\scshape SKA} collaboration, \emph{{Cosmology with Phase 1 of the Square Kilometre Array: Red Book 2018: Technical specifications and performance forecasts}}, \href{https://doi.org/10.1017/pasa.2019.51}{\emph{Publ. Astron. Soc. Austral.} {\bfseries 37} (2020) e007} [\href{https://arxiv.org/abs/1811.02743}{{\ttfamily 1811.02743}}].

\bibitem{DiValentino:2021izs}
E.~Di~Valentino, O.~Mena, S.~Pan, L.~Visinelli, W.~Yang, A.~Melchiorri et~al., \emph{{In the realm of the Hubble tension\textemdash{}a review of solutions}}, \href{https://doi.org/10.1088/1361-6382/ac086d}{\emph{Class. Quant. Grav.} {\bfseries 38} (2021) 153001} [\href{https://arxiv.org/abs/2103.01183}{{\ttfamily 2103.01183}}].

\bibitem{Perivolaropoulos:2021jda}
L.~Perivolaropoulos and F.~Skara, \emph{{Challenges for \ensuremath{\Lambda}CDM: An update}}, \href{https://doi.org/10.1016/j.newar.2022.101659}{\emph{New Astron. Rev.} {\bfseries 95} (2022) 101659} [\href{https://arxiv.org/abs/2105.05208}{{\ttfamily 2105.05208}}].

\bibitem{Abdalla:2022yfr}
E.~Abdalla et~al., \emph{{Cosmology intertwined: A review of the particle physics, astrophysics, and cosmology associated with the cosmological tensions and anomalies}}, \href{https://doi.org/10.1016/j.jheap.2022.04.002}{\emph{JHEAp} {\bfseries 34} (2022) 49} [\href{https://arxiv.org/abs/2203.06142}{{\ttfamily 2203.06142}}].

\bibitem{Gariazzo:2024sil}
S.~Gariazzo, W.~Giar\`e, O.~Mena and E.~Di~Valentino, \emph{{How robust are the parameter constraints extending the $\Lambda$CDM model?}},  \href{https://arxiv.org/abs/2404.11182}{{\ttfamily 2404.11182}}.

\bibitem{Bean:2001xy}
R.~Bean and A.~Melchiorri, \emph{{Current constraints on the dark energy equation of state}}, \href{https://doi.org/10.1103/PhysRevD.65.041302}{\emph{Phys. Rev. D} {\bfseries 65} (2002) 041302} [\href{https://arxiv.org/abs/astro-ph/0110472}{{\ttfamily astro-ph/0110472}}].

\bibitem{Hannestad:2002ur}
S.~Hannestad and E.~Mortsell, \emph{{Probing the dark side: Constraints on the dark energy equation of state from CMB, large scale structure and Type Ia supernovae}}, \href{https://doi.org/10.1103/PhysRevD.66.063508}{\emph{Phys. Rev. D} {\bfseries 66} (2002) 063508} [\href{https://arxiv.org/abs/astro-ph/0205096}{{\ttfamily astro-ph/0205096}}].

\bibitem{Said:2013jxa}
N.~Said, C.~Baccigalupi, M.~Martinelli, A.~Melchiorri and A.~Silvestri, \emph{{New Constraints On The Dark Energy Equation of State}}, \href{https://doi.org/10.1103/PhysRevD.88.043515}{\emph{Phys. Rev. D} {\bfseries 88} (2013) 043515} [\href{https://arxiv.org/abs/1303.4353}{{\ttfamily 1303.4353}}].

\bibitem{Shafer:2013pxa}
D.L.~Shafer and D.~Huterer, \emph{{Chasing the phantom: A closer look at Type Ia supernovae and the dark energy equation of state}}, \href{https://doi.org/10.1103/PhysRevD.89.063510}{\emph{Phys. Rev. D} {\bfseries 89} (2014) 063510} [\href{https://arxiv.org/abs/1312.1688}{{\ttfamily 1312.1688}}].

\bibitem{Zhang:2015uhk}
X.~Zhang, \emph{{Impacts of dark energy on weighing neutrinos after Planck 2015}}, \href{https://doi.org/10.1103/PhysRevD.93.083011}{\emph{Phys. Rev. D} {\bfseries 93} (2016) 083011} [\href{https://arxiv.org/abs/1511.02651}{{\ttfamily 1511.02651}}].

\bibitem{Xu:2016grp}
Y.-Y.~Xu and X.~Zhang, \emph{{Comparison of dark energy models after Planck 2015}}, \href{https://doi.org/10.1140/epjc/s10052-016-4446-5}{\emph{Eur. Phys. J. C} {\bfseries 76} (2016) 588} [\href{https://arxiv.org/abs/1607.06262}{{\ttfamily 1607.06262}}].

\bibitem{Wang:2016tsz}
S.~Wang, Y.-F.~Wang, D.-M.~Xia and X.~Zhang, \emph{{Impacts of dark energy on weighing neutrinos: mass hierarchies considered}}, \href{https://doi.org/10.1103/PhysRevD.94.083519}{\emph{Phys. Rev. D} {\bfseries 94} (2016) 083519} [\href{https://arxiv.org/abs/1608.00672}{{\ttfamily 1608.00672}}].

\bibitem{Vagnozzi:2017ovm}
S.~Vagnozzi, E.~Giusarma, O.~Mena, K.~Freese, M.~Gerbino, S.~Ho et~al., \emph{{Unveiling $\nu$ secrets with cosmological data: neutrino masses and mass hierarchy}}, \href{https://doi.org/10.1103/PhysRevD.96.123503}{\emph{Phys. Rev. D} {\bfseries 96} (2017) 123503} [\href{https://arxiv.org/abs/1701.08172}{{\ttfamily 1701.08172}}].

\bibitem{Zhang:2017rbg}
X.~Zhang, \emph{{Weighing neutrinos in dynamical dark energy models}}, \href{https://doi.org/10.1007/s11433-017-9025-7}{\emph{Sci. China Phys. Mech. Astron.} {\bfseries 60} (2017) 060431} [\href{https://arxiv.org/abs/1703.00651}{{\ttfamily 1703.00651}}].

\bibitem{Feng:2017mfs}
L.~Feng, J.-F.~Zhang and X.~Zhang, \emph{{Searching for sterile neutrinos in dynamical dark energy cosmologies}}, \href{https://doi.org/10.1007/s11433-017-9150-3}{\emph{Sci. China Phys. Mech. Astron.} {\bfseries 61} (2018) 050411} [\href{https://arxiv.org/abs/1706.06913}{{\ttfamily 1706.06913}}].

\bibitem{Wang:2018ahw}
D.~Wang, \emph{{Dark Energy Constraints in light of Pantheon SNe Ia, BAO, Cosmic Chronometers and CMB Polarization and Lensing Data}}, \href{https://doi.org/10.1103/PhysRevD.97.123507}{\emph{Phys. Rev. D} {\bfseries 97} (2018) 123507} [\href{https://arxiv.org/abs/1801.02371}{{\ttfamily 1801.02371}}].

\bibitem{Sprenger:2018tdb}
T.~Sprenger, M.~Archidiacono, T.~Brinckmann, S.~Clesse and J.~Lesgourgues, \emph{{Cosmology in the era of Euclid and the Square Kilometre Array}}, \href{https://doi.org/10.1088/1475-7516/2019/02/047}{\emph{JCAP} {\bfseries 02} (2019) 047} [\href{https://arxiv.org/abs/1801.08331}{{\ttfamily 1801.08331}}].

\bibitem{Poulin:2018zxs}
V.~Poulin, K.K.~Boddy, S.~Bird and M.~Kamionkowski, \emph{{Implications of an extended dark energy cosmology with massive neutrinos for cosmological tensions}}, \href{https://doi.org/10.1103/PhysRevD.97.123504}{\emph{Phys. Rev. D} {\bfseries 97} (2018) 123504} [\href{https://arxiv.org/abs/1803.02474}{{\ttfamily 1803.02474}}].

\bibitem{RoyChoudhury:2018vnm}
S.~Roy~Choudhury and A.~Naskar, \emph{{Strong Bounds on Sum of Neutrino Masses in a 12 Parameter Extended Scenario with Non-Phantom Dynamical Dark Energy ($w(z)\geq -1$) with CPL parameterization}}, \href{https://doi.org/10.1140/epjc/s10052-019-6762-z}{\emph{Eur. Phys. J. C} {\bfseries 79} (2019) 262} [\href{https://arxiv.org/abs/1807.02860}{{\ttfamily 1807.02860}}].

\bibitem{DES:2018ufa}
{\scshape DES} collaboration, \emph{{Dark Energy Survey Year 1 Results: Constraints on Extended Cosmological Models from Galaxy Clustering and Weak Lensing}}, \href{https://doi.org/10.1103/PhysRevD.99.123505}{\emph{Phys. Rev. D} {\bfseries 99} (2019) 123505} [\href{https://arxiv.org/abs/1810.02499}{{\ttfamily 1810.02499}}].

\bibitem{Wang:2019acf}
D.~Wang, \emph{{Exploring new physics beyond the standard cosmology with Dark Energy Survey Year 1 Data}}, \href{https://doi.org/10.1016/j.dark.2021.100810}{\emph{Phys. Dark Univ.} {\bfseries 32} (2021) 100810} [\href{https://arxiv.org/abs/1904.00657}{{\ttfamily 1904.00657}}].

\bibitem{RoyChoudhury:2019hls}
S.~Roy~Choudhury and S.~Hannestad, \emph{{Updated results on neutrino mass and mass hierarchy from cosmology with Planck 2018 likelihoods}}, \href{https://doi.org/10.1088/1475-7516/2020/07/037}{\emph{JCAP} {\bfseries 07} (2020) 037} [\href{https://arxiv.org/abs/1907.12598}{{\ttfamily 1907.12598}}].

\bibitem{Chudaykin:2020ghx}
A.~Chudaykin, K.~Dolgikh and M.M.~Ivanov, \emph{{Constraints on the curvature of the Universe and dynamical dark energy from the Full-shape and BAO data}}, \href{https://doi.org/10.1103/PhysRevD.103.023507}{\emph{Phys. Rev. D} {\bfseries 103} (2021) 023507} [\href{https://arxiv.org/abs/2009.10106}{{\ttfamily 2009.10106}}].

\bibitem{DAmico:2020kxu}
G.~D'Amico, L.~Senatore and P.~Zhang, \emph{{Limits on $w$CDM from the EFTofLSS with the PyBird code}}, \href{https://doi.org/10.1088/1475-7516/2021/01/006}{\emph{JCAP} {\bfseries 01} (2021) 006} [\href{https://arxiv.org/abs/2003.07956}{{\ttfamily 2003.07956}}].

\bibitem{Vagnozzi:2020dfn}
S.~Vagnozzi, A.~Loeb and M.~Moresco, \emph{{Eppur \`e piatto? The Cosmic Chronometers Take on Spatial Curvature and Cosmic Concordance}}, \href{https://doi.org/10.3847/1538-4357/abd4df}{\emph{Astrophys. J.} {\bfseries 908} (2021) 84} [\href{https://arxiv.org/abs/2011.11645}{{\ttfamily 2011.11645}}].

\bibitem{Yang:2021flj}
W.~Yang, E.~Di~Valentino, S.~Pan, Y.~Wu and J.~Lu, \emph{{Dynamical dark energy after Planck CMB final release and $H_0$ tension}}, \href{https://doi.org/10.1093/mnras/staa3914}{\emph{Mon. Not. Roy. Astron. Soc.} {\bfseries 501} (2021) 5845} [\href{https://arxiv.org/abs/2101.02168}{{\ttfamily 2101.02168}}].

\bibitem{DiValentino:2020vnx}
E.~Di~Valentino, \emph{{A combined analysis of the $H_0$ late time direct measurements and the impact on the Dark Energy sector}}, \href{https://doi.org/10.1093/mnras/stab187}{\emph{Mon. Not. Roy. Astron. Soc.} {\bfseries 502} (2021) 2065} [\href{https://arxiv.org/abs/2011.00246}{{\ttfamily 2011.00246}}].

\bibitem{Brieden:2022lsd}
S.~Brieden, H.~Gil-Mar\'\i{}n and L.~Verde, \emph{{Model-agnostic interpretation of 10 billion years of cosmic evolution traced by BOSS and eBOSS data}}, \href{https://doi.org/10.1088/1475-7516/2022/08/024}{\emph{JCAP} {\bfseries 08} (2022) 024} [\href{https://arxiv.org/abs/2204.11868}{{\ttfamily 2204.11868}}].

\bibitem{Grillo:2020yvj}
C.~Grillo, P.~Rosati, S.H.~Suyu, G.B.~Caminha, A.~Mercurio and A.~Halkola, \emph{{On the accuracy of time-delay cosmography in the Frontier Fields Cluster MACS J1149.5+2223 with supernova Refsdal}}, \href{https://doi.org/10.3847/1538-4357/ab9a4c}{\emph{Astrophys. J.} {\bfseries 898} (2020) 87} [\href{https://arxiv.org/abs/2001.02232}{{\ttfamily 2001.02232}}].

\bibitem{Cao:2021cix}
S.~Cao, J.~Ryan and B.~Ratra, \emph{{Cosmological constraints from H\,ii starburst galaxy, quasar angular size, and other measurements}}, \href{https://doi.org/10.1093/mnras/stab3304}{\emph{Mon. Not. Roy. Astron. Soc.} {\bfseries 509} (2022) 4745} [\href{https://arxiv.org/abs/2109.01987}{{\ttfamily 2109.01987}}].

\bibitem{Zhang:2021yof}
M.~Zhang, B.~Wang, P.-J.~Wu, J.-Z.~Qi, Y.~Xu, J.-F.~Zhang et~al., \emph{{Prospects for Constraining Interacting Dark Energy Models with 21 cm Intensity Mapping Experiments}}, \href{https://doi.org/10.3847/1538-4357/ac0ef5}{\emph{Astrophys. J.} {\bfseries 918} (2021) 56} [\href{https://arxiv.org/abs/2102.03979}{{\ttfamily 2102.03979}}].

\bibitem{Colgain:2021pmf}
E.O.~Colg\'ain, M.M.~Sheikh-Jabbari and L.~Yin, \emph{{Can dark energy be dynamical?}}, \href{https://doi.org/10.1103/PhysRevD.104.023510}{\emph{Phys. Rev. D} {\bfseries 104} (2021) 023510} [\href{https://arxiv.org/abs/2104.01930}{{\ttfamily 2104.01930}}].

\bibitem{Teng:2021cvy}
Y.-P.~Teng, W.~Lee and K.-W.~Ng, \emph{{Constraining the dark-energy equation of state with cosmological data}}, \href{https://doi.org/10.1103/PhysRevD.104.083519}{\emph{Phys. Rev. D} {\bfseries 104} (2021) 083519} [\href{https://arxiv.org/abs/2105.02667}{{\ttfamily 2105.02667}}].

\bibitem{Krishnan:2021dyb}
C.~Krishnan, R.~Mohayaee, E.O.~Colg\'ain, M.M.~Sheikh-Jabbari and L.~Yin, \emph{{Does Hubble tension signal a breakdown in FLRW cosmology?}}, \href{https://doi.org/10.1088/1361-6382/ac1a81}{\emph{Class. Quant. Grav.} {\bfseries 38} (2021) 184001} [\href{https://arxiv.org/abs/2105.09790}{{\ttfamily 2105.09790}}].

\bibitem{Nunes:2021ipq}
R.C.~Nunes and S.~Vagnozzi, \emph{{Arbitrating the S8 discrepancy with growth rate measurements from redshift-space distortions}}, \href{https://doi.org/10.1093/mnras/stab1613}{\emph{Mon. Not. Roy. Astron. Soc.} {\bfseries 505} (2021) 5427} [\href{https://arxiv.org/abs/2106.01208}{{\ttfamily 2106.01208}}].

\bibitem{Bernardo:2021cxi}
R.C.~Bernardo, D.~Grand\'on, J.~Said~Levi and V.H.~C\'ardenas, \emph{{Parametric and nonparametric methods hint dark energy evolution}}, \href{https://doi.org/10.1016/j.dark.2022.101017}{\emph{Phys. Dark Univ.} {\bfseries 36} (2022) 101017} [\href{https://arxiv.org/abs/2111.08289}{{\ttfamily 2111.08289}}].

\bibitem{Bargiacchi:2021hdp}
G.~Bargiacchi, M.~Benetti, S.~Capozziello, E.~Lusso, G.~Risaliti and M.~Signorini, \emph{{Quasar cosmology: dark energy evolution and spatial curvature}}, \href{https://doi.org/10.1093/mnras/stac1941}{\emph{Mon. Not. Roy. Astron. Soc.} {\bfseries 515} (2022) 1795} [\href{https://arxiv.org/abs/2111.02420}{{\ttfamily 2111.02420}}].

\bibitem{Semenaite:2022unt}
A.~Semenaite, A.G.~S\'anchez, A.~Pezzotta, J.~Hou, A.~Eggemeier, M.~Crocce et~al., \emph{{Beyond \textendash{} \ensuremath{\Lambda}CDM constraints from the full shape clustering measurements from BOSS and eBOSS}}, \href{https://doi.org/10.1093/mnras/stad849}{\emph{Mon. Not. Roy. Astron. Soc.} {\bfseries 521} (2023) 5013} [\href{https://arxiv.org/abs/2210.07304}{{\ttfamily 2210.07304}}].

\bibitem{Carrilho:2022mon}
P.~Carrilho, C.~Moretti and A.~Pourtsidou, \emph{{Cosmology with the EFTofLSS and BOSS: dark energy constraints and a note on priors}}, \href{https://doi.org/10.1088/1475-7516/2023/01/028}{\emph{JCAP} {\bfseries 01} (2023) 028} [\href{https://arxiv.org/abs/2207.14784}{{\ttfamily 2207.14784}}].

\bibitem{Wang:2022xdw}
D.~Wang, \emph{{Pantheon+ constraints on dark energy and modified gravity: An evidence of dynamical dark energy}}, \href{https://doi.org/10.1103/PhysRevD.106.063515}{\emph{Phys. Rev. D} {\bfseries 106} (2022) 063515} [\href{https://arxiv.org/abs/2207.07164}{{\ttfamily 2207.07164}}].

\bibitem{Koussour:2022jyk}
M.~Koussour, S.K.J.~Pacif, M.~Bennai and P.K.~Sahoo, \emph{{A New Parametrization of Hubble Parameter in f(Q)$f(Q)$ Gravity}}, \href{https://doi.org/10.1002/prop.202200172}{\emph{Fortsch. Phys.} {\bfseries 71} (2023) 2200172} [\href{https://arxiv.org/abs/2208.04723}{{\ttfamily 2208.04723}}].

\bibitem{Bernardo:2022pyz}
R.C.~Bernardo, D.~Grand\'on, J.~Levi~Said and V.H.~C\'ardenas, \emph{{Dark energy by natural evolution: Constraining dark energy using Approximate Bayesian Computation}}, \href{https://doi.org/10.1016/j.dark.2023.101213}{\emph{Phys. Dark Univ.} {\bfseries 40} (2023) 101213} [\href{https://arxiv.org/abs/2211.05482}{{\ttfamily 2211.05482}}].

\bibitem{Narawade:2022cgb}
S.A.~Narawade and B.~Mishra, \emph{{Phantom Cosmological Model with Observational Constraints in f(Q)$f(Q)$ Gravity}}, \href{https://doi.org/10.1002/andp.202200626}{\emph{Annalen Phys.} {\bfseries 535} (2023) 2200626} [\href{https://arxiv.org/abs/2211.09701}{{\ttfamily 2211.09701}}].

\bibitem{Hou:2022rvk}
W.-T.~Hou, J.-Z.~Qi, T.~Han, J.-F.~Zhang, S.~Cao and X.~Zhang, \emph{{Prospects for constraining interacting dark energy models from gravitational wave and gamma ray burst joint observation}}, \href{https://doi.org/10.1088/1475-7516/2023/05/017}{\emph{JCAP} {\bfseries 05} (2023) 017} [\href{https://arxiv.org/abs/2211.10087}{{\ttfamily 2211.10087}}].

\bibitem{Kumar:2023bqj}
S.~Kumar, R.C.~Nunes, S.~Pan and P.~Yadav, \emph{{New late-time constraints on f(R) gravity}}, \href{https://doi.org/10.1016/j.dark.2023.101281}{\emph{Phys. Dark Univ.} {\bfseries 42} (2023) 101281} [\href{https://arxiv.org/abs/2301.07897}{{\ttfamily 2301.07897}}].

\bibitem{Bhagat:2023ych}
R.~Bhagat, S.A.~Narawade and B.~Mishra, \emph{{Weyl type f(Q,T) gravity observational constrained cosmological models}}, \href{https://doi.org/10.1016/j.dark.2023.101250}{\emph{Phys. Dark Univ.} {\bfseries 41} (2023) 101250} [\href{https://arxiv.org/abs/2305.01659}{{\ttfamily 2305.01659}}].

\bibitem{Mussatayeva:2023aoa}
A.~Mussatayeva, N.~Myrzakulov and M.~Koussour, \emph{{Cosmological constraints on dark energy in f(Q) gravity: A parametrized perspective}}, \href{https://doi.org/10.1016/j.dark.2023.101276}{\emph{Phys. Dark Univ.} {\bfseries 42} (2023) 101276} [\href{https://arxiv.org/abs/2307.00281}{{\ttfamily 2307.00281}}].

\bibitem{Hogg:2023khs}
N.B.~Hogg, \emph{{Constraints on dark energy from TDCOSMO~\& SLACS lenses}}, \href{https://doi.org/10.1093/mnrasl/slae005}{\emph{Mon. Not. Roy. Astron. Soc.} {\bfseries 528} (2024) L95} [\href{https://arxiv.org/abs/2310.11977}{{\ttfamily 2310.11977}}].

\bibitem{DESI:2024mwx}
{\scshape DESI} collaboration, \emph{{DESI 2024 VI: Cosmological Constraints from the Measurements of Baryon Acoustic Oscillations}},  \href{https://arxiv.org/abs/2404.03002}{{\ttfamily 2404.03002}}.

\bibitem{Verde:2019ivm}
L.~Verde, T.~Treu and A.G.~Riess, \emph{{Tensions between the Early and the Late Universe}}, \href{https://doi.org/10.1038/s41550-019-0902-0}{\emph{Nature Astron.} {\bfseries 3} (2019) 891} [\href{https://arxiv.org/abs/1907.10625}{{\ttfamily 1907.10625}}].

\bibitem{Hu:2023jqc}
J.-P.~Hu and F.-Y.~Wang, \emph{{Hubble Tension: The Evidence of New Physics}}, \href{https://doi.org/10.3390/universe9020094}{\emph{Universe} {\bfseries 9} (2023) 94} [\href{https://arxiv.org/abs/2302.05709}{{\ttfamily 2302.05709}}].

\bibitem{Giare:2024akf}
W.~Giar\`e, \emph{{Inflation, the Hubble tension, and early dark energy: An alternative overview}}, \href{https://doi.org/10.1103/PhysRevD.109.123545}{\emph{Phys. Rev. D} {\bfseries 109} (2024) 123545} [\href{https://arxiv.org/abs/2404.12779}{{\ttfamily 2404.12779}}].

\bibitem{Giare:2023xoc}
W.~Giar\`e, \emph{{CMB Anomalies and the Hubble Tension}},  \href{https://arxiv.org/abs/2305.16919}{{\ttfamily 2305.16919}}.

\bibitem{Mainini:2003uf}
R.~Mainini, A.V.~Maccio, S.A.~Bonometto and A.~Klypin, \emph{{Modeling dynamical dark energy}}, \href{https://doi.org/10.1086/379236}{\emph{Astrophys. J.} {\bfseries 599} (2003) 24} [\href{https://arxiv.org/abs/astro-ph/0303303}{{\ttfamily astro-ph/0303303}}].

\bibitem{Alam:2004jy}
U.~Alam, V.~Sahni and A.A.~Starobinsky, \emph{{The Case for dynamical dark energy revisited}}, \href{https://doi.org/10.1088/1475-7516/2004/06/008}{\emph{JCAP} {\bfseries 06} (2004) 008} [\href{https://arxiv.org/abs/astro-ph/0403687}{{\ttfamily astro-ph/0403687}}].

\bibitem{Sola:2005nh}
J.~Sola and H.~Stefancic, \emph{{Dynamical dark energy or variable cosmological parameters?}}, \href{https://doi.org/10.1142/S0217732306019554}{\emph{Mod. Phys. Lett. A} {\bfseries 21} (2006) 479} [\href{https://arxiv.org/abs/astro-ph/0507110}{{\ttfamily astro-ph/0507110}}].

\bibitem{Antoniadis:2006wq}
I.~Antoniadis, P.O.~Mazur and E.~Mottola, \emph{{Cosmological dark energy: Prospects for a dynamical theory}}, \href{https://doi.org/10.1088/1367-2630/9/1/011}{\emph{New J. Phys.} {\bfseries 9} (2007) 11} [\href{https://arxiv.org/abs/gr-qc/0612068}{{\ttfamily gr-qc/0612068}}].

\bibitem{Zhao:2012aw}
G.-B.~Zhao, R.G.~Crittenden, L.~Pogosian and X.~Zhang, \emph{{Examining the evidence for dynamical dark energy}}, \href{https://doi.org/10.1103/PhysRevLett.109.171301}{\emph{Phys. Rev. Lett.} {\bfseries 109} (2012) 171301} [\href{https://arxiv.org/abs/1207.3804}{{\ttfamily 1207.3804}}].

\bibitem{SolaPeracaula:2016qlq}
J.~Sol\`a~Peracaula, J.~de~Cruz~P\'erez and A.~G\'omez-Valent, \emph{{Dynamical dark energy vs. $\Lambda$ = const in light of observations}}, \href{https://doi.org/10.1209/0295-5075/121/39001}{\emph{EPL} {\bfseries 121} (2018) 39001} [\href{https://arxiv.org/abs/1606.00450}{{\ttfamily 1606.00450}}].

\bibitem{Sola:2016hnq}
J.~Sola, A.~Gomez-Valent and J.~de~Cruz~P\'erez, \emph{{Dynamical dark energy: scalar fields and running vacuum}}, \href{https://doi.org/10.1142/S0217732317500547}{\emph{Mod. Phys. Lett. A} {\bfseries 32} (2017) 1750054} [\href{https://arxiv.org/abs/1610.08965}{{\ttfamily 1610.08965}}].

\bibitem{Zhao:2017cud}
G.-B.~Zhao et~al., \emph{{Dynamical dark energy in light of the latest observations}}, \href{https://doi.org/10.1038/s41550-017-0216-z}{\emph{Nature Astron.} {\bfseries 1} (2017) 627} [\href{https://arxiv.org/abs/1701.08165}{{\ttfamily 1701.08165}}].

\bibitem{Yang:2017yme}
W.~Yang, N.~Banerjee and S.~Pan, \emph{{Constraining a dark matter and dark energy interaction scenario with a dynamical equation of state}}, \href{https://doi.org/10.1103/PhysRevD.95.123527}{\emph{Phys. Rev. D} {\bfseries 95} (2017) 123527} [\href{https://arxiv.org/abs/1705.09278}{{\ttfamily 1705.09278}}].

\bibitem{SolaPeracaula:2018wwm}
J.~Sola~Peracaula, A.~Gomez-Valent and J.~de~Cruz~P\'erez, \emph{{Signs of Dynamical Dark Energy in Current Observations}}, \href{https://doi.org/10.1016/j.dark.2019.100311}{\emph{Phys. Dark Univ.} {\bfseries 25} (2019) 100311} [\href{https://arxiv.org/abs/1811.03505}{{\ttfamily 1811.03505}}].

\bibitem{Pan:2019gop}
S.~Pan, W.~Yang, E.~Di~Valentino, E.N.~Saridakis and S.~Chakraborty, \emph{{Interacting scenarios with dynamical dark energy: Observational constraints and alleviation of the $H_0$ tension}}, \href{https://doi.org/10.1103/PhysRevD.100.103520}{\emph{Phys. Rev. D} {\bfseries 100} (2019) 103520} [\href{https://arxiv.org/abs/1907.07540}{{\ttfamily 1907.07540}}].

\bibitem{Escamilla-Rivera:2021boq}
C.~Escamilla-Rivera and A.~N\'ajera, \emph{{Dynamical dark energy models in the light of gravitational-wave transient catalogues}}, \href{https://doi.org/10.1088/1475-7516/2022/03/060}{\emph{JCAP} {\bfseries 03} (2022) 060} [\href{https://arxiv.org/abs/2103.02097}{{\ttfamily 2103.02097}}].

\bibitem{Zhao:2020ole}
Z.-W.~Zhao, Z.-X.~Li, J.-Z.~Qi, H.~Gao, J.-F.~Zhang and X.~Zhang, \emph{{Cosmological parameter estimation for dynamical dark energy models with future fast radio burst observations}}, \href{https://doi.org/10.3847/1538-4357/abb8ce}{\emph{Astrophys. J.} {\bfseries 903} (2020) 83} [\href{https://arxiv.org/abs/2006.01450}{{\ttfamily 2006.01450}}].

\bibitem{DiValentino:2017zyq}
E.~Di~Valentino, A.~Melchiorri, E.V.~Linder and J.~Silk, \emph{{Constraining Dark Energy Dynamics in Extended Parameter Space}}, \href{https://doi.org/10.1103/PhysRevD.96.023523}{\emph{Phys. Rev. D} {\bfseries 96} (2017) 023523} [\href{https://arxiv.org/abs/1704.00762}{{\ttfamily 1704.00762}}].

\bibitem{DiValentino:2019dzu}
E.~Di~Valentino, A.~Melchiorri and J.~Silk, \emph{{Cosmological constraints in extended parameter space from the Planck 2018 Legacy release}}, \href{https://doi.org/10.1088/1475-7516/2020/01/013}{\emph{JCAP} {\bfseries 01} (2020) 013} [\href{https://arxiv.org/abs/1908.01391}{{\ttfamily 1908.01391}}].

\bibitem{DiValentino:2022oon}
E.~Di~Valentino, W.~Giar\`e, A.~Melchiorri and J.~Silk, \emph{{Health checkup test of the standard cosmological model in view of recent cosmic microwave background anisotropies experiments}}, \href{https://doi.org/10.1103/PhysRevD.106.103506}{\emph{Phys. Rev. D} {\bfseries 106} (2022) 103506} [\href{https://arxiv.org/abs/2209.12872}{{\ttfamily 2209.12872}}].

\bibitem{Rubin:2023ovl}
D.~Rubin et~al., \emph{{Union Through UNITY: Cosmology with 2,000 SNe Using a Unified Bayesian Framework}},  \href{https://arxiv.org/abs/2311.12098}{{\ttfamily 2311.12098}}.

\bibitem{Scolnic:2021amr}
D.~Scolnic et~al., \emph{{The Pantheon+ Analysis: The Full Data Set and Light-curve Release}}, \href{https://doi.org/10.3847/1538-4357/ac8b7a}{\emph{Astrophys. J.} {\bfseries 938} (2022) 113} [\href{https://arxiv.org/abs/2112.03863}{{\ttfamily 2112.03863}}].

\bibitem{Cortes:2024lgw}
M.~Cort\^es and A.R.~Liddle, \emph{{Interpreting DESI's evidence for evolving dark energy}},  \href{https://arxiv.org/abs/2404.08056}{{\ttfamily 2404.08056}}.

\bibitem{Patel:2024odo}
V.~Patel and L.~Amendola, \emph{{Comments on the prior dependence of the DESI results}},  \href{https://arxiv.org/abs/2407.06586}{{\ttfamily 2407.06586}}.

\bibitem{Orchard:2024bve}
L.~Orchard and V.H.~C\'ardenas, \emph{{Probing Dark Energy Evolution Post-DESI 2024}},  \href{https://arxiv.org/abs/2407.05579}{{\ttfamily 2407.05579}}.

\bibitem{Liu:2024gfy}
G.~Liu, Y.~Wang and W.~Zhao, \emph{{Impact of LRG1 and LRG2 in DESI 2024 BAO data on dark energy evolution}},  \href{https://arxiv.org/abs/2407.04385}{{\ttfamily 2407.04385}}.

\bibitem{Chudaykin:2024gol}
A.~Chudaykin and M.~Kunz, \emph{{Modified gravity interpretation of the evolving dark energy in light of DESI data}},  \href{https://arxiv.org/abs/2407.02558}{{\ttfamily 2407.02558}}.

\bibitem{Notari:2024rti}
A.~Notari, M.~Redi and A.~Tesi, \emph{{Consistent Theories for the DESI dark energy fit}},  \href{https://arxiv.org/abs/2406.08459}{{\ttfamily 2406.08459}}.

\bibitem{Gialamas:2024lyw}
I.D.~Gialamas, G.~H\"utsi, K.~Kannike, A.~Racioppi, M.~Raidal, M.~Vasar et~al., \emph{{Interpreting DESI 2024 BAO: late-time dynamical dark energy or a local effect?}},  \href{https://arxiv.org/abs/2406.07533}{{\ttfamily 2406.07533}}.

\bibitem{Wang:2024hwd}
H.~Wang, Z.-Y.~Peng and Y.-S.~Piao, \emph{{Can recent DESI BAO measurements accommodate a negative cosmological constant?}},  \href{https://arxiv.org/abs/2406.03395}{{\ttfamily 2406.03395}}.

\bibitem{Wang:2024dka}
H.~Wang and Y.-S.~Piao, \emph{{Dark energy in light of recent DESI BAO and Hubble tension}},  \href{https://arxiv.org/abs/2404.18579}{{\ttfamily 2404.18579}}.

\bibitem{Carloni:2024zpl}
Y.~Carloni, O.~Luongo and M.~Muccino, \emph{{Does dark energy really revive using DESI 2024 data?}},  \href{https://arxiv.org/abs/2404.12068}{{\ttfamily 2404.12068}}.

\bibitem{Colgain:2024xqj}
E.O.~Colg\'ain, M.G.~Dainotti, S.~Capozziello, S.~Pourojaghi, M.M.~Sheikh-Jabbari and D.~Stojkovic, \emph{{Does DESI 2024 Confirm $\Lambda$CDM?}},  \href{https://arxiv.org/abs/2404.08633}{{\ttfamily 2404.08633}}.

\bibitem{Tada:2024znt}
Y.~Tada and T.~Terada, \emph{{Quintessential interpretation of the evolving dark energy in light of DESI observations}}, \href{https://doi.org/10.1103/PhysRevD.109.L121305}{\emph{Phys. Rev. D} {\bfseries 109} (2024) L121305} [\href{https://arxiv.org/abs/2404.05722}{{\ttfamily 2404.05722}}].

\bibitem{Yin:2024hba}
W.~Yin, \emph{{Cosmic clues: DESI, dark energy, and the cosmological constant problem}}, \href{https://doi.org/10.1007/JHEP05(2024)327}{\emph{JHEP} {\bfseries 05} (2024) 327} [\href{https://arxiv.org/abs/2404.06444}{{\ttfamily 2404.06444}}].

\bibitem{Luongo:2024fww}
O.~Luongo and M.~Muccino, \emph{{Model independent cosmographic constraints from DESI 2024}},  \href{https://arxiv.org/abs/2404.07070}{{\ttfamily 2404.07070}}.

\bibitem{Park:2024jns}
C.-G.~Park, J.~de~Cruz~Perez and B.~Ratra, \emph{{Using non-DESI data to confirm and strengthen the DESI 2024 spatially-flat $w_0w_a$CDM cosmological parameterization result}},  \href{https://arxiv.org/abs/2405.00502}{{\ttfamily 2405.00502}}.

\bibitem{Shlivko:2024llw}
D.~Shlivko and P.J.~Steinhardt, \emph{{Assessing observational constraints on dark energy}}, \href{https://doi.org/10.1016/j.physletb.2024.138826}{\emph{Phys. Lett. B} {\bfseries 855} (2024) 138826} [\href{https://arxiv.org/abs/2405.03933}{{\ttfamily 2405.03933}}].

\bibitem{Ye:2024ywg}
G.~Ye, M.~Martinelli, B.~Hu and A.~Silvestri, \emph{{Non-minimally coupled gravity as a physically viable fit to DESI 2024 BAO}},  \href{https://arxiv.org/abs/2407.15832}{{\ttfamily 2407.15832}}.

\bibitem{Wang:2024pui}
Z.~Wang, S.~Lin, Z.~Ding and B.~Hu, \emph{{The role of LRG1 and LRG2's monopole in inferring the DESI 2024 BAO cosmology}},  \href{https://arxiv.org/abs/2405.02168}{{\ttfamily 2405.02168}}.

\bibitem{Naredo-Tuero:2024sgf}
D.~Naredo-Tuero, M.~Escudero, E.~Fern\'andez-Mart\'\i{}nez, X.~Marcano and V.~Poulin, \emph{{Living at the Edge: A Critical Look at the Cosmological Neutrino Mass Bound}},  \href{https://arxiv.org/abs/2407.13831}{{\ttfamily 2407.13831}}.

\bibitem{dePutter:2008wt}
R.~de~Putter and E.V.~Linder, \emph{{Calibrating Dark Energy}}, \href{https://doi.org/10.1088/1475-7516/2008/10/042}{\emph{JCAP} {\bfseries 10} (2008) 042} [\href{https://arxiv.org/abs/0808.0189}{{\ttfamily 0808.0189}}].

\bibitem{Hernandez-Almada:2024ost}
A.~Hern\'andez-Almada, M.L.~Mendoza-Mart\'\i{}nez, M.A.~Garc\'\i{}a-Aspeitia and V.~Motta, \emph{{Phenomenological emergent dark energy in the light of DESI Data Release 1}},  \href{https://arxiv.org/abs/2407.09430}{{\ttfamily 2407.09430}}.

\bibitem{Pourojaghi:2024tmw}
S.~Pourojaghi, M.~Malekjani and Z.~Davari, \emph{{Cosmological constraints on dark energy parametrizations after DESI 2024: Persistent deviation from standard $\Lambda$CDM cosmology}},  \href{https://arxiv.org/abs/2407.09767}{{\ttfamily 2407.09767}}.

\bibitem{Ramadan:2024kmn}
O.F.~Ramadan, J.~Sakstein and D.~Rubin, \emph{{DESI Constraints on Exponential Quintessence}},  \href{https://arxiv.org/abs/2405.18747}{{\ttfamily 2405.18747}}.

\bibitem{Berghaus:2024kra}
K.V.~Berghaus, J.A.~Kable and V.~Miranda, \emph{{Quantifying Scalar Field Dynamics with DESI 2024 Y1 BAO measurements}},  \href{https://arxiv.org/abs/2404.14341}{{\ttfamily 2404.14341}}.

\bibitem{Qu:2024lpx}
F.J.~Qu, K.M.~Surrao, B.~Bolliet, J.C.~Hill, B.D.~Sherwin and H.T.~Jense, \emph{{Accelerated inference on accelerated cosmic expansion: New constraints on axion-like early dark energy with DESI BAO and ACT DR6 CMB lensing}},  \href{https://arxiv.org/abs/2404.16805}{{\ttfamily 2404.16805}}.

\bibitem{Adolf:2024twn}
P.~Adolf, M.~Hirsch, S.~Krieg, H.~P\"as and M.~Tabet, \emph{{Fitting the DESI BAO Data with Dark Energy Driven by the Cohen--Kaplan--Nelson Bound}},  \href{https://arxiv.org/abs/2406.09964}{{\ttfamily 2406.09964}}.

\bibitem{Ma:1995ey}
C.-P.~Ma and E.~Bertschinger, \emph{{Cosmological perturbation theory in the synchronous and conformal Newtonian gauges}}, \href{https://doi.org/10.1086/176550}{\emph{Astrophys. J.} {\bfseries 455} (1995) 7} [\href{https://arxiv.org/abs/astro-ph/9506072}{{\ttfamily astro-ph/9506072}}].

\bibitem{Dimakis:2016mip}
N.~Dimakis, A.~Karagiorgos, A.~Zampeli, A.~Paliathanasis, T.~Christodoulakis and P.A.~Terzis, \emph{{General Analytic Solutions of Scalar Field Cosmology with Arbitrary Potential}}, \href{https://doi.org/10.1103/PhysRevD.93.123518}{\emph{Phys. Rev. D} {\bfseries 93} (2016) 123518} [\href{https://arxiv.org/abs/1604.05168}{{\ttfamily 1604.05168}}].

\bibitem{Pan:2019brc}
S.~Pan, W.~Yang and A.~Paliathanasis, \emph{{Imprints of an extended Chevallier\textendash{}Polarski\textendash{}Linder parametrization on the large scale of our universe}}, \href{https://doi.org/10.1140/epjc/s10052-020-7832-y}{\emph{Eur. Phys. J. C} {\bfseries 80} (2020) 274} [\href{https://arxiv.org/abs/1902.07108}{{\ttfamily 1902.07108}}].

\bibitem{Lewis:1999bs}
A.~Lewis, A.~Challinor and A.~Lasenby, \emph{{Efficient computation of CMB anisotropies in closed FRW models}}, \href{https://doi.org/10.1086/309179}{\emph{Astrophys. J.} {\bfseries 538} (2000) 473} [\href{https://arxiv.org/abs/astro-ph/9911177}{{\ttfamily astro-ph/9911177}}].

\bibitem{Howlett:2012mh}
C.~Howlett, A.~Lewis, A.~Hall and A.~Challinor, \emph{{CMB power spectrum parameter degeneracies in the era of precision cosmology}}, \href{https://doi.org/10.1088/1475-7516/2012/04/027}{\emph{JCAP} {\bfseries 04} (2012) 027} [\href{https://arxiv.org/abs/1201.3654}{{\ttfamily 1201.3654}}].

\bibitem{Lewis:2002ah}
A.~Lewis and S.~Bridle, \emph{{Cosmological parameters from CMB and other data: A Monte Carlo approach}}, \href{https://doi.org/10.1103/PhysRevD.66.103511}{\emph{Phys. Rev. D} {\bfseries 66} (2002) 103511} [\href{https://arxiv.org/abs/astro-ph/0205436}{{\ttfamily astro-ph/0205436}}].

\bibitem{Lewis:2013hha}
A.~Lewis, \emph{{Efficient sampling of fast and slow cosmological parameters}}, \href{https://doi.org/10.1103/PhysRevD.87.103529}{\emph{Phys. Rev. D} {\bfseries 87} (2013) 103529} [\href{https://arxiv.org/abs/1304.4473}{{\ttfamily 1304.4473}}].

\bibitem{Neal:2005}
R.M.~{Neal}, \emph{{Taking Bigger Metropolis Steps by Dragging Fast Variables}}, {\emph{ArXiv Mathematics e-prints} (2005) } [\href{https://arxiv.org/abs/math/0502099}{{\ttfamily math/0502099}}].

\bibitem{Gelman:1992zz}
A.~Gelman and D.B.~Rubin, \emph{{Inference from Iterative Simulation Using Multiple Sequences}}, \href{https://doi.org/10.1214/ss/1177011136}{\emph{Statist. Sci.} {\bfseries 7} (1992) 457}.

\bibitem{Planck:2019nip}
{\scshape Planck} collaboration, \emph{{Planck 2018 results. V. CMB power spectra and likelihoods}}, \href{https://doi.org/10.1051/0004-6361/201936386}{\emph{Astron. Astrophys.} {\bfseries 641} (2020) A5} [\href{https://arxiv.org/abs/1907.12875}{{\ttfamily 1907.12875}}].

\bibitem{Carron:2022eyg}
J.~Carron, M.~Mirmelstein and A.~Lewis, \emph{{CMB lensing from Planck PR4~maps}}, \href{https://doi.org/10.1088/1475-7516/2022/09/039}{\emph{JCAP} {\bfseries 09} (2022) 039} [\href{https://arxiv.org/abs/2206.07773}{{\ttfamily 2206.07773}}].

\bibitem{DESI:2024lzq}
{\scshape DESI} collaboration, \emph{{DESI 2024 IV: Baryon Acoustic Oscillations from the Lyman Alpha Forest}},  \href{https://arxiv.org/abs/2404.03001}{{\ttfamily 2404.03001}}.

\bibitem{Huterer:2000mj}
D.~Huterer and M.S.~Turner, \emph{{Probing the dark energy: Methods and strategies}}, \href{https://doi.org/10.1103/PhysRevD.64.123527}{\emph{Phys. Rev. D} {\bfseries 64} (2001) 123527} [\href{https://arxiv.org/abs/astro-ph/0012510}{{\ttfamily astro-ph/0012510}}].

\bibitem{Albrecht:2006um}
A.~Albrecht et~al., \emph{{Report of the Dark Energy Task Force}},  \href{https://arxiv.org/abs/astro-ph/0609591}{{\ttfamily astro-ph/0609591}}.

\end{thebibliography}\endgroup

\end{document}